\documentclass[a4paper,10pt]{article}

\usepackage{fontspec}
\usepackage[left=2.25cm,right=2.25cm,top=2.25cm,bottom=2.35cm]{geometry}
\usepackage{microtype}
\usepackage{setspace}
\usepackage{indentfirst}
\usepackage{amsmath,amssymb,bm}
\usepackage{graphicx}
\usepackage{xcolor}
\usepackage{booktabs,longtable,array,tabularx,multirow,makecell,threeparttable}
\usepackage{calc}
\usepackage{adjustbox}
\usepackage{float}
\usepackage[section]{placeins}
\usepackage{caption}
\usepackage{subcaption}
\usepackage{enumitem}
\usepackage{etoolbox}
\usepackage{lineno}
\usepackage{xurl}
\usepackage[hyperfootnotes=false]{hyperref}
\usepackage{bookmark}
\usepackage{fancyhdr}
\usepackage{titlesec}
\usepackage{chngcntr}

\definecolor{SciBlue}{HTML}{24527A}
\definecolor{SciGray}{HTML}{5F6B7A}
\hypersetup{unicode=true,colorlinks=true,linkcolor=SciBlue,citecolor=SciBlue,urlcolor=SciBlue,pdfborder={0 0 0}}
\renewcommand{\arraystretch}{1.16}
\AtBeginEnvironment{longtable}{\footnotesize\setlength{\tabcolsep}{3pt}}
\setlist{nosep,leftmargin=2em}
\counterwithin{figure}{section}
\counterwithin{table}{section}

\titleformat{\section}{\large\bfseries\color{SciBlue}}{\thesection.}{0.6em}{}
\titleformat{\subsection}{\normalsize\bfseries}{\thesubsection}{0.6em}{}
\titleformat{\subsubsection}{\normalsize\bfseries}{\thesubsubsection}{0.6em}{}
\titlespacing*{\section}{0pt}{2.0ex plus .4ex}{.9ex}
\titlespacing*{\subsection}{0pt}{1.5ex plus .3ex}{.6ex}
\titlespacing*{\subsubsection}{0pt}{1.2ex plus .2ex}{.4ex}
\pretocmd{\subsection}{\FloatBarrier}{}{}
\pretocmd{\subsubsection}{\FloatBarrier}{}{}
\newcommand{\tightlist}{\setlength{\itemsep}{0pt}\setlength{\parskip}{0pt}}
\newif\ifreviewcopy
\reviewcopyfalse 
\newcommand{\PaperTitle}{Joint Effects of GPU Server Topology, Parallelism, and Congestion Control on MoE Inference: A Controlled Simulation Study}
\newcommand{\PaperAuthors}{Kaikai Yuan\textsuperscript{1}, Rui Xi\textsuperscript{1}, and Yu Liu\textsuperscript{1,*}}
\newcommand{\PaperAuthorsMetadata}{Kaikai Yuan; Rui Xi; Yu Liu}
\newcommand{\PaperAffiliations}{\textsuperscript{1}\textit{Inspur Group, Jinan, Shandong, China.}}
\newcommand{\CorrespondingAuthorFootnote}{Corresponding author: Yu Liu, \href{mailto:liuyubj@inspur.com}{liuyubj@inspur.com}}

\AtBeginDocument{\hypersetup{pdftitle={\PaperTitle},pdfauthor={\PaperAuthorsMetadata},pdfsubject={MoE inference topology, parallelism, and congestion-control simulation}}}

\begin{document}
\begin{titlepage}
  \centering
  \vspace{1.6cm}
  {\LARGE\bfseries\PaperTitle\par}
  \vspace{1.3cm}
  {\large\PaperAuthors\par}
  \vspace{0.45cm}
  {\normalsize\PaperAffiliations\par}
  \begingroup
    \renewcommand{\thefootnote}{*}
    \footnotetext{\CorrespondingAuthorFootnote}
  \endgroup
  \vfill
  \begin{minipage}{0.92\textwidth}
    \noindent\textbf{Abstract}\par\smallskip
    Mixture-of-experts (MoE) models expand capacity through sparse activation, but inference across multiple graphics processing units (GPUs) introduces tensor-parallel (TP) collectives and expert-parallel (EP) dispatch and combine operations. Completion time depends not only on communication volume but also on how logical groups map onto intra-server interconnects, GPU--network-interface-card (NIC) connections, and the inter-node network. Using ASTRA-sim with the NS-3 discrete-event network backend, we construct a controlled matrix with 32 GPU ranks and data and pipeline parallelism fixed at one. The workloads are fixed-length, 4096-token, prefill-like synthetic Chakra execution traces generated from four MoE configurations. The experiments cover six server topologies, four TP/EP partitions, two TP collective algorithms, and four network/congestion-control modes, yielding 768 deterministic simulations. In the 144-configuration feedback-enabled subset for each model, exposed communication accounts for 89.9\%--95.8\% of mean completion time. TP16EP2 requires 3.68--4.35$\times$ the mean completion time of TP2EP16. Under the evaluated ASTRA-sim Double Binary Tree (DBT) semantics and fixed rank mapping, DBT/Direct incurs 28.3\%--83.2\% more time than Ring/Direct. After excluding the topology without an external network and equally aggregating the per-model conditional means, InfiniBand-like High Precision Congestion Control (IB-like HPCC) is approximately 0.9\% lower than HPCC over RDMA over Converged Ethernet (RoCE), whereas RoCE with Data Center Quantized Congestion Notification (DCQCN) is 23.8\%--35.7\% slower than RoCE HPCC. Topology effects are conditional: Topology~6 leads at low TP degrees but loses its advantage at high TP degrees, and additional GPUs or NICs improve performance only when rank mapping balances traffic across available injection paths. Under uniform 32-way sharding, the largest checkpoint-weight shard is approximately 48.75~GB per rank; all four configurations therefore satisfy a weight-residency criterion of at least 64~GB per accelerator. Within the evaluated workload and simulator semantics, server topology, parallelism, collective implementation, and congestion control jointly determine exposed communication and completion time.

  \end{minipage}
  \vfill
\end{titlepage}

\ifreviewcopy\linenumbers\fi
\hypertarget{ux5f15ux8a00}{%
\section{Introduction}\label{ux5f15ux8a00}}

\hypertarget{moe-ux63a8ux7406ux7684ux7cfbux7edfux4e0eux901aux4fe1ux6311ux6218}{%
\subsection{System and Communication Challenges in MoE Inference}\label{moe-ux63a8ux7406ux7684ux7cfbux7edfux4e0eux901aux4fe1ux6311ux6218}}

Large-model inference is evolving toward greater model capacity, longer contexts, and higher serving concurrency. A single graphics processing unit (GPU) is often unable to meet model-residency, computational-throughput, and memory-capacity requirements simultaneously; multi-GPU, multi-server execution has therefore become common. Mixture-of-experts (MoE) architectures selectively activate a subset of experts for each token, increasing model capacity while controlling per-token computation. Representative designs include GShard\cite{ref1}, Switch Transformer\cite{ref2}, DeepSeekMoE\cite{ref3}, and Mixtral\cite{ref4}. Sparse activation, however, does not eliminate system overhead; it shifts part of the bottleneck to expert routing and inter-device communication. Tokens must be dispatched to and aggregated from experts on different GPUs or nodes according to the gating decisions, producing All-to-All or point-to-point transfers and potentially additional stalls under expert-load imbalance.

In tensor-parallel inference, inter-rank communication in dense Transformers arises primarily from AllReduce, ReduceScatter, or AllGather operations within tensor-parallel (TP) groups. MoE models additionally require token dispatch and combine operations within expert-parallel (EP) groups. Prior measurements have characterized communication behavior in distributed large-model inference\cite{ref5}, while expert-affinity placement\cite{ref6} and architectural modifications\cite{ref7} have been proposed to reduce MoE communication overhead. These findings indicate that communication is no longer merely an implementation detail; it constrains parallelization strategy, expert placement, and system architecture.

\hypertarget{ux73b0ux6709ux7814ux7a76ux4e0eux672aux89e3ux51b3ux95eeux9898}{%
\subsection{Related Work and Open Questions}\label{ux73b0ux6709ux7814ux7a76ux4e0eux672aux89e3ux51b3ux95eeux9898}}

Existing research on large-model inference systems has improved serving efficiency through memory management, request scheduling, and stage-specific resource organization. vLLM employs PagedAttention to reduce key-value (KV) cache fragmentation and redundancy, thereby increasing batchability\cite{ref8}; SGLang uses mechanisms such as RadixAttention to optimize structured language-model programs\cite{ref9}. To address the different computational characteristics and latency objectives of prefill and decode, DistServe jointly optimizes resource allocation for the two stages subject to time-to-first-token (TTFT) and time-per-output-token (TPOT) constraints\cite{ref10}, whereas Splitwise maps the stages to machines better suited to their execution characteristics\cite{ref11}. Although these systems substantially advance service-level memory utilization and runtime scheduling, they do not directly establish how logical parallelism, physical topology, collective algorithms, and network congestion feedback jointly determine exposed communication time for a fixed model and logical workload.

MegaScale-Infer disaggregates attention and expert computation and introduces specialized communication for expert-parallel inference\cite{ref12}. Tensor parallelism, expert parallelism, expert placement, and communication-efficient model architectures have also been studied individually\cite{ref1,ref5,ref6,ref7,ref13,ref14}. MoX further demonstrates that MoE routing and physical direct-connect topology must be considered jointly\cite{ref15}. These lines of work operate at different levels: model and parallelism studies usually emphasize communication volume or logical communication groups, inference-serving studies emphasize request-level scheduling and resource management, and networking studies emphasize flow control and queue dynamics. A central question therefore remains to be quantified systematically: does a locally optimal choice at one level remain optimal under different intra-server interconnects, rank-to-device mappings, NIC configurations, and inter-node congestion-control schemes? In particular, when TP, EP, collective algorithms, and physical paths interact, isolated increases in GPU count, NIC count, or nominal link bandwidth may fail to predict completion time.

\hypertarget{ux62d3ux6251ux5e76ux884cux4e0eux7f51ux7edcux673aux5236ux7684ux8026ux5408}{%
\subsection{Coupling among Topology, Parallelism, and Network Mechanisms}\label{ux62d3ux6251ux5e76ux884cux4e0eux7f51ux7edcux673aux5236ux7684ux8026ux5408}}

Logical communication groups in MoE inference must be mapped onto specific GPUs, central processing units (CPUs), Peripheral Component Interconnect Express (PCIe) switches, network interface cards (NICs), and external switching fabrics. Ultra Path Interconnect (UPI) links between dual CPUs, multiple PCIe switches, GPU/NIC affinities, and InfiniBand (IB) or RDMA over Converged Ethernet (RoCE) links jointly affect hop count, shared egress points, available injection bandwidth, and queue locations. Increasing TP generally enlarges the number of ranks and synchronization scope of a collective, but its actual cost depends on the algorithm and on whether the TP group crosses servers, CPUs, or PCIe switches. Increasing EP changes the size and peer range of expert communication groups, with costs that likewise depend on token routing, the communication implementation, and physical placement. Megatron-LM describes typical Transformer tensor-parallel communication patterns\cite{ref13}, while GShard and DeepSpeed-MoE establish implementation foundations for expert parallelism\cite{ref1,ref14}.

Hardware quantity and performance therefore have no simple monotonic relationship. Additional NICs may raise theoretical injection capacity, but an imbalanced rank-to-NIC mapping can leave ports idle or create local hot spots. Consolidating nodes can reduce the fraction of inter-node traffic while concentrating flows onto a small number of inter-switch PCIe links. A single-node deployment eliminates external-network traffic but can remain constrained by shared intra-node links and synchronization-critical paths at large TP degrees. The decisive issue is not simply whether an individual component is faster, but which physical paths logical communication edges traverse and whether concurrent synchronized flows compete at the same egress. We therefore place server topology, TP/EP partitioning, collective algorithms, and network congestion control in a common experimental matrix and focus on their interactions rather than evaluating hardware metrics in isolation.

\hypertarget{ux7814ux7a76ux8303ux56f4ux4e0eux4effux771fux65b9ux6cd5}{%
\subsection{Scope and Simulation Methodology}\label{ux7814ux7a76ux8303ux56f4ux4e0eux4effux771fux65b9ux6cd5}}

Exhaustively evaluating server organizations, parallelization strategies, collective algorithms, and network configurations on a physical cluster requires substantial heterogeneous hardware, repeated deployment, and a stable measurement environment, making controlled experimentation costly. ASTRA-sim is a discrete-event simulation framework for hardware--software co-design of distributed machine-learning systems. Its separation of workload, system policy, and network model permits parallel strategies or interconnect configurations to be exchanged while other factors remain fixed\cite{ref16}; ASTRA-sim~2.0 further supports graph-based workloads and hierarchical heterogeneous networks\cite{ref17}. 
To provide executable workloads, we use Chakra Execution Trace (Chakra ET), an open graph-based representation\cite{ref19}. The trace for each rank is a directed acyclic graph (DAG): nodes describe compute, memory, or communication operations, while edges encode data and control dependencies. ASTRA-sim schedules nodes according to these dependencies, and its system layer decomposes high-level collectives into point-to-point transfers. These transfers are then submitted to the discrete-event network simulator NS-3\cite{ref20}, which models physical links, switch queues, remote direct memory access (RDMA) flows, and congestion feedback. The inter-node network employs High Precision Congestion Control (HPCC)\cite{ref21}, Data Center Quantized Congestion Notification (DCQCN)\cite{ref22}, or a controlled mode without rate feedback, enabling relative comparisons across network and congestion-control combinations.

We evaluate four MoE model configurations---Kimi K3, GLM-5.3, DeepSeek V4 Pro 0813, and DeepSeek V4 Flash 0731---under six GPU server topologies, four TP/EP strategies, two collective-algorithm combinations, and four network/congestion-control modes. The total number of ranks and the model-specific workload are fixed so that hardware and system configurations for the same model can be compared on a common basis. The objective is to identify relative performance trends, ranking reversals, and their physical causes, rather than to reproduce a complete online inference service. The ASTRA-sim simulation completion time reported herein is not equivalent to end-to-end request latency, which additionally includes request queuing, continuous batching, the KV-cache lifecycle, runtime scheduling, and software-stack overhead.

\hypertarget{ux7814ux7a76ux95eeux9898ux4e0eux4e3bux8981ux8d21ux732e}{%
\subsection{Research Questions and Contributions}\label{ux7814ux7a76ux95eeux9898ux4e0eux4e3bux8981ux8d21ux732e}}

Against this background, we investigate how node density, intra-node interconnects, and the GPU/NIC ratio alter exposed communication time while holding total GPU count and model workload constant. We further examine how interactions among TP/EP partitioning, server topology, and rank mapping affect simulation completion, computation, and communication times. We construct a unified MoE-inference simulation matrix that controls the workload while comparing server topology, TP/EP partitioning, collective algorithms, and network/congestion-control modes. This design identifies both main effects and interactions that cannot be explained by GPU count, NIC count, or nominal bandwidth alone. Cross-model comparisons relate the observed trends to TP and EP communication proxies, while queue and flow statistics for representative configurations help explain performance differences. These contributions are explicitly limited to synthetic execution traces, IB-like network modeling, and the present ASTRA-sim/NS-3 implementation; correlations in this finite sample are not interpreted as universal causal effects of individual architectural parameters.

\hypertarget{ux80ccux666f}{%
\section{Background}\label{ux80ccux666f}}

This section establishes three foundations for the study: how MoE inference generates TP and EP communication, how logical communication groups map onto hierarchical physical interconnects, and how ASTRA-sim, Chakra ET, and NS-3 model the resulting execution and network behavior.

\hypertarget{moe-ux63a8ux7406ux7684ux6267ux884cux4e0eux901aux4fe1ux673aux5236}{%
\subsection{Execution and Communication in MoE Inference}\label{moe-ux63a8ux7406ux7684ux6267ux884cux4e0eux901aux4fe1ux673aux5236}}

A typical MoE Transformer layer comprises an attention module, shared dense computation, and sparsely activated expert modules. A router selects the top-$k$ experts from each token representation, after which token dispatch, expert computation, and result aggregation are performed in sequence. GShard, Switch Transformer, DeepSeekMoE, and Mixtral exemplify different sparse-activation and expert-organization schemes\cite{ref1,ref2,ref3,ref4}. Although these designs increase parameter capacity while controlling per-token computation, they also make expert routing a major source of inter-device communication.

In distributed execution, tensors associated with attention and multilayer perceptron (MLP) operations can be partitioned along the hidden dimension, triggering AllReduce, ReduceScatter, or AllGather at operator boundaries. When experts are distributed across ranks, the router must also perform dispatch and combine through All-to-All or point-to-point transfers. The critical path of an MoE layer therefore commonly follows ``local computation---TP collective---EP token dispatch---expert computation---EP token combine---subsequent TP collective.'' Increased latency in any synchronous communication phase may stall downstream operators.

Communication intensity is jointly determined by model architecture, token count, and parallel partitioning. For a batch of $N$ tokens, hidden width $h$, and $k$ activated experts per token, the activation volume for EP dispatch and combine scales as $N\times k\times h$ in the absence of capacity padding, quantization effects, and load skew. TP traffic depends on tensor dimensions, the partitioning scheme, TP-group size, and collective algorithm. Expert-selection skew further shapes the actual message distribution: even at comparable total byte counts, uneven destinations can create hot spots at particular ranks, NICs, or switch egress ports\cite{ref5,ref6,ref7}.

The contribution of communication to completion time depends on its overlap with computation. Here, \emph{exposed communication time} denotes communication not hidden by computation and therefore present on the execution critical path. Batch composition, prefill/decode phases, KV caching, continuous batching, and runtime scheduling in a production service alter token counts, message granularity, and overlap opportunities. A workload generated from a model configuration captures principal compute and communication dependencies but does not automatically reproduce these dynamic serving behaviors.

\hypertarget{tpep-ux4e0eux903bux8f91ux901aux4fe1ux7ec4}{%
\subsection{TP/EP and Logical Communication Groups}\label{tpep-ux4e0eux903bux8f91ux901aux4fe1ux7ec4}}

Tensor parallelism shards the weights or activations of an operator along selected tensor dimensions so that multiple ranks jointly execute the operator\cite{ref13}. This reduces per-GPU storage and computation, but requires collectives within the TP group to combine partial results. A larger TP group increases the number of synchronizing ranks and may increase either the number of communication phases or the number of logical edges crossing physical domains. The actual cost nevertheless depends on message segmentation, the collective algorithm, and rank mapping, and cannot be inferred from the TP degree alone.

Expert parallelism assigns different experts to different ranks, such that each device stores or executes only a subset of experts\cite{ref1,ref14}. The router dispatches tokens within an EP group and combines their results after expert computation. A larger EP group expands the expert-placement space but also increases the set of potential peers. Direct communication allows ranks to exchange data with their destinations without forcing traffic between logically nonadjacent experts through intermediate Ring ranks; its performance still depends on physical links and egress contention.

Hybrid parallelism commonly also includes data parallelism (DP) and pipeline parallelism (PP). Excluding other sharding dimensions, the total rank count is $DP\times PP\times TP\times EP$. The main experiments fix DP=1 and PP=1 to isolate the tradeoff between TP and EP. The resulting conclusions do not directly cover multiple data replicas, pipeline bubbles, or disaggregated prefill/decode deployments.

TP and EP define logical communication groups, not physical transfer paths. The ranks in a group may share one PCIe switch, span non-uniform memory access (NUMA) domains within a server, or reside on different servers. Only the joint consideration of logical groups, rank numbering, and physical device placement determines whether a collective traverses UPI, a shared NIC, or an external switch, and thereby explains differences in bandwidth utilization, queuing, and critical paths.

\hypertarget{astra-simchakrans-3-ux4effux771fux6808}{%
\subsection{ASTRA-sim--Chakra--NS-3 Simulation Stack}\label{astra-simchakrans-3-ux4effux771fux6808}}

ASTRA-sim adopts a discrete-event architecture that decouples workload, system, and network layers\cite{ref16}. ASTRA-sim~2.0 extends this design with graph-based workloads, hybrid parallelism, and hierarchical heterogeneous networks\cite{ref17}. The workload layer describes compute, memory, and communication events; the system layer determines event scheduling and collective implementation; and the network layer returns point-to-point transfer completion times. This separation permits the same workload to be reused across system policies and network models.

Chakra ET is a standardized graph representation connecting model workloads to the simulator\cite{ref19}. Each rank has an execution trace whose nodes record compute, memory-access, or communication operations and their attributes, while edges record data or control dependencies. ASTRA-sim schedules a successor only after its predecessors have completed and the required resources are available. An ET therefore specifies not only communication volume but also where communication lies on the critical path. Conversely, it expresses only explicitly recorded events: dynamic batching, the KV-cache lifecycle, kernel gaps, or compute--communication overlap absent from the trace are not inferred automatically.
SimAI provides a complementary high-fidelity simulation stack that integrates framework, kernel, and collective-communication models for large-scale language-model systems\cite{ref18}.

The system layer reads high-level communication nodes such as ALL\_REDUCE and ALL\_TO\_ALL and decomposes them into point-to-point transfers according to Ring, Direct, or Double Binary Tree (DBT) implementations. The collective algorithm specifies the transmission order among logical ranks, whereas the network backend determines the actual routes and completion times on the physical topology. Logical collective topology and physical network topology are thus distinct layers: the same TP Ring can produce markedly different link loads and synchronization delays when mapped onto a single PCIe switch, a path crossing UPI, or inter-node NICs.

ASTRA-sim's analytical backend rapidly estimates transfer times from bandwidth, latency, and simplified contention relationships and is well suited to initial screening of large design spaces. NS-3 provides finer-grained discrete-event simulation of physical links, switching nodes, queues, RDMA flows, and congestion feedback\cite{ref20}. We select NS-3 as the primary network backend because the research questions involve not only message size but also shared egress among flows, queue buildup, and congestion control; HPCC\cite{ref21} and DCQCN\cite{ref22} subsequently represent different rate-feedback mechanisms. The results should therefore be interpreted as completion times under specified ETs, system configurations, and network models, not as direct reproduction of end-to-end latency in an online service.

\hypertarget{ux5b9eux9a8cux65b9ux6cd5}{%
\section{Experimental Methodology}\label{ux5b9eux9a8cux65b9ux6cd5}}

To ensure comparability across experimental combinations, we construct a controlled design space: total rank count, workload-generation rules, and baseline hardware parameters are fixed, while model configuration, TP/EP partitioning, physical topology, collective algorithm, and network congestion-control mode are varied. This section describes the rationale, parameterization, statistical conventions, and validation boundaries for each factor.

\hypertarget{ux6a21ux578bux9009ux62e9}{%
\subsection{Model Selection}\label{ux6a21ux578bux9009ux62e9}}

The four model configurations were selected to provide both architectural and communication diversity. They span 284B--2.8T total parameters, 43--93 layers, and workload dimensions of 3584--7168, allowing us to test whether conclusions are specific to a single model scale. The number of routed experts ranges from 256 to 896 and the number activated per token from 6 to 16, so the sample includes configurations with high TP communication proxies as well as those with high EP communication proxies. The models also form a deliberately non-collinear comparison: Kimi K3 has the largest parameter count but not the widest workload dimension, whereas DeepSeek V4 Pro 0813 has the widest workload dimension but a smaller top-$k$. This variation helps distinguish model capacity, TP message size, and EP routing intensity rather than predicting communication performance from total parameters alone.

Table~\ref{tbl:model-parameters} lists the principal architectural parameters used to generate the workloads. Values were obtained from the experimental configuration files and official model descriptions\cite{ref23,ref24,ref25,ref26,ref27} and converted into common inputs, including workload dimension, layer count, number of routed experts, and number of experts activated per token. Kimi K3, GLM-5.3, DeepSeek V4 Pro 0813 (hereafter V4 Pro), and DeepSeek V4 Flash 0731 (hereafter V4 Flash) constitute four fixed configurations whose structural parameters remain unchanged in all subsequent experiments.

\begin{longtable}[]{@{}
  >{\raggedright\arraybackslash}p{(\columnwidth - 10\tabcolsep) * \real{0.1304}}
  >{\raggedleft\arraybackslash}p{(\columnwidth - 10\tabcolsep) * \real{0.1739}}
  >{\raggedleft\arraybackslash}p{(\columnwidth - 10\tabcolsep) * \real{0.1739}}
  >{\raggedleft\arraybackslash}p{(\columnwidth - 10\tabcolsep) * \real{0.1739}}
  >{\raggedleft\arraybackslash}p{(\columnwidth - 10\tabcolsep) * \real{0.1739}}
  >{\raggedleft\arraybackslash}p{(\columnwidth - 10\tabcolsep) * \real{0.1739}}@{}}
\caption{Principal architectural parameters of the four MoE models}\label{tbl:model-parameters}\tabularnewline
\toprule\noalign{}
\begin{minipage}[b]{\linewidth}\raggedright
Model
\end{minipage} & \begin{minipage}[b]{\linewidth}\raggedleft
Total parameters
\end{minipage} & \begin{minipage}[b]{\linewidth}\raggedleft
Workload dimension h
\end{minipage} & \begin{minipage}[b]{\linewidth}\raggedleft
Layers L
\end{minipage} & \begin{minipage}[b]{\linewidth}\raggedleft
Routed experts E
\end{minipage} & \begin{minipage}[b]{\linewidth}\raggedleft
Experts activated per token k
\end{minipage} \\
\midrule\noalign{}
\endfirsthead
\toprule\noalign{}
\begin{minipage}[b]{\linewidth}\raggedright
Model
\end{minipage} & \begin{minipage}[b]{\linewidth}\raggedleft
Total parameters
\end{minipage} & \begin{minipage}[b]{\linewidth}\raggedleft
Workload dimension h
\end{minipage} & \begin{minipage}[b]{\linewidth}\raggedleft
Layers L
\end{minipage} & \begin{minipage}[b]{\linewidth}\raggedleft
Routed experts E
\end{minipage} & \begin{minipage}[b]{\linewidth}\raggedleft
Experts activated per token k
\end{minipage} \\
\midrule\noalign{}
\endhead
\bottomrule\noalign{}
\endlastfoot
Kimi K3 & 2.8T & 3584 & 93 & 896 & 16 \\
GLM-5.3 & 753B & 6144 & 78 & 256 & 8 \\
DeepSeek V4 Pro 0813 & 1.7T & 7168 & 61 & 384 & 6 \\
DeepSeek V4 Flash 0731 & 284B & 4096 & 43 & 256 & 6 \\
\end{longtable}

For Kimi K3, the value $h=3584$ is the Latent MoE Dimension reported in Table~1 of the official technical report, rather than the model attention hidden dimension of 7168\cite{ref23}. The workload generator uses this latent dimension as the Kimi K3 communication-dimension input; the other rows use the corresponding configured hidden dimension. This distinction is retained explicitly so that the simulated traffic parameters are traceable to their architectural definitions.

\begin{figure}[tbp]
  \centering
  \includegraphics[width=0.78\textwidth,height=0.72\textheight,keepaspectratio]{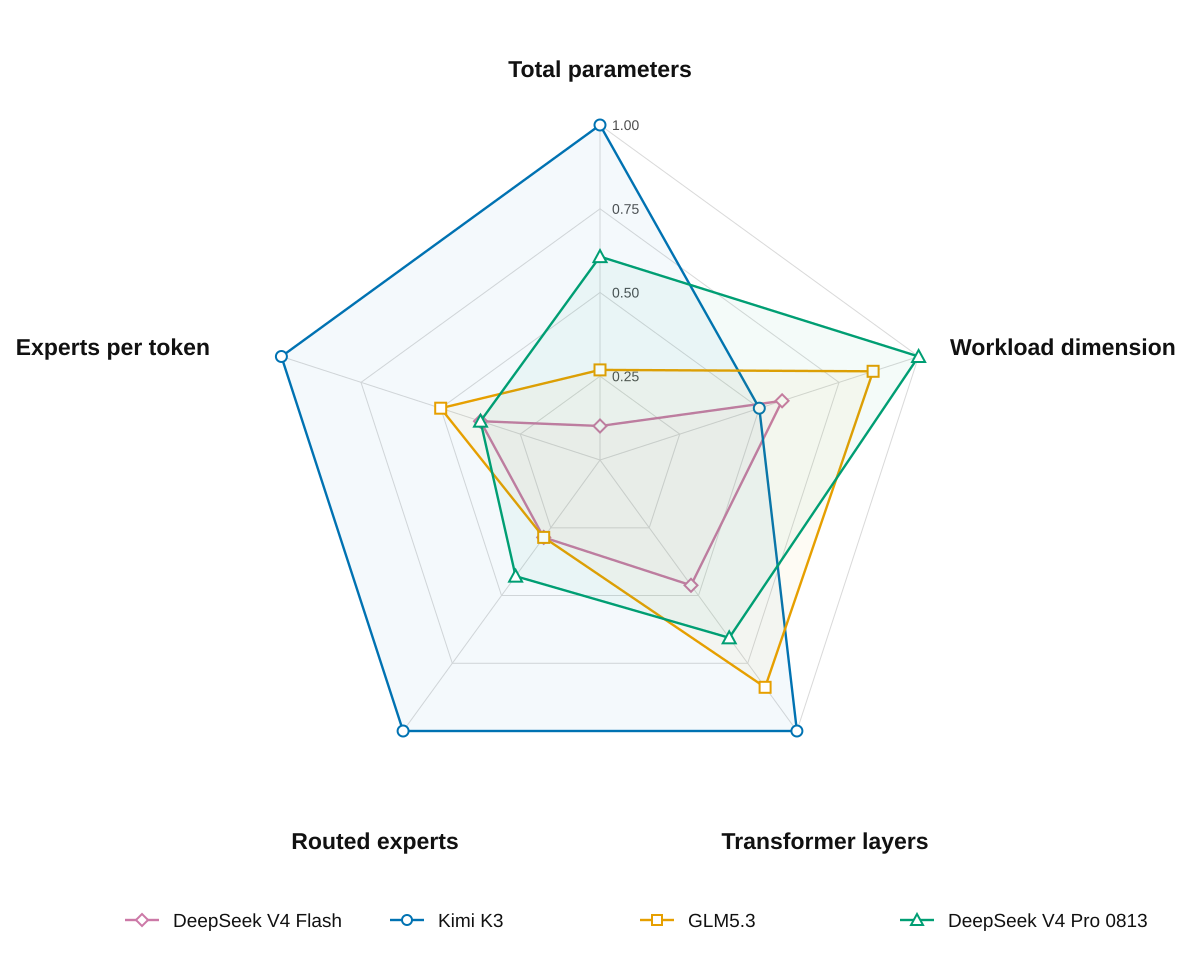}
  \caption{Normalized comparison of architectural parameters across the four MoE models}
  \label{fig:model-radar}
\end{figure}

In the workload model, workload dimension and layer count jointly affect TP collective message size and frequency, whereas the number of activated experts and top-$k$ affect EP dispatch/combine volume. Total parameter count primarily reflects capacity and sharding requirements and is not equivalent to bytes transferred in an individual collective. Figure~\ref{fig:model-radar} normalizes total parameters, workload dimension, layer count, routed-expert count, and experts activated per token by the maximum of each dimension across the four configurations. Kimi K3 represents a deep configuration with high total capacity, many experts, and a large top-$k$; GLM-5.3 combines a relatively wide workload dimension with many layers; V4 Pro provides the widest workload dimension and a large parameter count; and V4 Flash is a smaller control in both parameter count and depth. This structured sample spans different TP message scales and EP routing intensities, providing a common basis for analyzing interactions among model architecture, parallelism, and physical topology.

\subsection{Model-Residency Capacity Based on Actual Weight Files}\label{sec:capacity-feasibility}

We estimate weight-residency capacity from the actual storage footprints of the officially released checkpoints. Because weight representations differ across models, actual storage cannot be derived solely from total parameter count and a uniform bytes-per-parameter assumption. Kimi K3 uses mixed-precision representations including MXFP4; the official GLM-5.3 weights mix BF16, FP8 E4M3, and F32; and V4 Pro and V4 Flash combine FP4 expert weights with other formats including FP8.
The study adopts a weight-residency criterion of at least 64~GB per accelerator. Model weights are evenly sharded over 32 ranks, with one accelerator per rank; the per-rank weight requirement is therefore the checkpoint size divided by 32. As shown in Table~\ref{tbl:model-residency-capacity}, Kimi K3 has the largest footprint, approximately 1560~GB, or 48.75~GB per rank after even sharding. V4 Pro requires approximately 27.91~GB per rank, GLM-5.3 approximately 23.63~GB, and V4 Flash approximately 5.22~GB. All four models therefore satisfy the weight-residency criterion under uniform 32-way sharding.
This criterion accounts only for checkpoint weights. Runtime execution must additionally accommodate the KV cache, activations and intermediate tensors, collective buffers, runtime workspaces, and other framework and system overheads. The table therefore establishes weight residency only; it does not directly determine batch size, context length, or maximum KV-cache capacity.

\begin{table}[tbp]
\centering
\caption{Estimated 32-accelerator weight-residency requirements based on official weight files}
\label{tbl:model-residency-capacity}
\scriptsize
\setlength{\tabcolsep}{2.5pt}
\renewcommand{\arraystretch}{1.18}
\begin{tabularx}{\textwidth}{>{\raggedright\arraybackslash}p{2.20cm} >{\centering\arraybackslash}p{0.75cm} >{\raggedright\arraybackslash}X >{\centering\arraybackslash}p{1.20cm} >{\centering\arraybackslash}p{1.30cm} >{\centering\arraybackslash}p{1.35cm}}
\toprule
Model & Params. & Actual weight format & \makecell{Total\\weights (GB)} & \makecell{Weights\\per rank (GB)} & \makecell{Meets $\geq$64-GB\\criterion} \\
\midrule
Kimi K3 & 2.8T & Mixed precision including MXFP4 & $\approx 1\,560$ & \textbf{48.75} & Yes \\
GLM-5.3 & 753B & BF16 / FP8 E4M3 / F32 & $\approx 756$ & \textbf{23.63} & Yes \\
DeepSeek V4 Pro 0813 & 1.7T & FP4 experts + mixed precision including FP8 & $\approx 893$ & \textbf{27.91} & Yes \\
DeepSeek V4 Flash 0731 & 284B & FP4 experts + mixed precision including FP8 & $\approx 166.9$ & \textbf{5.22} & Yes \\
\bottomrule
\end{tabularx}
\end{table}

\hypertarget{ux5e76ux884cux7b56ux7565ux4e0e-rank-ux6620ux5c04}{%
\subsection{Parallelization Strategies and Rank Mapping}\label{ux5e76ux884cux7b56ux7565ux4e0e-rank-ux6620ux5c04}}

The main experiments fix the total scale at 32 ranks, covering servers with 1, 2, 8, 16, or 32 GPUs per node while preserving an identical logical workload across topologies. We additionally set DP=1 and PP=1, such that $TP\times EP=32$ and the design variable is restricted to the TP/EP partition. This control excludes additional data replicas, pipeline stages, and pipeline bubbles; accordingly, the conclusions do not directly apply to deployments with DP, PP, or prefill/decode disaggregation.

Both TP and EP use powers of two, producing four strategies from low-TP/high-EP to high-TP/low-EP. TP2EP16, TP4EP8, TP8EP4, and TP16EP2 satisfy the regular group-size requirements of common collective implementations, while each adjacent comparison doubles TP and halves EP. This structure exposes the tradeoff between an expanded TP synchronization scope and a contracted EP group. We omit the degenerate TP1EP32 and TP32EP1 endpoints because the former eliminates TP collectives and the latter eliminates multi-rank EP communication, preventing joint evaluation of both mechanisms.

\begin{longtable}[]{@{}lrrrl@{}}
\caption{TP/EP parallelization strategies evaluated in this study}\label{tbl:parallel-strategies}\tabularnewline
\toprule\noalign{}
Strategy & TP & EP & Total ranks & Purpose \\
\midrule\noalign{}
\endfirsthead
\toprule\noalign{}
Strategy & TP & EP & Total ranks & Purpose \\
\midrule\noalign{}
\endhead
\bottomrule\noalign{}
\endlastfoot
TP2EP16 & 2 & 16 & 32 & Low-TP, high-EP baseline \\
TP4EP8 & 4 & 8 & 32 & Effect of doubling TP \\
TP8EP4 & 8 & 4 & 32 & Cross-domain mapping of a larger TP group \\
TP16EP2 & 16 & 2 & 32 & High-TP, low-EP control \\
\end{longtable}

The logical topology uniformly uses the two dimensions {[}TP,EP{]}, with $rank=ep_{index}\times TP+tp_{index}$. A TP group at a fixed EP coordinate consequently comprises contiguous ranks, whereas an EP group at a fixed TP coordinate is \{tp\_index,\ tp\_index+\allowbreak{}TP,\ ...\}. This numbering is unchanged across models, topologies, network modes, and collective algorithms, so experimental differences can be attributed to the varied design factor rather than an implicit rank permutation.

\hypertarget{ux7269ux7406ux62d3ux6251ux8bbeux8ba1}{%
\subsection{Physical Topology Design}\label{ux7269ux7406ux62d3ux6251ux8bbeux8ba1}}

The six physical topologies systematically vary node density, GPU/NIC ratio, and intra-node interconnect hierarchy; they are not intended to reproduce six specific commercial servers. Every topology contains 32 GPUs, ranging from 32 single-GPU nodes to one 32-GPU node. Consolidation progressively reduces the fraction of inter-node traffic while exposing local contention at PCIe switches, UPI links, and shared NICs, enabling a comparison between distributed network injection and dense intra-node consolidation.

\begin{longtable}[]{@{}
  >{\raggedright\arraybackslash}p{(\columnwidth - 12\tabcolsep) * \real{0.1200}}
  >{\raggedright\arraybackslash}p{(\columnwidth - 12\tabcolsep) * \real{0.1200}}
  >{\raggedleft\arraybackslash}p{(\columnwidth - 12\tabcolsep) * \real{0.1600}}
  >{\raggedleft\arraybackslash}p{(\columnwidth - 12\tabcolsep) * \real{0.1600}}
  >{\raggedleft\arraybackslash}p{(\columnwidth - 12\tabcolsep) * \real{0.1600}}
  >{\raggedleft\arraybackslash}p{(\columnwidth - 12\tabcolsep) * \real{0.1600}}
  >{\raggedright\arraybackslash}p{(\columnwidth - 12\tabcolsep) * \real{0.1200}}@{}}
\caption{Six physical GPU server topologies}\label{tbl:physical-topologies}\tabularnewline
\toprule\noalign{}
\begin{minipage}[b]{\linewidth}\raggedright
Topology
\end{minipage} & \begin{minipage}[b]{\linewidth}\raggedright
Per-node organization
\end{minipage} & \begin{minipage}[b]{\linewidth}\raggedleft
Nodes
\end{minipage} & \begin{minipage}[b]{\linewidth}\raggedleft
Total GPUs
\end{minipage} & \begin{minipage}[b]{\linewidth}\raggedleft
Total NICs
\end{minipage} & \begin{minipage}[b]{\linewidth}\raggedleft
GPU/NIC
\end{minipage} & \begin{minipage}[b]{\linewidth}\raggedright
Primary comparison
\end{minipage} \\
\midrule\noalign{}
\endfirsthead
\toprule\noalign{}
\begin{minipage}[b]{\linewidth}\raggedright
Topology
\end{minipage} & \begin{minipage}[b]{\linewidth}\raggedright
Per-node organization
\end{minipage} & \begin{minipage}[b]{\linewidth}\raggedleft
Nodes
\end{minipage} & \begin{minipage}[b]{\linewidth}\raggedleft
Total GPUs
\end{minipage} & \begin{minipage}[b]{\linewidth}\raggedleft
Total NICs
\end{minipage} & \begin{minipage}[b]{\linewidth}\raggedleft
GPU/NIC
\end{minipage} & \begin{minipage}[b]{\linewidth}\raggedright
Primary comparison
\end{minipage} \\
\midrule\noalign{}
\endhead
\bottomrule\noalign{}
\endlastfoot
Topology 1 & 1 CPU, 1 GPU, 1 NIC & 32 & 32 & 32 & 1 & Distributed baseline with one NIC per GPU \\
Topology 2 & 1 CPU, 2 GPUs, 1 NIC, 1 PCIe switch & 16 & 32 & 16 & 2 & Dual-GPU, single-PCIe-domain baseline \\
Topology 3 & 2 CPUs, 2 GPUs, 1 NIC; inter-CPU UPI & 16 & 32 & 16 & 2 & Isolate NUMA/UPI effects at fixed node scale \\
Topology 4 & 2 CPUs, 8 GPUs, 2 NICs, 2 PCIe switches & 4 & 32 & 8 & 4 & Approximate dense 8-GPU node \\
Topology 5 & 2 CPUs, 16 GPUs, 4 NICs, 4 PCIe switches & 2 & 32 & 8 & 4 & Further consolidation at the same GPU/NIC ratio \\
Topology 6 & 2 CPUs, 32 GPUs, 8 PCIe switches, no external NIC & 1 & 32 & 0 & N/A & Single-node intra-node-only baseline \\
\end{longtable}

In Topologies~1--5, all NICs connect to a single external 400-Gbps switch. A single switch avoids additional variables from top-of-rack (ToR) and spine hierarchies or multipath routing, retaining the focus on endpoint injection, intra-node paths, and congestion control. In Topologies~4 and 5, each group of four GPUs and one NIC shares a PCIe switch; in Topology~6, each group of four GPUs shares a PCIe switch. The PCIe switches in Topologies~5 and 6 are approximated as fully connected to test whether logical communication edges can exploit links to nonadjacent switches.

The six topologies form three progressively deeper comparisons. Topologies~1 and 2 primarily vary GPUs per NIC and server count. Topologies~2 and 3 hold GPU, node, and NIC counts constant and introduce only dual CPUs and a UPI path. Topologies~4 and 5 hold total NIC count and GPU/NIC ratio constant while varying GPUs per node, node count, and the number of intra-node switches. Topology~6 removes the external network, separating inter-node networking costs from purely intra-node interconnect costs. These paired comparisons identify physical causes more effectively than ordering systems by GPU or NIC count alone.

Physical placement follows contiguous rank order: Topology~1 places one rank per node; Topologies~2 and 3 place two; Topology~4 places eight; Topology~5 places 16; and Topology~6 places all 32 ranks on one node. On nodes with multiple PCIe switches, each consecutive group of four local ranks is assigned to one switch. Fixing this mapping prevents implicit placement changes across experiments, but also makes the conclusions mapping-dependent and precludes direct generalization to arbitrary rank permutations.

From an engineering-cost perspective, Topology~1 uses the most hosts, NICs, and switch ports to obtain independent injection capacity. Topologies~4 and 5 reduce external device count through higher node density at the expense of shared egress and more complex intra-node mapping, whereas Topology~6 provides an extreme consolidation baseline without an external network. We do not model acquisition cost, energy, rack space, or reliability; the configuration with the shortest simulated time is therefore not necessarily that with the lowest total cost of ownership. PCIe servers generally offer flexible CPU/NIC combinations and a lower deployment barrier than dedicated scale-up fabrics. Accordingly, the main experiments use a common reference PCIe accelerator as the compute baseline.
Detailed node and cluster diagrams for the six physical organizations are provided in Figures~\ref{fig:topology1}--\ref{fig:topology6} in Appendix~A.

\hypertarget{ux7f51ux7edcux4e0eux62e5ux585eux63a7ux5236ux8bbeux7f6e}{%
\subsection{Network and Congestion-Control Settings}\label{ux7f51ux7edcux4e0eux62e5ux585eux63a7ux5236ux8bbeux7f6e}}

MoE collectives inject multiple flows synchronously near layer boundaries and can therefore affect queue length, sending rate, and straggler flow completion time simultaneously. To distinguish baseline link costs from feedback-control behavior, we define one no-rate-feedback baseline and three feedback-enabled modes. DCQCN adjusts sending rates using Explicit Congestion Notification (ECN) and Congestion Notification Packets (CNPs)\cite{ref22}, whereas HPCC uses In-band Network Telemetry (INT) to provide per-hop load information\cite{ref21}. The two schemes compare indirect congestion marking with telemetry-based feedback under synchronized bursts. The ENABLE\_QCN variable controls feedback paths related to Quantized Congestion Notification (QCN).

\begin{longtable}[]{@{}
  >{\raggedright\arraybackslash}p{(\columnwidth - 4\tabcolsep) * \real{0.3333}}
  >{\raggedright\arraybackslash}p{(\columnwidth - 4\tabcolsep) * \real{0.3333}}
  >{\raggedright\arraybackslash}p{(\columnwidth - 4\tabcolsep) * \real{0.3333}}@{}}
\caption{Network and congestion-control modes}\label{tbl:network-modes}\tabularnewline
\toprule\noalign{}
\begin{minipage}[b]{\linewidth}\raggedright
Network mode
\end{minipage} & \begin{minipage}[b]{\linewidth}\raggedright
Key settings
\end{minipage} & \begin{minipage}[b]{\linewidth}\raggedright
Experimental role
\end{minipage} \\
\midrule\noalign{}
\endfirsthead
\toprule\noalign{}
\begin{minipage}[b]{\linewidth}\raggedright
Network mode
\end{minipage} & \begin{minipage}[b]{\linewidth}\raggedright
Key settings
\end{minipage} & \begin{minipage}[b]{\linewidth}\raggedright
Experimental role
\end{minipage} \\
\midrule\noalign{}
\endhead
\bottomrule\noalign{}
\endlastfoot
IB CC0 & CC\_MODE=0, ENABLE\_QCN=0 & Controlled baseline without rate feedback \\
IB-like HPCC & CC\_MODE=3, ENABLE\_QCN=0 & Low-latency HPCC/INT-style approximation without QCN \\
RoCE HPCC & CC\_MODE=3, ENABLE\_QCN=1 & HPCC-style control on a RoCE data path \\
RoCE DCQCN & CC\_MODE=1, ENABLE\_QCN=1 & Conventional RoCEv2 ECN/CNP-feedback baseline \\
\end{longtable}

All IB CC0 experiments use CC\_MODE=0. For Topology~6, which does not traverse an external network, all four network modes are run as independent experimental units and their original outputs are analyzed directly; values are neither copied nor minimized across modes. Network-mode labels for Topology~6 preserve the matrix structure and are not used to infer the relative merits of inter-node IB and RoCE protocols.

The NS-3 backend models packets, queues, and congestion feedback through the project's Qbb RDMA data path. It does not fully implement native InfiniBand credit-based flow control, service levels/virtual lanes, switch arbitration, or vendor NIC firmware. We therefore interpret IB\ HPCC in the original experiment directories as \textbf{IB-like HPCC}; tables and figures retain the shorter label ``IB HPCC.'' This mode is not a reproduction of a native InfiniBand stack or of ``HPCC over native IB.'' Similarly, IB CC0 is only a no-rate-feedback control, not a performance upper bound for all workloads.

\begin{longtable}[]{@{}
  >{\raggedright\arraybackslash}p{(\columnwidth - 6\tabcolsep) * \real{0.2143}}
  >{\raggedleft\arraybackslash}p{(\columnwidth - 6\tabcolsep) * \real{0.2857}}
  >{\raggedleft\arraybackslash}p{(\columnwidth - 6\tabcolsep) * \real{0.2857}}
  >{\raggedright\arraybackslash}p{(\columnwidth - 6\tabcolsep) * \real{0.2143}}@{}}
\caption{Principal link-bandwidth and propagation-delay parameters}\label{tbl:link-parameters}\tabularnewline
\toprule\noalign{}
\begin{minipage}[b]{\linewidth}\raggedright
Link type
\end{minipage} & \begin{minipage}[b]{\linewidth}\raggedleft
Bandwidth
\end{minipage} & \begin{minipage}[b]{\linewidth}\raggedleft
One-way propagation delay
\end{minipage} & \begin{minipage}[b]{\linewidth}\raggedright
Applicable topology or mode
\end{minipage} \\
\midrule\noalign{}
\endfirsthead
\toprule\noalign{}
\begin{minipage}[b]{\linewidth}\raggedright
Link type
\end{minipage} & \begin{minipage}[b]{\linewidth}\raggedleft
Bandwidth
\end{minipage} & \begin{minipage}[b]{\linewidth}\raggedleft
One-way propagation delay
\end{minipage} & \begin{minipage}[b]{\linewidth}\raggedright
Applicable topology or mode
\end{minipage} \\
\midrule\noalign{}
\endhead
\bottomrule\noalign{}
\endlastfoot
GPU to local CPU/PCIe-switch/NIC attachment & 512 Gbps & 0.2 µs & All main experiments \\
NIC to external switch & 400 Gbps & 0.7 µs for IB-like; 1 µs for RoCE & Topologies 1--5 \\
Inter-PCIe-switch link & 400 Gbps & 0.1 µs & Topologies 4--6 \\
UPI link & 400 Gbps & 2 µs & Topology 3 \\
\end{longtable}

The modeled local accelerator attachment provides 512~Gbps of nominal unidirectional bandwidth with a propagation delay of 0.2~µs; encoding, transaction-layer, and protocol-efficiency losses are not modeled separately.

UPI bandwidth is set to 400~Gbps to match the nominal bandwidth of one external NIC link and an inter-PCIe-switch link. Topologies~2 and 3 have identical GPU, node, and NIC counts; after link bandwidths are equalized, their differences primarily reflect the overhead of a dual-socket NUMA/UPI path and its 2-µs propagation delay. The 400-Gbps value is an upper bound for the UPI link, not the sustainable end-to-end bandwidth of inter-socket GPU communication. Multi-GPU inter-socket DMA traverses the local PCIe root complex, UPI, and the remote I/O path, and its efficiency is further affected by protocol overhead, transaction granularity, bidirectional traffic, and contention. Effective UPI utilization can therefore remain below the nominal rate, producing higher latency and lower sustained throughput. The setting avoids a comparison confounded by different nominal bandwidths while allowing inter-socket flows to share and queue on the same UPI path in NS-3; additional hops, 2-µs latency, and contention express the lower effective utilization of inter-socket DMA.

The IB-like mode assigns 0.7~µs to each NIC--switch edge, making the sum of the two link delays consistent with the approximately 1.4~µs measured for two IB NICs connected through one switch. This is an equivalent allocation of end-to-end link latency; NIC processing, switch forwarding, and measurement-software overhead are not decomposed further. All modes share the same physical topology except for the protocol and latency settings in the table, reducing extraneous variation.

\hypertarget{ux96c6ux5408ux901aux4fe1ux7b97ux6cd5ux7ec4ux5408}{%
\subsection{Collective-Algorithm Combinations}\label{ux96c6ux5408ux901aux4fe1ux7b97ux6cd5ux7ec4ux5408}}

TP and EP have different communication semantics, so we select implementations suited to AllReduce and All-to-All, respectively. Ring AllReduce divides a message into chunks and performs reduce-scatter and all-gather along a logical ring. For $N$ ranks and a total message of size $M$, a typical implementation has $2(N-1)$ phases and transfers approximately M/N per phase. The phase count grows linearly with group size, but link utilization is regular. Double Binary Tree (DBT) performs reduction and broadcast over two complementary trees, with theoretical phase complexity O(log\ N), representing a tree-based TP collective with fewer phases. Direct All-to-All allows ranks in an EP group to send shards directly to their destinations and better matches the many-to-many semantics of token dispatch and combine.

The main matrix fixes Direct for EP and varies only the TP AllReduce algorithm, producing the following two configurations. Names follow ``TP collective\_EP collective''; for example, ring\_direct uses Ring for TP and Direct for EP.

\begin{longtable}[]{@{}llll@{}}
\caption{Collective-algorithm combinations}\label{tbl:collective-combinations}\tabularnewline
\toprule\noalign{}
Combination & TP collective & EP collective & Purpose \\
\midrule\noalign{}
\endfirsthead
\toprule\noalign{}
Combination & TP collective & EP collective & Purpose \\
\midrule\noalign{}
\endhead
\bottomrule\noalign{}
\endlastfoot
ring\_direct & Ring & Direct & Bandwidth-oriented TP baseline \\
dbt\_direct & Double Binary Tree & Direct & Low-phase-count tree control \\
\end{longtable}

This choice reflects both communication semantics and preliminary screening. Early Kimi K3 exploration covered EP Ring, EP Direct, TP Halving-Doubling, and TP DBT. In a 20-run EP-algorithm screen aggregated at fixed model, topology, and TP/EP strategy, mean completion time was 2.388~s for ring\_ring and 1.036~s for ring\_direct, a 56.6\% reduction. Preliminary differences between TP Halving-Doubling and TP DBT were smaller. The main matrix therefore fixes EP Direct and compares TP Ring with TP DBT, controlling experimental scale and avoiding simultaneous variation of TP and EP algorithms. This internal exploratory screen was used to select factors for the confirmatory experiments; it is excluded from the four-model matrix means and is not treated as independent confirmatory evidence.

In the ASTRA-sim version used here, DBT transfers the full data\_size on each tree edge, whereas Ring shards the message by ring size. We therefore compare these specific implementations and their physical mappings, not all Tree/DBT and Ring implementations in the NVIDIA Collective Communications Library (NCCL)\cite{ref28}. DBT's lower phase count may coincide with larger per-edge messages and aggregation pressure near tree roots; this implementation difference is addressed explicitly in the results.

\hypertarget{ux5de5ux4f5cux8d1fux8f7dux6784ux5efa}{%
\subsection{Workload Construction}\label{ux5de5ux4f5cux8d1fux8f7dux6784ux5efa}}

All four workloads are generated from model configuration files by the same program, ensuring that models differ only in architectural parameters. The generator reads hidden\_size, num\_hidden\_layers, num\_attention\_heads, num\_key\_value\_heads, n\_routed\_experts, num\_experts\_per\_tok, moe\_intermediate\_size, intermediate\_size, and vocab\_size, while fixing sequence length at $T=4096$ tokens, data type, and other simulation assumptions. One Chakra ET file is generated per rank, recording compute and communication nodes and their dependencies.

The generator constructs a serial critical path in layer order: attention computation is followed by TP AllReduce; an MoE layer then performs EP-dispatch All-to-All, expert computation, EP-combine All-to-All, shared feed-forward-network (FFN) computation, and TP-output AllReduce. The volume of one TP communication event is approximated as

\begin{equation}
V_{\mathrm{TP}} = T h \times 2~\mathrm{bytes}.
\label{eq:tp-volume}
\end{equation}

The per-rank volume of one EP dispatch or combine is approximated as

\begin{equation}
V_{\mathrm{EP}} = \frac{T h k \times 2~\mathrm{bytes}}{EP}.
\label{eq:ep-volume}
\end{equation}

The factor $2~\mathrm{bytes}$ is the 2-byte storage width of an FP16/BF16 activation element; this traffic model is independent of the mixed precision used for checkpoint weights. Here, $T$ is the token count represented in each TP lane, $h$ is the workload communication dimension defined in Table~\ref{tbl:model-parameters}, and $k$ is the number of experts activated per token. Uniform routing creates $Tk$ token--expert assignments within each EP group and assigns $Tk/EP$ assignments to each rank. Because the $[TP,EP]$ layout contains $TP$ independent EP groups, the aggregate activation proxy over all ranks is $TP\times T h k\times2~\mathrm{bytes}$ per dispatch or combine event. Compute nodes derive floating-point-operation (FLOP) counts and tensor sizes from the model configuration, and execution time is estimated by a roofline model using peak GPU performance and memory bandwidth. The workload captures principal architecture-induced compute and communication dependencies and resembles a fixed 4096-token, prefill-like forward segment rather than a complete online inference request.

The generator uses a fixed 4096-token input and serializes principal events according to their dependencies. It assumes uniform expert routing, equal traffic across EP ranks, no capacity drops, and no routing skew. Channel count, protocol thresholds, and kernel implementations in NCCL and the ROCm Communication Collectives Library (RCCL) are abstracted by ASTRA-sim's collective model. Per-token decoding, dynamic batching, expert-load skew, and compute--communication overlap are not experimental variables, ensuring common execution semantics across all models and topologies. Model-residency capacity is independently calculated from the official weight footprints in Section~\ref{sec:capacity-feasibility}; all four models satisfy the 32-GPU weight-capacity constraint. Simulation completion, computation, and communication times are the comparison metrics for the controlled design space.

\hypertarget{ux4effux771fux5668ux4e0eux57faux7840ux53c2ux6570}{%
\subsection{Simulator and Baseline Parameters}\label{ux4effux771fux5668ux4e0eux57faux7840ux53c2ux6570}}

The main experiments couple ASTRA-sim with NS-3\cite{ref16,ref17,ref20}. ASTRA-sim schedules compute, memory, and collective events from the Chakra ET and submits decomposed point-to-point transfers to NS-3, which models physical links, switching nodes, queues, RDMA flows, and congestion feedback. Every main-matrix combination uses the same software version and baseline parameters; only explicitly listed experimental factors vary.

\begin{longtable}[]{@{}
  >{\raggedright\arraybackslash}p{(\columnwidth - 6\tabcolsep) * \real{0.2308}}
  >{\raggedright\arraybackslash}p{(\columnwidth - 6\tabcolsep) * \real{0.2308}}
  >{\raggedleft\arraybackslash}p{(\columnwidth - 6\tabcolsep) * \real{0.3077}}
  >{\raggedright\arraybackslash}p{(\columnwidth - 6\tabcolsep) * \real{0.2308}}@{}}
\caption{Baseline simulator and workload parameters}\label{tbl:simulator-parameters}\tabularnewline
\toprule\noalign{}
\begin{minipage}[b]{\linewidth}\raggedright
Category
\end{minipage} & \begin{minipage}[b]{\linewidth}\raggedright
Parameter
\end{minipage} & \begin{minipage}[b]{\linewidth}\raggedleft
Setting
\end{minipage} & \begin{minipage}[b]{\linewidth}\raggedright
Description
\end{minipage} \\
\midrule\noalign{}
\endfirsthead
\toprule\noalign{}
\begin{minipage}[b]{\linewidth}\raggedright
Category
\end{minipage} & \begin{minipage}[b]{\linewidth}\raggedright
Parameter
\end{minipage} & \begin{minipage}[b]{\linewidth}\raggedleft
Setting
\end{minipage} & \begin{minipage}[b]{\linewidth}\raggedright
Description
\end{minipage} \\
\midrule\noalign{}
\endhead
\bottomrule\noalign{}
\endlastfoot
Workload format & Chakra ET & One .et file per rank & Preserves nodes and dependencies \\
Network backend & NS-3 & All main experiments & Models links, queues, and RDMA flows \\
GPU compute model & Roofline & Configuration driven & Estimated from FLOPs, peak performance, and memory bandwidth \\
Scheduling policy & scheduling-policy & Last-in, first-out (LIFO) & Matches main-matrix system.json \\
Active chunks & active-chunks-per-dimension & 1 & Matches main-matrix configuration \\
Dataset splits & preferred-dataset-splits & 4 & Required by the current collective interface; not an experimental factor \\
Fixed endpoint delay & endpoint-delay & 10 ns & ASTRA-sim injection delay; injection scale is 1 \\
EP Direct concurrency & Direct & EP-1 peers & Covers all other group members \\
Sequence length & T & 4096 & Affects both computation and communication volume \\
\end{longtable}

The main system.json uses common roofline assumptions for the reference PCIe accelerator: a dense-BF16-equivalent peak throughput of 1.0~PFLOP/s and a local-memory bandwidth of approximately 1.6~TB/s per accelerator. These rounded reference values characterize the target performance class without implying device-specific measurement: 1.0~PFLOP/s defines the compute ceiling for compute-bound operations, whereas 1.6~TB/s defines the data-movement ceiling for memory-bound operations. The same pair is applied uniformly across all configurations to avoid false precision and to prevent accelerator variation from confounding comparisons of topology, TP/EP, collectives, and network modes; GPU computation time is therefore an estimate under this fixed roofline assumption.

The network configuration additionally fixes PACKET\_PAYLOAD\_SIZE=1000, L2\_CHUNK\_SIZE=4000, L2\_ACK\_INTERVAL=1, BUFFER\_SIZE=32~MiB, HEADROOM\_FACTOR=3, U\_TARGET=0.95, and ERROR\_RATE\_PER\_LINK=0. HEADROOM\_FACTOR is a dimensionless multiplier applied to the bandwidth--delay headroom estimate. We thus examine algorithmic responses under one fixed set of queue and feedback parameters, not fully tuned HPCC or DCQCN.

The original NS-3 interface initializes every RDMA queue pair with the same congestion-control state from a global CC\_MODE, potentially placing intra-node GPU flows on a NIC-oriented feedback path. To better represent the boundary in which intra-node PCIe traffic is not governed by NIC congestion control but inter-node RDMA traffic is, we use a per-flow policy: enable\_cc=0 when source and destination ranks share a server, and enable\_cc=1 otherwise. Each experiment stores this mapping in cc\_policy.txt, which the run script passes to NS-3 queue pairs through ASTRA\_NS3\_CC\_POLICY.

Disabling queue-pair rate updates alone does not isolate the global HPCC path. The strict local-no-CC modification therefore attaches an internal disable\_cc flag to local flows and skips them during CNP generation, sender-side HPCC/DCQCN state updates, and switch INT sampling. This modification does not remove intra-node communication: local packets are still segmented within the NS-3 topology, traverse GPU/PCIe-switch/UPI links, and contend for links and queues. Only rate-feedback and telemetry states intended for inter-node RDMA are bypassed; inter-node flows still execute the selected DCQCN or HPCC state machine.

\hypertarget{ux6570ux636eux63d0ux53d6ux4e0eux7edfux8ba1ux65b9ux6cd5}{%
\subsection{Data Extraction and Statistical Methods}\label{ux6570ux636eux63d0ux53d6ux4e0eux7edfux8ba1ux65b9ux6cd5}}

For each experimental combination, we parse all sys{[}rank{]} statistics in the ASTRA-sim output and define the rank with the longest completion time as the distributed job's bottleneck rank. GPU computation and communication times are taken from the same rank and converted to seconds. The slowest rather than a fixed rank is used because synchronous execution completes only after its slowest participant. Within output precision and rounding, all configurations satisfy $T_{\mathrm{wall}}\approx T_{\mathrm{GPU}}+T_{\mathrm{comm}}$.

The main four-model matrix contains six topologies, four TP/EP partitions, two collective combinations, and four network modes: 192 configurations per model and 768 model--configuration units in total. We denote this full set by $S_{\mathrm{full}}$ ($6\times4\times2\times4=192$ per model). Three additional matched subsets are used: $S_{\mathrm{feedback}}$ contains the three feedback-enabled modes ($6\times4\times2\times3=144$); $S_{\mathrm{RoCE}}$ contains the two RoCE modes ($6\times4\times2\times2=96$); and $S_{\mathrm{network}}$ excludes Topology~6 and contains $5\times4\times2=40$ configurations per network mode. When aggregating by topology, parallel strategy, algorithm, or network mode, we take the arithmetic mean over combinations of the remaining factors within the stated set. These are descriptive means over the design space, not expected performance for a particular deployment. To prevent means from obscuring interactions, the results also report conditional rankings, representative paired comparisons, and the minimum and maximum completion-time combinations.

For representative congestion diagnoses, we additionally parse queue-length files (qlen.txt), flow-completion-time (FCT) files (fct.txt), and Priority Flow Control (PFC) event files (pfc.txt) to relate straggling flows, peak queues, and flow-control events to the critical path. Because these network statistics do not cover every combination, they explain selected cases only and are excluded from full-matrix means.

Four derived metrics are used: communication fraction, $T_{\mathrm{comm}}/T_{\mathrm{wall}}$; relative overhead of configuration $B$ against baseline $A$, $(B-A)/A$; relative reduction of $A$ against $B$, $(B-A)/B$; and configuration sensitivity, $T_{\mathrm{wall,max}}/T_{\mathrm{wall,min}}$. All conditional rankings are restricted to the same model and statistical subset. Configuration tables without performance ordering list Topologies~1--6, whereas explicitly ranked tables are ordered from best to worst on the stated metric.

\hypertarget{ux4effux771fux51c6ux786eux6027ux9a8cux8bc1}{%
\subsection{Validation of Communication-Performance Credibility}\label{ux4effux771fux51c6ux786eux6027ux9a8cux8bc1}}

When a physical GPU cluster is unavailable, ASTRA-sim provides a practical means of evaluating cluster communication performance from parameterized topology, link, collective, and congestion-control models. We assess the credibility of this approach by comparing the simulated communication time with an independently measured collective microbenchmark. Relative error is calculated as $|T_{\mathrm{sim}}-T_{\mathrm{measured}}|/T_{\mathrm{measured}}\times100\%$.

\begin{longtable}[]{@{}
  >{\raggedright\arraybackslash}p{(\columnwidth - 8\tabcolsep) * \real{0.1667}}
  >{\raggedright\arraybackslash}p{(\columnwidth - 8\tabcolsep) * \real{0.1667}}
  >{\raggedleft\arraybackslash}p{(\columnwidth - 8\tabcolsep) * \real{0.2222}}
  >{\raggedleft\arraybackslash}p{(\columnwidth - 8\tabcolsep) * \real{0.2222}}
  >{\raggedleft\arraybackslash}p{(\columnwidth - 8\tabcolsep) * \real{0.2222}}@{}}
\caption{ASTRA-sim communication-performance validation against a collective microbenchmark}\label{tbl:validation-results}\tabularnewline
\toprule\noalign{}
\begin{minipage}[b]{\linewidth}\raggedright
Validation level
\end{minipage} & \begin{minipage}[b]{\linewidth}\raggedright
Platform and workload
\end{minipage} & \begin{minipage}[b]{\linewidth}\raggedleft
Simulated
\end{minipage} & \begin{minipage}[b]{\linewidth}\raggedleft
Measured
\end{minipage} & \begin{minipage}[b]{\linewidth}\raggedleft
Relative error
\end{minipage} \\
\midrule\noalign{}
\endfirsthead
\toprule\noalign{}
\begin{minipage}[b]{\linewidth}\raggedright
Validation level
\end{minipage} & \begin{minipage}[b]{\linewidth}\raggedright
Platform and workload
\end{minipage} & \begin{minipage}[b]{\linewidth}\raggedleft
Simulated
\end{minipage} & \begin{minipage}[b]{\linewidth}\raggedleft
Measured
\end{minipage} & \begin{minipage}[b]{\linewidth}\raggedleft
Relative error
\end{minipage} \\
\midrule\noalign{}
\endhead
\bottomrule\noalign{}
\endlastfoot
Communication microbenchmark & 32-accelerator validation cluster; four 8-accelerator scale-up domains; four 400-Gbps RoCE NICs per node; NCCL-test All-to-All with 2-GB messages & 107788 µs & 102716 µs & 4.9\% \\
\end{longtable}

The simulated communication time differs from the measured value by only 4.9\%, demonstrating that ASTRA-sim reproduces the communication timescale of the evaluated cluster configuration with close agreement. This result supports the use of ASTRA-sim as a credible platform for assessing cluster communication performance and comparing topology, parallelism, collective algorithms, and congestion-control schemes when the corresponding physical GPU clusters are unavailable.

\hypertarget{ux5b9eux9a8cux7ed3ux679c}{%
\section{Experimental Results}\label{ux5b9eux9a8cux7ed3ux679c}}

\hypertarget{deepseek-v4-flash-ux7ed3ux679c}{%
\subsection{DeepSeek V4 Flash 0731}\label{deepseek-v4-flash-ux7ed3ux679c}}

\hypertarget{ux603bux4f53ux7edfux8ba1}{%
\subsubsection{Overall Statistics}\label{ux603bux4f53ux7edfux8ba1}}

DeepSeek V4 Flash 0731 has approximately 284B parameters, and all 192 simulations completed. Its configuration is hidden\_size=4096, num\_hidden\_layers=43, num\_attention\_heads=64, num\_key\_value\_heads=1, n\_routed\_experts=256, num\_experts\_per\_tok=6, and moe\_intermediate\_size=2048.

\begin{longtable}[]{@{}lr@{}}
\caption{Overall statistics for DeepSeek V4 Flash 0731}\label{tbl:flash-summary}\tabularnewline
\toprule\noalign{}
Metric & Value \\
\midrule\noalign{}
\endfirsthead
\toprule\noalign{}
Metric & Value \\
\midrule\noalign{}
\endhead
\bottomrule\noalign{}
\endlastfoot
Sample count & 192 \\
Wall avg(s) & 0.550729 \\
GPU avg(s) & 0.024281 \\
Comm avg(s) & 0.526448 \\
Comm avg / Wall avg & 95.6\% \\
Wall min(s) & 0.105609 \\
Wall max(s) & 2.172640 \\
\end{longtable}

V4 Flash has the shortest absolute time but remains communication dominated: mean communication time is 21.7$\times$ mean GPU time, and the ratio of these aggregated means gives a 95.6\% communication fraction. Averaging Comm/Wall over configurations instead yields 91.9\%. Because these definitions differ, subsequent aggregate decompositions consistently use Comm avg / Wall avg.

Its maximum completion time is 20.6$\times$ its minimum, showing that a lightweight model is not necessarily insensitive to network configuration. A shorter GPU baseline instead makes an unfavorable TP degree, collective, or physical path amplify total time more strongly in relative terms.

\begin{figure}[H]
  \centering
  \includegraphics[width=\textwidth,height=0.72\textheight,keepaspectratio]{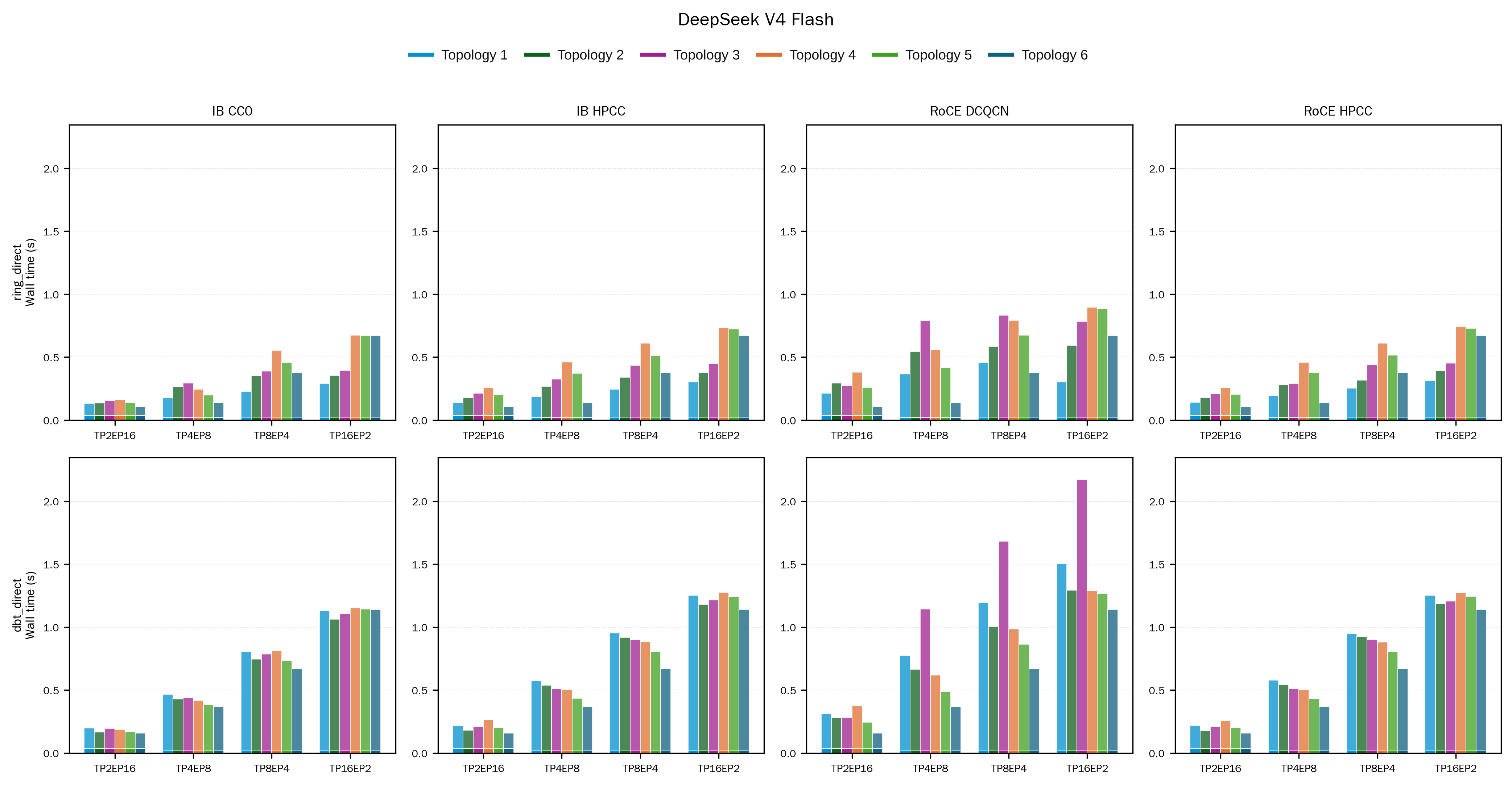}
  \caption{Simulation completion, GPU computation, and communication times for DeepSeek V4 Flash 0731 across 192 configurations}
  \label{fig:flash-matrix}
\end{figure}

Figure~\ref{fig:flash-matrix} presents all 192 Flash combinations. The horizontal axis shows the four TP/EP strategies, each containing bars for Topologies~1--6. Each bar stacks GPU computation (dark) and communication (light), with total height equal to completion time. The two rows correspond to ring\_direct and dbt\_direct; columns show IB CC0, IB HPCC, RoCE DCQCN, and RoCE HPCC. Communication dominates almost every combination, and variation across topology--algorithm combinations substantially exceeds variation in GPU time.

\hypertarget{ux62d3ux6251ux5bf9ux6bd4}{%
\subsubsection{Topology Comparison}\label{ux62d3ux6251ux5bf9ux6bd4}}

Topology ranking depends on the parallel strategy. Under IB CC0 + ring\_direct at TP2EP16, Topology~6 is fastest (0.1056~s), 34.6\% below the slowest, Topology~4 (0.1615~s). At TP16EP2, however, Topology~6 rises to 0.6721~s, 2.32$\times$ Topology~1 (0.2896~s), and joins Topologies~4 (0.6737~s) and 5 (0.6709~s) in the slow group. Its single-node, multi-switch consolidation advantage holds only at low TP; at high TP, concentrated intra-node Ring paths become the bottleneck.

Topology~6 reduces mean completion time by 28.1\% relative to the slowest topology (Topology~3), but is only 11.0\% faster than Topology~1. Eliminating the external network is beneficial but does not dominate every other factor. Although all Topology~1 traffic crosses an external switch, each GPU has a dedicated NIC and avoids intra-node contention for a shared injection point, bringing it close to Topology~6.

Topology~2 is approximately 17.0\% faster than Topology~3. Both use 16 nodes with two GPUs and one NIC per node; their principal difference is Topology~3's dual-CPU/UPI path versus the single PCIe switch aggregating both GPUs in Topology~2. This comparison supports the interpretation that an additional host-bridging layer increases synchronous-path cost, although the data do not separately quantify UPI bytes and queue time and cannot attribute the entire difference to UPI.

Topology~4 is approximately 19.8\% slower than Topology~2, demonstrating that total NIC count is insufficient. In Topology~4, eight GPUs share two NICs per node and may traverse inter-PCIe-switch links; local hot spots can offset the benefits of fewer nodes when multiple collective edges map concurrently to a small set of egresses. Topology~5 lies between them, indicating that additional per-node GPUs and NICs help only when rank-to-NIC mapping uses them evenly.

\hypertarget{tpep-ux7b56ux7565ux5bf9ux6bd4}{%
\subsubsection{TP/EP Strategy Comparison}\label{tpep-ux7b56ux7565ux5bf9ux6bd4}}

The cost of increasing TP varies with topology and algorithm. For Topology 6 + IB CC0 + ring\_direct, completion time increases monotonically from 0.1056~s at TP2EP16 to 0.6721~s at TP16EP2 (6.36$\times$). With dbt\_direct, TP16EP2 reaches 1.1273~s, 7.10$\times$ TP2EP16 (0.1589~s). From TP2EP16 to TP16EP2, mean completion and communication times rise by factors of 4.51 and 5.36. Relative to TP2, TP4, TP8, and TP16 increase completion time by 106.4\%, 232.7\%, and 350.9\%, respectively, revealing a consistent high-TP penalty.

GPU time falls from 0.036198~s at TP2 to 0.017360~s at TP8, confirming that tensor sharding reduces some per-rank computation, but the concurrent communication increase greatly exceeds this saving. The aggregate communication fraction rises from 82.1\% at TP2 to 97.6\% at TP16, making collectives almost solely responsible for the high-TP bottleneck. High-TP optimization should therefore prioritize shortening the AllReduce critical path and improving rank mapping, rather than only increasing peak GPU throughput.

\hypertarget{ux7f51ux7edcux4e0eux62e5ux585eux63a7ux5236ux5bf9ux6bd4}{%
\subsubsection{Network and Congestion-Control Comparison}\label{ux7f51ux7edcux4e0eux62e5ux585eux63a7ux5236ux5bf9ux6bd4}}

Congestion-control penalties concentrate on hot-spot paths. For TP16EP2 + ring\_direct, Topology~3 increases from 0.3946~s under IB CC0 to 0.7830~s under RoCE DCQCN (98\% slower), whereas Topology~1 at the same TP is only 0.3020~s under DCQCN and is barely affected. DCQCN's penalty is therefore concentrated on hot-spot injection paths such as Topology~3 rather than uniformly distributed across topologies.

After excluding Topology~6, which has no external network, IB CC0 has a mean completion time of 0.469332~s, approximately 13.0\% lower than the 0.539725~s of RoCE HPCC. However, IB CC0 contains neither rate feedback nor its convergence process and therefore serves only as a no-feedback baseline. IB HPCC and RoCE HPCC have mean completion times of 0.539281 and 0.539725~s, respectively, a difference of approximately 0.1\%. This similarity primarily reflects their comparable 400-Gbps links and shared HPCC state machine and does not imply equivalence between native IB and RoCE.

After excluding Topology~6, RoCE DCQCN is 35.7\% (0.192850~s) slower than RoCE HPCC, entirely due to communication; GPU time is unchanged. Under the simulated synchronous collective bursts, DCQCN's ECN/CNP feedback and recovery therefore expose more network time than HPCC's telemetry-based feedback. Because Topology~6 has no external NIC, its four modes characterize intra-node communication only and are excluded from inter-node protocol comparisons.

\hypertarget{collective-ux7b97ux6cd5ux5bf9ux6bd4}{%
\subsubsection{Collective-Algorithm Comparison}\label{collective-ux7b97ux6cd5ux5bf9ux6bd4}}

DBT's disadvantage grows with parallel scale. Under Topology 6 + IB CC0, dbt\_direct (0.1589~s) is 50\% slower than ring\_direct (0.1056~s) at TP2EP16; at TP16, it is 1.70$\times$ slower (1.1273 versus 0.6721~s), and the absolute gap grows from 0.053 to 0.455~s.

Across the full matrix, dbt\_direct is 86.0\% (0.331142~s) slower than ring\_direct, entirely due to communication. DBT's theoretically lower phase count benefits startup-latency-dominated cases when tree edges map evenly onto the network. Here, ASTRA-sim DBT sends the full data\_size over each tree edge, whereas Ring shards by rank count; on a hierarchical topology, parent--child tree edges consequently concentrate more readily on a few PCIe/NIC egresses.

This ranking is therefore specific to the present implementations and rank mapping and cannot be generalized to all NCCL Tree/DBT implementations. Stronger mechanistic attribution would require joint analysis of collective timelines, per-port bytes, peak queues, and straggler FCTs.

\hypertarget{ux6700ux4f18ux7ec4ux5408}{%
\subsubsection{Minimum-Completion-Time Combination}\label{ux6700ux4f18ux7ec4ux5408}}

\begin{longtable}[]{@{}
  >{\raggedright\arraybackslash}p{(\columnwidth - 10\tabcolsep) * \real{0.1356}}
  >{\raggedright\arraybackslash}p{(\columnwidth - 10\tabcolsep) * \real{0.0847}}
  >{\raggedright\arraybackslash}p{(\columnwidth - 10\tabcolsep) * \real{0.1186}}
  >{\raggedright\arraybackslash}p{(\columnwidth - 10\tabcolsep) * \real{0.3559}}
  >{\raggedright\arraybackslash}p{(\columnwidth - 10\tabcolsep) * \real{0.1864}}
  >{\raggedleft\arraybackslash}p{(\columnwidth - 10\tabcolsep) * \real{0.1186}}@{}}
\caption{Minimum-completion-time configurations for DeepSeek V4 Flash 0731}\label{tbl:flash-best}\tabularnewline
\toprule\noalign{}
\begin{minipage}[b]{\linewidth}\raggedright
Scope
\end{minipage} & \begin{minipage}[b]{\linewidth}\raggedright
Topology
\end{minipage} & \begin{minipage}[b]{\linewidth}\raggedright
TP/EP
\end{minipage} & \begin{minipage}[b]{\linewidth}\raggedright
Network
\end{minipage} & \begin{minipage}[b]{\linewidth}\raggedright
Algorithm
\end{minipage} & \begin{minipage}[b]{\linewidth}\raggedleft
Wall(s)
\end{minipage} \\
\midrule\noalign{}
\endfirsthead
\toprule\noalign{}
\begin{minipage}[b]{\linewidth}\raggedright
Scope
\end{minipage} & \begin{minipage}[b]{\linewidth}\raggedright
Topology
\end{minipage} & \begin{minipage}[b]{\linewidth}\raggedright
TP/EP
\end{minipage} & \begin{minipage}[b]{\linewidth}\raggedright
Network
\end{minipage} & \begin{minipage}[b]{\linewidth}\raggedright
Algorithm
\end{minipage} & \begin{minipage}[b]{\linewidth}\raggedleft
Wall(s)
\end{minipage} \\
\midrule\noalign{}
\endhead
\bottomrule\noalign{}
\endlastfoot
Global minimum (all four network modes tied) & Topology 6 & TP2EP16 & All four modes & ring\_direct & 0.105609 \\
Minimum with external network & Topology 1 & TP2EP16 & IB CC0 & ring\_direct & 0.131895 \\
\end{longtable}

Both minimum categories identify Topology 6 + TP2EP16 + ring\_direct, showing that topology, parallelism, and the collective have more stable effects than Topology~6's network label. Because this topology has no external network, IB/RoCE labels do not represent inter-node protocol differences. Among external-network designs, Topologies~1, 2, and 5 with TP2EP16 + IB CC0 reach 0.131895, 0.134670, and 0.137794~s, respectively, and provide more meaningful multi-node controls.

\hypertarget{ux4ea4ux4e92ux6548ux5e94ux5c0fux7ed3}{%
\subsubsection{Summary of Interaction Effects}\label{ux4ea4ux4e92ux6548ux5e94ux5c0fux7ed3}}

Flash exhibits three pronounced interactions. First, \textbf{topology $\times$ TP}: Topology~6 moves from fastest at TP2EP16 (0.1056~s) to the slow group at TP16EP2 (0.6721~s), 2.32$\times$ slower than Topology~1 and nearly tied with Topologies~4 and 5. Second, \textbf{algorithm $\times$ TP}: DBT's disadvantage over Ring grows from 1.50$\times$ at TP2 to 1.70$\times$ at TP16. Third, \textbf{network $\times$ topology}: under TP16EP2 + ring\_direct, DCQCN increases the completion time of Topology~3 by 98\% relative to IB CC0, whereas Topology~1 changes little. Communication fraction ranges from 65.7\% for the fastest combination to 99.0\% for the slowest. Flash's short GPU baseline produces the largest relative configuration effect, with a global max/min ratio of 20.6.

\hypertarget{kimi-k3-ux7ed3ux679c}{%
\subsection{Kimi K3}\label{kimi-k3-ux7ed3ux679c}}

\hypertarget{ux603bux4f53ux7edfux8ba1-1}{%
\subsubsection{Overall Statistics}\label{ux603bux4f53ux7edfux8ba1-1}}

\begin{figure}[H]
  \centering
  \includegraphics[width=\textwidth,height=0.72\textheight,keepaspectratio]{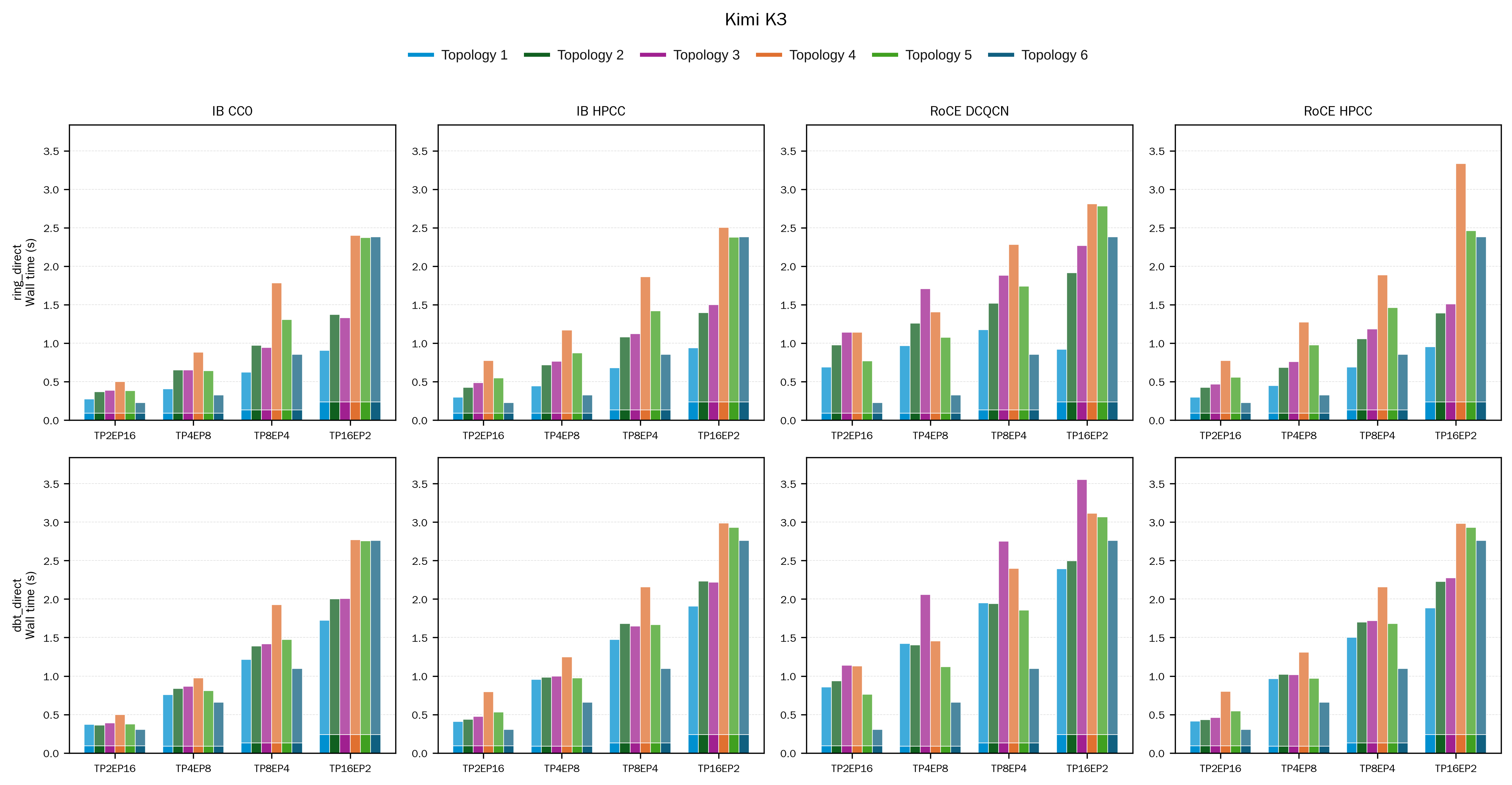}
  \caption{Simulation completion, GPU computation, and communication times for Kimi K3 across 192 configurations}
  \label{fig:kimi-matrix}
\end{figure}

Kimi K3 has approximately 2.8T parameters, a modeled Latent MoE Dimension of 3584\cite{ref23}, 93 layers, 896 routed experts, and top-$k=16$. Its complete $6\times4\times2\times4=192$ simulations cover six topologies, four TP/EP partitions, two collective algorithms, and four network modes. Table~\ref{tbl:kimi-summary} and Figure~\ref{fig:kimi-matrix} use the complete matrix. Cross-model marginal comparisons in Section~4.5 use the common 144-configuration feedback-enabled subset (IB HPCC, RoCE HPCC, and RoCE DCQCN), and network/congestion-control comparisons additionally exclude Topology~6.

\begin{longtable}[]{@{}lr@{}}
\caption{Overall full-matrix statistics for Kimi K3}\label{tbl:kimi-summary}\tabularnewline
\toprule\noalign{}
Metric & Value \\
\midrule\noalign{}
\endfirsthead
\toprule\noalign{}
Metric & Value \\
\midrule\noalign{}
\endhead
\bottomrule\noalign{}
\endlastfoot
Sample count & 192 \\
Wall avg(s) & 1.293425 \\
GPU avg(s) & 0.137182 \\
Comm avg(s) & 1.156243 \\
Comm avg / Wall avg & 89.4\% \\
Wall min(s) & 0.227998 \\
Wall max(s) & 3.554893 \\
\end{longtable}

Kimi K3 remains communication dominated, although its 89.4\% aggregate communication fraction is the lowest of the four models. Its mean GPU time, 0.137182~s, is the highest, so the fixed computational cost reduces relative network differences. Nevertheless, the maximum is 15.6$\times$ the minimum, and configuration still strongly affects completion time.

Figure~\ref{fig:kimi-matrix} shows all 192 Kimi K3 configurations, using normalized local results for every Topology~6 network mode. On this common basis, Kimi K3's mean completion time is approximately 2.35$\times$ that of V4 Flash, while its communication fraction is lower (89.4\% versus 95.6\%). In both RoCE panels of the ring\_direct row, TP2EP16 and TP4EP8 bars are lower than TP8EP4 and TP16EP2, demonstrating the consistent advantage of lower TP under these RoCE conditions.

\hypertarget{ux62d3ux6251ux5bf9ux6bd4-1}{%
\subsubsection{Topology Comparison}\label{ux62d3ux6251ux5bf9ux6bd4-1}}

Topology advantages again depend on parallelism. Under Topology~6 with IB CC0 + ring\_direct, TP2EP16 is fastest (0.2280~s), but TP16EP2 rises to 2.3814~s, 2.63$\times$ Topology~1 (0.9072~s), and joins Topologies~4 (2.3973~s) and 5 (2.3725~s) in the slow group. The reversal matches Flash, although Kimi K3 has higher absolute times.

Within the same 144-configuration feedback-enabled subset used in Table~\ref{tbl:topology-ranking}, Topology~1 has a mean completion time of 1.0256~s, lower than Topology~6's 1.0756~s; relative to Topology~1, Topology~6 is approximately 4.9\% slower. The two achieve low completion times through different mechanisms: Topology~1 distributes inter-node flows over 32 independent NIC injection points, whereas Topology~6 eliminates the external network. Topology~2 averages 1.2636~s, approximately 23.2\% slower than Topology~1, but remains faster than Topologies~3--5.

Topology~4 is approximately 77.7\% slower than Topology~1: eight GPUs share two NICs per node and may span two local PCIe domains, allowing a few heavily loaded egresses to determine the collective tail. Topology~3 is 15.8\% slower than Topology~2, again exposing the additional cost of a dual-CPU/UPI path relative to aggregation on one PCIe switch. Despite more NICs and intra-node links, Topology~5 remains 18.9\% slower than Topology~2 on average; physical ports provide effective bandwidth only when used evenly by the rank mapping.

\hypertarget{tpep-ux7b56ux7565ux5bf9ux6bd4-1}{%
\subsubsection{TP/EP Strategy Comparison}\label{tpep-ux7b56ux7565ux5bf9ux6bd4-1}}

For Topology 6 + IB CC0 + ring\_direct, completion time grows from 0.2280~s at TP2EP16 to 2.3814~s at TP16EP2 (10.4$\times$), exceeding Flash's 6.36$\times$ increase and reflecting Kimi K3's higher communication baseline. In $S_{\mathrm{RoCE}}$, TP4, TP8, and TP16 are 60.2\%, 143.8\%, and 265.2\% slower than TP2, and TP16 communication time is 3.83$\times$ TP2. Although top-$k=16$ raises the EP communication proxy, EP uses Direct; enlarging the EP group does not add serial propagation phases in the manner of a larger TP Ring, so TP2EP16 remains superior.

GPU time does not decrease monotonically with TP and rebounds to 0.236797~s at TP16. This reflects the joint effects of the roofline model, operator partitioning, and critical path; computation does not simply scale as 1/TP. The common token-routing distribution makes this ranking a direct consequence of TP/EP partitioning on the controlled workload's critical path.

\hypertarget{ux7f51ux7edcux4e0eux62e5ux585eux63a7ux5236ux5bf9ux6bd4-1}{%
\subsubsection{Network and Congestion-Control Comparison}\label{ux7f51ux7edcux4e0eux62e5ux585eux63a7ux5236ux5bf9ux6bd4-1}}

For TP16EP2 + ring\_direct, Topology~3 takes 1.3276~s under IB CC0 and 2.2666~s under DCQCN (71\% slower). The relative penalty is below Flash's 98\%, but the absolute increase is larger (0.939 versus 0.388~s).

Excluding Topology~6, RoCE DCQCN is 32.3\% (0.416637~s) slower than RoCE HPCC; unchanged GPU time means the difference is entirely communicational. With Kimi K3's long event sequence and synchronized bursts, feedback and recovery differences accumulate across layers. This percentage quantifies the two RoCE control modes; IB--RoCE comparisons use separate statistics for their corresponding modes.

\hypertarget{collective-ux7b97ux6cd5ux5bf9ux6bd4-1}{%
\subsubsection{Collective-Algorithm Comparison}\label{collective-ux7b97ux6cd5ux5bf9ux6bd4-1}}

Kimi K3's algorithm $\times$ TP interaction is weaker than for the other models. In $S_{\mathrm{RoCE}}$, the DBT/Ring ratio grows from 1.05 at TP2 to 1.32 at TP8, then declines slightly to 1.29 at TP16. Its larger EP communication fraction partially dilutes the TP AllReduce difference in total time. The relationship is weakly nonmonotonic and differs from Flash's monotonic amplification.

dbt\_direct is 26.2\% (0.330850~s) slower than ring\_direct, again entirely due to communication. Its smaller relative penalty than for Flash, GLM-5.3, and V4 Pro shows that DBT cost varies with message sequence, size, and tree-edge mapping, although it still provides no average benefit for Kimi K3.

\hypertarget{ux6700ux4f18ux7ec4ux5408-1}{%
\subsubsection{Minimum-Completion-Time Combination}\label{ux6700ux4f18ux7ec4ux5408-1}}

\begin{longtable}[]{@{}
  >{\raggedright\arraybackslash}p{(\columnwidth - 10\tabcolsep) * \real{0.1633}}
  >{\raggedright\arraybackslash}p{(\columnwidth - 10\tabcolsep) * \real{0.1020}}
  >{\raggedright\arraybackslash}p{(\columnwidth - 10\tabcolsep) * \real{0.1429}}
  >{\raggedright\arraybackslash}p{(\columnwidth - 10\tabcolsep) * \real{0.2245}}
  >{\raggedright\arraybackslash}p{(\columnwidth - 10\tabcolsep) * \real{0.2245}}
  >{\raggedleft\arraybackslash}p{(\columnwidth - 10\tabcolsep) * \real{0.1429}}@{}}
\caption{Minimum-completion-time configurations for Kimi K3}\label{tbl:kimi-best}\tabularnewline
\toprule\noalign{}
\begin{minipage}[b]{\linewidth}\raggedright
Scope
\end{minipage} & \begin{minipage}[b]{\linewidth}\raggedright
Topology
\end{minipage} & \begin{minipage}[b]{\linewidth}\raggedright
TP/EP
\end{minipage} & \begin{minipage}[b]{\linewidth}\raggedright
Network
\end{minipage} & \begin{minipage}[b]{\linewidth}\raggedright
Algorithm
\end{minipage} & \begin{minipage}[b]{\linewidth}\raggedleft
Wall(s)
\end{minipage} \\
\midrule\noalign{}
\endfirsthead
\toprule\noalign{}
\begin{minipage}[b]{\linewidth}\raggedright
Scope
\end{minipage} & \begin{minipage}[b]{\linewidth}\raggedright
Topology
\end{minipage} & \begin{minipage}[b]{\linewidth}\raggedright
TP/EP
\end{minipage} & \begin{minipage}[b]{\linewidth}\raggedright
Network
\end{minipage} & \begin{minipage}[b]{\linewidth}\raggedright
Algorithm
\end{minipage} & \begin{minipage}[b]{\linewidth}\raggedleft
Wall(s)
\end{minipage} \\
\midrule\noalign{}
\endhead
\bottomrule\noalign{}
\endlastfoot
Global minimum (all four network modes tied) & Topology 6 & TP2EP16 & All four modes & ring\_direct & 0.227998 \\
Minimum RoCE with external network & Topology 1 & TP2EP16 & RoCE HPCC & ring\_direct & 0.298447 \\
\end{longtable}

The second row of Table~\ref{tbl:kimi-best} reports only the minimum RoCE configuration with an external network to compare multi-node RoCE designs; it does not indicate that Kimi K3 lacks IB experiments.

The minimum is Topology 6 + TP2EP16 + ring\_direct; its four independent network-mode runs agree because no external network is traversed. Within external-network RoCE runs, Topology~1 with HPCC is fastest, and HPCC is also faster in aggregates over all five external-network topologies. Kimi K3 has the most parameters and highest top-$k$ but a smaller modeled workload dimension, so its TP communication proxy is below those of GLM-5.3 and V4 Pro; total parameter count alone cannot predict completion time.

\hypertarget{ux4ea4ux4e92ux6548ux5e94ux5c0fux7ed3-1}{%
\subsubsection{Summary of Interaction Effects}\label{ux4ea4ux4e92ux6548ux5e94ux5c0fux7ed3-1}}

Kimi K3 shows two notable interactions. First, \textbf{topology $\times$ TP} follows Flash: Topology~6 moves from fastest at TP2 (0.2280~s) to the slow group at TP16 (2.3814~s), 2.63$\times$ Topology~1, although Topology~4 remains slightly slower (2.3973~s). Second, \textbf{algorithm $\times$ TP} is weaker and nonmonotonic: DBT/Ring rises from 1.05 at TP2 to 1.32 at TP8 and falls to 1.29 at TP16, showing that a larger EP communication fraction attenuates the relative effect of TP AllReduce choice on total time.

\hypertarget{glm5.3-ux7ed3ux679c}{%
\subsection{GLM-5.3}\label{glm5.3-ux7ed3ux679c}}

\hypertarget{ux603bux4f53ux7edfux8ba1-2}{%
\subsubsection{Overall Statistics}\label{ux603bux4f53ux7edfux8ba1-2}}

GLM-5.3 has approximately 753B parameters, hidden\_size=6144, 78 layers, 256 routed experts, and top-$k=8$; all 192 simulations completed.

\begin{longtable}[]{@{}lr@{}}
\caption{Overall statistics for GLM-5.3}\label{tbl:glm-summary}\tabularnewline
\toprule\noalign{}
Metric & Value \\
\midrule\noalign{}
\endfirsthead
\toprule\noalign{}
Metric & Value \\
\midrule\noalign{}
\endhead
\bottomrule\noalign{}
\endlastfoot
Sample count & 192 \\
Wall avg(s) & 1.597997 \\
GPU avg(s) & 0.090173 \\
Comm avg(s) & 1.507824 \\
Comm avg / Wall avg & 94.4\% \\
Wall min(s) & 0.339574 \\
Wall max(s) & 5.102144 \\
\end{longtable}

GLM-5.3's mean communication time is 16.7$\times$ its GPU time, and its maximum completion time is 15.0$\times$ its minimum. The 753B parameters primarily determine capacity and sharding requirements; they are not transferred in full by each collective. Completion time is more directly governed by hidden dimension, layer count, activation traffic, and TP/EP mapping.

\begin{figure}[H]
  \centering
  \includegraphics[width=\textwidth,height=0.72\textheight,keepaspectratio]{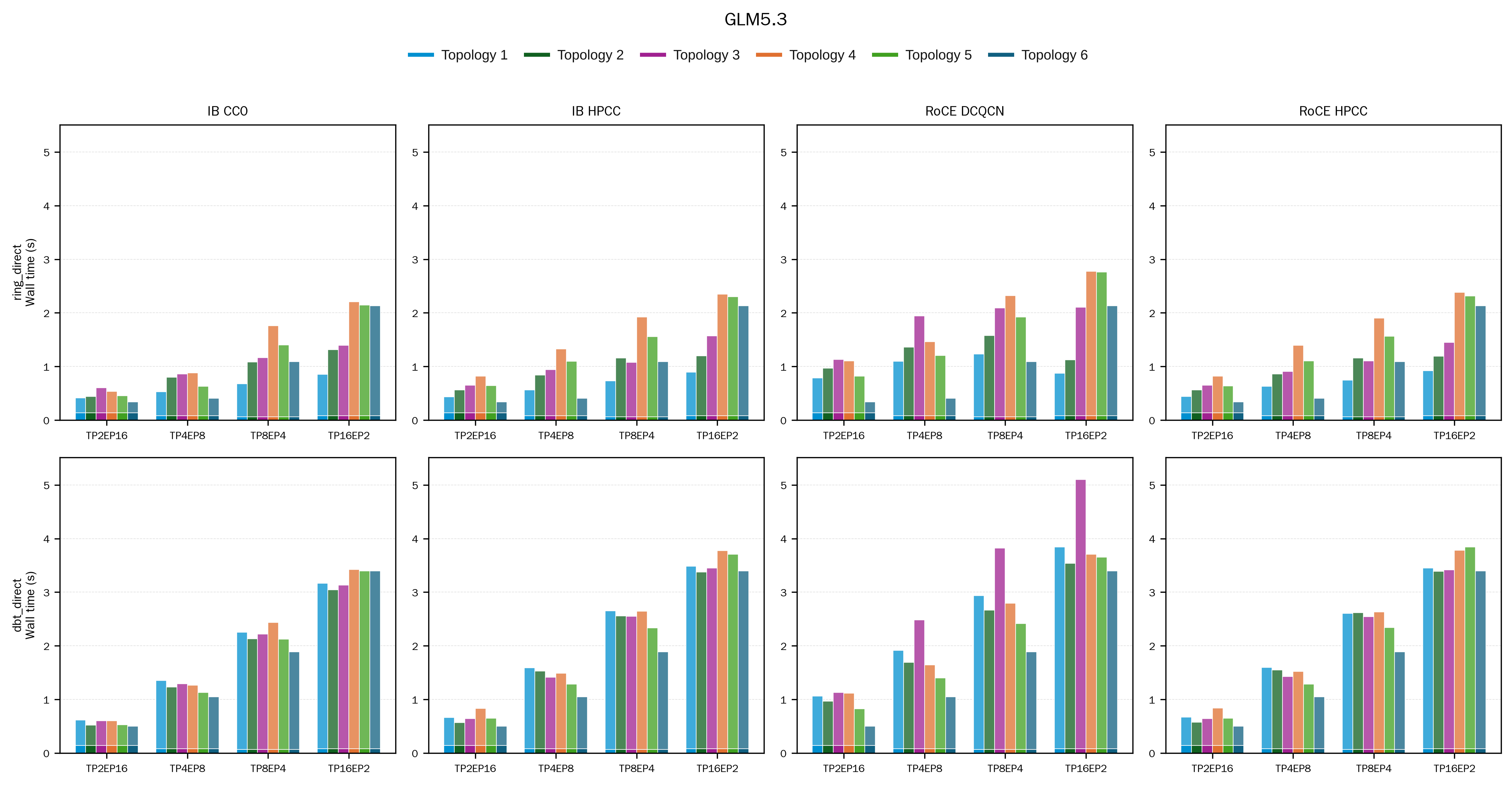}
  \caption{Simulation completion, GPU computation, and communication times for GLM-5.3 across 192 configurations}
  \label{fig:glm-matrix}
\end{figure}

Figure~\ref{fig:glm-matrix} presents the complete GLM-5.3 matrix. Although its parameter count is below Kimi K3 and V4 Pro and above Flash, it has the highest completion time. This agrees with the communication proxy formed by its large hidden dimension, many layers, and top-$k$, confirming that total parameters alone do not predict exposed communication.

\hypertarget{ux62d3ux6251ux5bf9ux6bd4-2}{%
\subsubsection{Topology Comparison}\label{ux62d3ux6251ux5bf9ux6bd4-2}}

Topology benefits again depend on parallelism. Under Topology~6 with IB CC0 + ring\_direct, TP2EP16 is fastest (0.3396~s), whereas TP16EP2 reaches 2.1282~s, 2.50$\times$ Topology~1 (0.8511~s), and joins Topologies~4 (2.2087~s) and 5 (2.1422~s) in the slow group. The pattern matches Flash and Kimi K3.

Topology~6 is 28.5\% faster than Topology~4 (equivalently, Topology~4 is 39.9\% slower) and 5.3\% faster than Topology~1. Both eliminating the external network and providing one NIC per GPU reduce exposed communication, although the latter requires more servers, switch ports, and power.

Topology~2 is 20.4\% faster than Topology~4, indicating that 16 distributed NIC injection points outperform eight shared egresses for this workload. Topologies~5 and 3 are 12.4\% and 15.4\% slower than Topology~2, respectively: the former is affected by multiple GPUs sharing local PCIe/NIC paths, and the latter by UPI and a larger inter-node fraction.

\hypertarget{tpep-ux7b56ux7565ux5bf9ux6bd4-2}{%
\subsubsection{TP/EP Strategy Comparison}\label{tpep-ux7b56ux7565ux5bf9ux6bd4-2}}

For Topology 6 + IB CC0 + ring\_direct, completion time grows from 0.3396~s at TP2EP16 to 2.1282~s at TP16EP2 (6.27$\times$). GLM-5.3's large hidden dimension (6144) gives a high TP-message proxy and stronger high-TP AllReduce amplification.

Relative to TP2, TP4, TP8, and TP16 increase completion time by 79.1\%, 190.4\%, and 306.8\%; TP16 communication time is 5.00$\times$ TP2. GPU time falls from 0.137765~s at TP2 to 0.063432~s at TP8, but the maximum 0.074-s saving is far smaller than the concurrent 1.321-s communication increase.

TP16 GPU time rebounds slightly to 0.079016~s, reflecting operator partitioning, the memory roofline, and dependency critical paths. The overall ranking remains clear: for GLM-5.3, communication cost grows faster than the computational benefit of increasing TP.

\hypertarget{ux7f51ux7edcux4e0eux62e5ux585eux63a7ux5236ux5bf9ux6bd4-2}{%
\subsubsection{Network and Congestion-Control Comparison}\label{ux7f51ux7edcux4e0eux62e5ux585eux63a7ux5236ux5bf9ux6bd4-2}}

For TP16EP2 + ring\_direct, Topology~3 takes 1.3961~s under IB CC0 and 2.1021~s under DCQCN (51\% slower). This relative penalty is lower than those of V4 Flash and Kimi K3 but higher than that of V4 Pro; GLM-5.3's high baseline still yields a substantial 0.706-s absolute increase.

Consistent with Table~\ref{tbl:cc-comparison}, this network-mode comparison excludes Topology~6, which has no external network. IB CC0 has a mean completion time of 1.413744~s, approximately 11.7\% lower than the 1.601189~s of RoCE HPCC, reflecting the difference between the no-rate-feedback baseline and a feedback-enabled mode. IB HPCC averages 1.593600~s, approximately 0.5\% lower than RoCE HPCC. This similarity does not establish performance equivalence between the native protocols; it only indicates that the present parameterizations are close.

Excluding Topology~6, RoCE DCQCN is 23.8\% (0.381462~s) slower than RoCE HPCC, entirely due to communication. Its relative penalty is below Flash's 35.7\%, but the absolute gap remains nearly 0.4~s; either percentage or absolute difference alone omits part of the deployment implication.

\hypertarget{collective-ux7b97ux6cd5ux5bf9ux6bd4-2}{%
\subsubsection{Collective-Algorithm Comparison}\label{collective-ux7b97ux6cd5ux5bf9ux6bd4-2}}

Under Topology 6 + IB CC0, the DBT/Ring ratio grows from 1.47 at TP2 to 1.60 at TP16, matching Flash's direction.

dbt\_direct is 74.1\% (0.863681~s) slower than ring\_direct, approximately 2.3$\times$ the RoCE DCQCN--HPCC gap after excluding Topology~6. Thus, TP collective choice and rank mapping matter more than the congestion-control switch for this GLM-5.3 configuration. Its larger DBT penalty than Kimi K3 also supports an interaction between algorithm and model message sequence.

\hypertarget{ux6700ux4f18ux7ec4ux5408-2}{%
\subsubsection{Minimum-Completion-Time Combination}\label{ux6700ux4f18ux7ec4ux5408-2}}

\begin{longtable}[]{@{}
  >{\raggedright\arraybackslash}p{(\columnwidth - 10\tabcolsep) * \real{0.1356}}
  >{\raggedright\arraybackslash}p{(\columnwidth - 10\tabcolsep) * \real{0.0847}}
  >{\raggedright\arraybackslash}p{(\columnwidth - 10\tabcolsep) * \real{0.1186}}
  >{\raggedright\arraybackslash}p{(\columnwidth - 10\tabcolsep) * \real{0.3559}}
  >{\raggedright\arraybackslash}p{(\columnwidth - 10\tabcolsep) * \real{0.1864}}
  >{\raggedleft\arraybackslash}p{(\columnwidth - 10\tabcolsep) * \real{0.1186}}@{}}
\caption{Minimum-completion-time configurations for GLM-5.3}\label{tbl:glm-best}\tabularnewline
\toprule\noalign{}
\begin{minipage}[b]{\linewidth}\raggedright
Scope
\end{minipage} & \begin{minipage}[b]{\linewidth}\raggedright
Topology
\end{minipage} & \begin{minipage}[b]{\linewidth}\raggedright
TP/EP
\end{minipage} & \begin{minipage}[b]{\linewidth}\raggedright
Network
\end{minipage} & \begin{minipage}[b]{\linewidth}\raggedright
Algorithm
\end{minipage} & \begin{minipage}[b]{\linewidth}\raggedleft
Wall(s)
\end{minipage} \\
\midrule\noalign{}
\endfirsthead
\toprule\noalign{}
\begin{minipage}[b]{\linewidth}\raggedright
Scope
\end{minipage} & \begin{minipage}[b]{\linewidth}\raggedright
Topology
\end{minipage} & \begin{minipage}[b]{\linewidth}\raggedright
TP/EP
\end{minipage} & \begin{minipage}[b]{\linewidth}\raggedright
Network
\end{minipage} & \begin{minipage}[b]{\linewidth}\raggedright
Algorithm
\end{minipage} & \begin{minipage}[b]{\linewidth}\raggedleft
Wall(s)
\end{minipage} \\
\midrule\noalign{}
\endhead
\bottomrule\noalign{}
\endlastfoot
Global minimum (all four network modes tied) & Topology 6 & TP2EP16 & All four modes & ring\_direct & 0.339574 \\
Minimum with external network & Topology 1 & TP2EP16 & IB CC0 & ring\_direct & 0.412796 \\
\end{longtable}

The minimum again occurs at Topology 6 + TP2EP16 + ring\_direct. Its four independent network-mode runs agree, but do not explain inter-node protocol differences because no external network is traversed. Although GLM-5.3 has fewer parameters than Kimi K3, it is slower in the comparable RoCE ring\_direct subset; the cross-model analysis relates this counterintuitive result to hidden dimension, layer count, and communication proxies.

\hypertarget{ux4ea4ux4e92ux6548ux5e94ux5c0fux7ed3-2}{%
\subsubsection{Summary of Interaction Effects}\label{ux4ea4ux4e92ux6548ux5e94ux5c0fux7ed3-2}}

GLM-5.3's interactions match Flash: under topology $\times$ TP, Topology~6 reverses rank and is 2.50$\times$ slower than Topology~1 at TP16; under algorithm $\times$ TP, DBT/Ring rises from 1.47 to 1.60; and, under TP16EP2 + ring\_direct, DCQCN increases Topology~3's completion time by 51\% relative to IB CC0. The slowest combination, Topology 3 + TP16EP2 + dbt + DCQCN, reaches 5.1021~s, the largest individual value among the models, and global sensitivity is 15.0$\times$. Despite fewer parameters than Kimi K3 and V4 Pro, GLM-5.3 has the highest mean completion time, consistent with the communication proxy based on hidden dimension, layers, and top-$k$ rather than total parameters.

\hypertarget{deepseek-v4-pro-0813-ux7ed3ux679c}{%
\subsection{DeepSeek V4 Pro 0813}\label{deepseek-v4-pro-0813-ux7ed3ux679c}}

\hypertarget{ux603bux4f53ux7edfux8ba1-3}{%
\subsubsection{Overall Statistics}\label{ux603bux4f53ux7edfux8ba1-3}}

DeepSeek V4 Pro 0813 has approximately 1.7T parameters, hidden\_size=7168, 61 layers, 384 routed experts, and top-$k=6$; all 192 simulations completed.

\begin{longtable}[]{@{}lr@{}}
\caption{Overall statistics for DeepSeek V4 Pro 0813}\label{tbl:pro-summary}\tabularnewline
\toprule\noalign{}
Metric & Value \\
\midrule\noalign{}
\endfirsthead
\toprule\noalign{}
Metric & Value \\
\midrule\noalign{}
\endhead
\bottomrule\noalign{}
\endlastfoot
Sample count & 192 \\
Wall avg(s) & 1.362831 \\
GPU avg(s) & 0.105370 \\
Comm avg(s) & 1.257461 \\
Comm avg / Wall avg & 92.3\% \\
Wall min(s) & 0.342243 \\
Wall max(s) & 4.259608 \\
\end{longtable}

Its mean communication time is 11.9$\times$ GPU time, and maximum completion time is 12.4$\times$ the minimum. It has the largest hidden dimension but fewer layers and a lower top-$k$ than GLM-5.3 and Kimi K3, so hidden dimension alone does not determine total communication.

\begin{figure}[H]
  \centering
  \includegraphics[width=\textwidth,height=0.72\textheight,keepaspectratio]{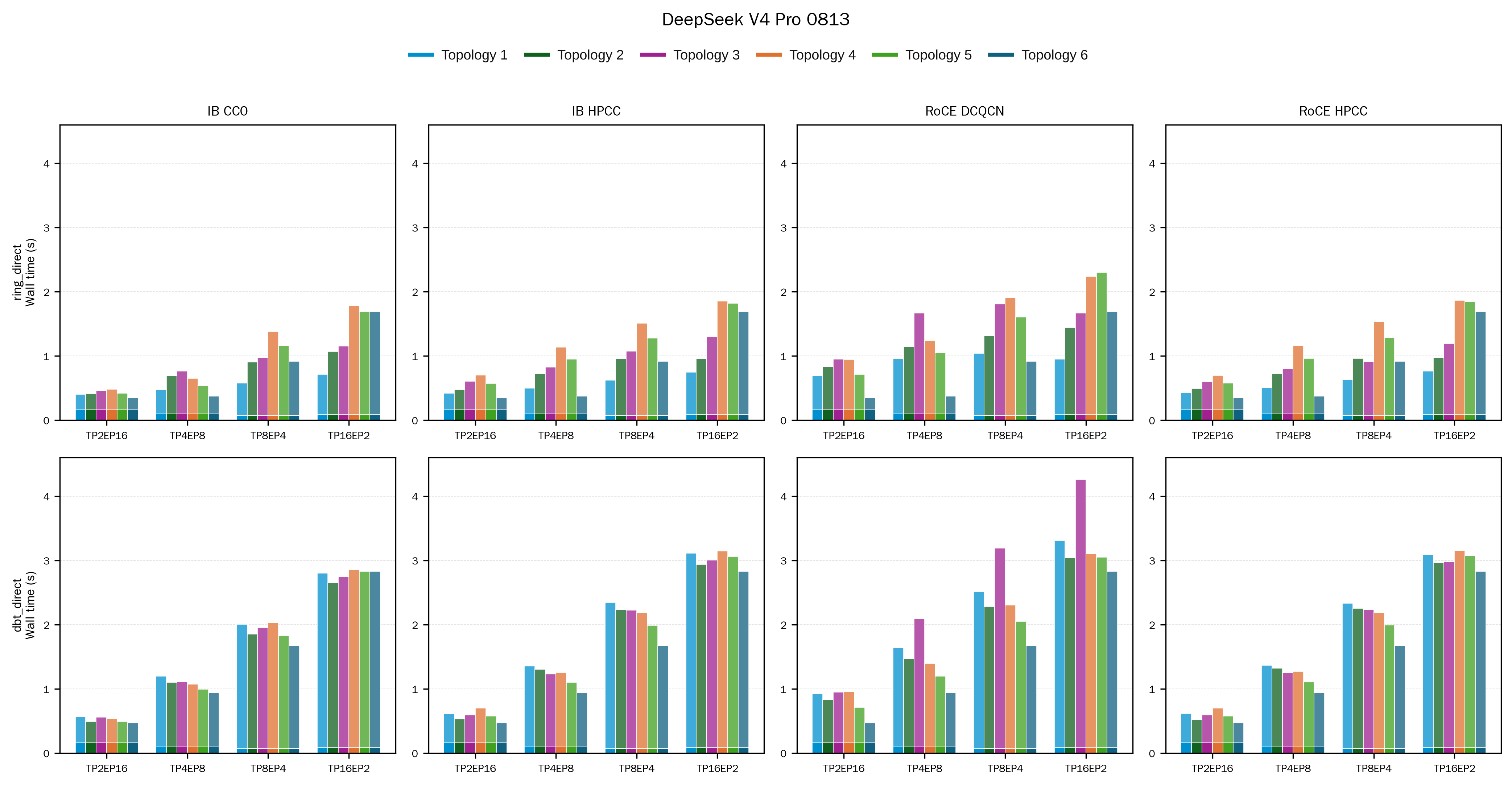}
  \caption{Simulation completion, GPU computation, and communication times for DeepSeek V4 Pro 0813 across 192 configurations}
  \label{fig:pro-matrix}
\end{figure}

Figure~\ref{fig:pro-matrix} shows the complete V4 Pro matrix. Despite its largest hidden dimension (7168), fewer layers and lower top-$k$ place its completion times between Kimi K3 and GLM-5.3, consistent with the strictly comparable means in Section~4.5.

\hypertarget{ux62d3ux6251ux5bf9ux6bd4-3}{%
\subsubsection{Topology Comparison}\label{ux62d3ux6251ux5bf9ux6bd4-3}}

Under Topology~6 with IB CC0 + ring\_direct, TP2EP16 is fastest (0.3422~s), whereas TP16EP2 reaches 1.6863~s, 2.37$\times$ Topology~1 (0.7120~s), and joins Topologies~4 (1.7796~s) and 5 (1.6880~s) in the slow group. This matches the other models.

Topology~6 is 26.1\% faster than Topology~4. Topologies~1 and 2 are 8.9\% and 13.3\% slower than Topology~6, respectively. The better topologies thus share no single physical form; they either eliminate the external network or reduce GPU contention for a single NIC egress.

Topologies~5, 3, and 4 are 22.9\%, 29.3\%, and 35.2\% slower than Topology~6. Topology~5's four NICs do not automatically minimize time, because rank mapping and GPU-to-NIC affinity constrain parallel injection. The UPI path in Topology~3 and shared egresses in Topology~4 create distinct bottlenecks.

\hypertarget{tpep-ux7b56ux7565ux5bf9ux6bd4-3}{%
\subsubsection{TP/EP Strategy Comparison}\label{tpep-ux7b56ux7565ux5bf9ux6bd4-3}}

For Topology 6 + IB CC0 + ring\_direct, completion time grows from 0.3422~s at TP2EP16 to 1.6863~s at TP16EP2 (4.93$\times$). Pro's fewer layers and lower top-$k$ reduce absolute communication relative to GLM-5.3 but do not change the strategy ranking.

Relative to TP2, TP4, TP8, and TP16 increase completion time by 72.6\%, 176.7\%, and 282.9\%; TP16 communication time is 5.15$\times$ TP2. GPU time falls from 0.167020~s at TP2 to 0.072703~s at TP8 (a 0.094-s saving), while communication rises by 1.128~s and continues to determine the ranking.

This pattern closely matches GLM-5.3: increasing TP reduces some computation but increases exposure of repeated TP collectives more rapidly. V4 Pro's fewer layers and lower top-$k$ reduce absolute communication without changing this direction.

\hypertarget{ux7f51ux7edcux4e0eux62e5ux585eux63a7ux5236ux5bf9ux6bd4-3}{%
\subsubsection{Network and Congestion-Control Comparison}\label{ux7f51ux7edcux4e0eux62e5ux585eux63a7ux5236ux5bf9ux6bd4-3}}

For TP16EP2 + ring\_direct, Topology~3 takes 1.1526~s under IB CC0 and 1.6648~s under DCQCN (44\% slower), the smallest relative penalty among the models.

After excluding Topology~6, IB CC0 has a mean completion time of 1.207237~s, approximately 11.3\% lower than the 1.360284~s of RoCE HPCC. RoCE HPCC and IB HPCC average 1.360284 and 1.361341~s, respectively, a difference of approximately 0.1\%. This similarity arises from their comparable modeled links and feedback mechanisms and does not imply equivalence between the native protocols.

Excluding Topology~6, RoCE DCQCN is 24.3\% (0.330629~s) slower than RoCE HPCC, entirely due to communication. The result is close to GLM-5.3's 23.8\%, suggesting similar relative sensitivity to the modeled RoCE control switch for these wide-hidden, lower-top-$k$ models.

\hypertarget{collective-ux7b97ux6cd5ux5bf9ux6bd4-3}{%
\subsubsection{Collective-Algorithm Comparison}\label{collective-ux7b97ux6cd5ux5bf9ux6bd4-3}}

Under Topology 6 + IB CC0, DBT/Ring grows from 1.37 at TP2 to 1.68 at TP16.

dbt\_direct is 77.8\% (0.763545~s) slower than ring\_direct, substantially exceeding the 0.330629-s HPCC--DCQCN gap after excluding Topology~6. Network tuning cannot replace collective-mapping optimization: if logical tree edges remain concentrated on a few egresses, faster feedback can mitigate queues but cannot remove structural hot spots.

\hypertarget{ux6700ux4f18ux7ec4ux5408-3}{%
\subsubsection{Minimum-Completion-Time Combination}\label{ux6700ux4f18ux7ec4ux5408-3}}

\begin{longtable}[]{@{}
  >{\raggedright\arraybackslash}p{(\columnwidth - 10\tabcolsep) * \real{0.1356}}
  >{\raggedright\arraybackslash}p{(\columnwidth - 10\tabcolsep) * \real{0.0847}}
  >{\raggedright\arraybackslash}p{(\columnwidth - 10\tabcolsep) * \real{0.1186}}
  >{\raggedright\arraybackslash}p{(\columnwidth - 10\tabcolsep) * \real{0.3559}}
  >{\raggedright\arraybackslash}p{(\columnwidth - 10\tabcolsep) * \real{0.1864}}
  >{\raggedleft\arraybackslash}p{(\columnwidth - 10\tabcolsep) * \real{0.1186}}@{}}
\caption{Minimum-completion-time configurations for DeepSeek V4 Pro 0813}\label{tbl:pro-best}\tabularnewline
\toprule\noalign{}
\begin{minipage}[b]{\linewidth}\raggedright
Scope
\end{minipage} & \begin{minipage}[b]{\linewidth}\raggedright
Topology
\end{minipage} & \begin{minipage}[b]{\linewidth}\raggedright
TP/EP
\end{minipage} & \begin{minipage}[b]{\linewidth}\raggedright
Network
\end{minipage} & \begin{minipage}[b]{\linewidth}\raggedright
Algorithm
\end{minipage} & \begin{minipage}[b]{\linewidth}\raggedleft
Wall(s)
\end{minipage} \\
\midrule\noalign{}
\endfirsthead
\toprule\noalign{}
\begin{minipage}[b]{\linewidth}\raggedright
Scope
\end{minipage} & \begin{minipage}[b]{\linewidth}\raggedright
Topology
\end{minipage} & \begin{minipage}[b]{\linewidth}\raggedright
TP/EP
\end{minipage} & \begin{minipage}[b]{\linewidth}\raggedright
Network
\end{minipage} & \begin{minipage}[b]{\linewidth}\raggedright
Algorithm
\end{minipage} & \begin{minipage}[b]{\linewidth}\raggedleft
Wall(s)
\end{minipage} \\
\midrule\noalign{}
\endhead
\bottomrule\noalign{}
\endlastfoot
Global minimum (all four network modes tied) & Topology 6 & TP2EP16 & All four modes & ring\_direct & 0.342243 \\
Minimum with external network & Topology 1 & TP2EP16 & IB CC0 & ring\_direct & 0.401587 \\
\end{longtable}

As for Flash, Kimi K3, and GLM-5.3, the minimum is Topology 6 + TP2EP16 + ring\_direct. Under these synthetic workloads, reducing inter-node paths, high-TP Ring phases, and DBT tree-edge hot spots is therefore a more stable performance factor than total model parameters. Topology~6's network labels, however, provide no evidence about native IB versus RoCE.

\hypertarget{ux4ea4ux4e92ux6548ux5e94ux5c0fux7ed3-3}{%
\subsubsection{Summary of Interaction Effects}\label{ux4ea4ux4e92ux6548ux5e94ux5c0fux7ed3-3}}

V4 Pro's interactions match Flash and GLM-5.3: under topology $\times$ TP, Topology~6 reverses rank and is 2.37$\times$ slower than Topology~1 at TP16; under algorithm $\times$ TP, DBT/Ring rises from 1.37 to 1.68; and, under TP16EP2 + ring\_direct, DCQCN increases Topology~3's completion time by 44\% relative to IB CC0. Global sensitivity is 12.4$\times$, the lowest among the models. Its widest hidden dimension and high completion-time baseline make configuration differences smaller in relative terms.

\hypertarget{ux56dbux6a21ux578bux6a2aux5411ux5bf9ux6bd4}{%
\subsection{Cross-Model Comparison}\label{ux56dbux6a21ux578bux6a2aux5411ux5bf9ux6bd4}}

\hypertarget{ux6bd4ux8f83ux53e3ux5f84ux4e0eux603bux4f53ux7edfux8ba1}{%
\subsubsection{Comparison Basis and Overall Statistics}\label{ux6bd4ux8f83ux53e3ux5f84ux4e0eux603bux4f53ux7edfux8ba1}}

For consistent cross-model comparison, model, topology, parallelism, and collective statistics use $S_{\mathrm{feedback}}$: the three feedback-enabled modes (IB HPCC, RoCE HPCC, and RoCE DCQCN), six topologies, four TP/EP partitions, and Ring/Direct and DBT/Direct, giving $3\times6\times4\times2=144$ configurations per model. Topology~6 has no external network; its three labels remain in matched model/topology/parallelism/algorithm comparisons but are not used to infer network-mode differences. Network and congestion-control comparisons use $S_{\mathrm{network}}$, which excludes Topology~6 and leaves $5\times4\times2=40$ configurations per mode and model. The same subsets are extracted from every model's 192-run matrix.

The four configurations differ substantially in communication-related architectural parameters:

\begin{longtable}[]{@{}
  >{\raggedright\arraybackslash}p{(\columnwidth - 18\tabcolsep) * \real{0.1980}}
  >{\raggedleft\arraybackslash}p{(\columnwidth - 18\tabcolsep) * \real{0.0891}}
  >{\raggedleft\arraybackslash}p{(\columnwidth - 18\tabcolsep) * \real{0.0693}}
  >{\raggedleft\arraybackslash}p{(\columnwidth - 18\tabcolsep) * \real{0.0693}}
  >{\raggedleft\arraybackslash}p{(\columnwidth - 18\tabcolsep) * \real{0.0792}}
  >{\raggedleft\arraybackslash}p{(\columnwidth - 18\tabcolsep) * \real{0.1485}}
  >{\raggedleft\arraybackslash}p{(\columnwidth - 18\tabcolsep) * \real{0.0594}}
  >{\raggedleft\arraybackslash}p{(\columnwidth - 18\tabcolsep) * \real{0.1089}}
  >{\raggedleft\arraybackslash}p{(\columnwidth - 18\tabcolsep) * \real{0.0891}}
  >{\raggedleft\arraybackslash}p{(\columnwidth - 18\tabcolsep) * \real{0.0891}}@{}}
\caption{Architectural parameters and communication proxies for the four models}\label{tbl:model-proxies}\tabularnewline
\toprule\noalign{}
\begin{minipage}[b]{\linewidth}\raggedright
Model
\end{minipage} & \begin{minipage}[b]{\linewidth}\raggedleft
Total parameters (B)
\end{minipage} & \begin{minipage}[b]{\linewidth}\raggedleft
workload dim.
\end{minipage} & \begin{minipage}[b]{\linewidth}\raggedleft
layers
\end{minipage} & \begin{minipage}[b]{\linewidth}\raggedleft
Estimated MoE layers
\end{minipage} & \begin{minipage}[b]{\linewidth}\raggedleft
routed experts
\end{minipage} & \begin{minipage}[b]{\linewidth}\raggedleft
top-k
\end{minipage} & \begin{minipage}[b]{\linewidth}\raggedleft
moe hidden
\end{minipage} & \begin{minipage}[b]{\linewidth}\raggedleft
TP proxy
\end{minipage} & \begin{minipage}[b]{\linewidth}\raggedleft
EP proxy
\end{minipage} \\
\midrule\noalign{}
\endfirsthead
\toprule\noalign{}
\begin{minipage}[b]{\linewidth}\raggedright
Model
\end{minipage} & \begin{minipage}[b]{\linewidth}\raggedleft
Total parameters (B)
\end{minipage} & \begin{minipage}[b]{\linewidth}\raggedleft
workload dim.
\end{minipage} & \begin{minipage}[b]{\linewidth}\raggedleft
layers
\end{minipage} & \begin{minipage}[b]{\linewidth}\raggedleft
Estimated MoE layers
\end{minipage} & \begin{minipage}[b]{\linewidth}\raggedleft
routed experts
\end{minipage} & \begin{minipage}[b]{\linewidth}\raggedleft
top-k
\end{minipage} & \begin{minipage}[b]{\linewidth}\raggedleft
moe hidden
\end{minipage} & \begin{minipage}[b]{\linewidth}\raggedleft
TP proxy
\end{minipage} & \begin{minipage}[b]{\linewidth}\raggedleft
EP proxy
\end{minipage} \\
\midrule\noalign{}
\endhead
\bottomrule\noalign{}
\endlastfoot
DeepSeek V4 Flash 0731 & 284 & 4096 & 43 & 43 & 256 & 6 & 2048 & 176128 & 1056768 \\
Kimi K3 & 2800 & 3584 & 93 & 92 & 896 & 16 & 3072 & 333312 & 5275648 \\
GLM-5.3 & 753 & 6144 & 78 & 75 & 256 & 8 & 2048 & 479232 & 3686400 \\
DeepSeek V4 Pro 0813 & 1700 & 7168 & 61 & 61 & 384 & 6 & 3072 & 437248 & 2623488 \\
\end{longtable}

MoE layers are calculated as $\mathrm{num\_hidden\_layers}-\mathrm{first\_k\_dense\_replace}$; models without leading dense layers use their total layer count. Kimi K3 thus has 92 MoE layers after subtracting one dense layer from 93, and GLM-5.3 has 75 after subtracting three from 78.

The TP proxy is $h\times layers$ and the EP proxy is $h\times\mathrm{top\text{-}k}\times\mathrm{MoE\ layers}$, where $h$ is the workload dimension defined in Table~\ref{tbl:model-parameters}. They explain architectural trends rather than actual per-operator bytes. Total parameters primarily characterize capacity and sharding, whereas a collective message depends more directly on token count, workload dimension, layer count, top-$k$, parallel degree, and algorithm.

Overall results for the three-mode comparable subset, ordered by completion time, are:

\begin{longtable}[]{@{}
  >{\raggedright\arraybackslash}p{(\columnwidth - 14\tabcolsep) * \real{0.2174}}
  >{\raggedleft\arraybackslash}p{(\columnwidth - 14\tabcolsep) * \real{0.0435}}
  >{\raggedleft\arraybackslash}p{(\columnwidth - 14\tabcolsep) * \real{0.1304}}
  >{\raggedleft\arraybackslash}p{(\columnwidth - 14\tabcolsep) * \real{0.1196}}
  >{\raggedleft\arraybackslash}p{(\columnwidth - 14\tabcolsep) * \real{0.1304}}
  >{\raggedleft\arraybackslash}p{(\columnwidth - 14\tabcolsep) * \real{0.0978}}
  >{\raggedleft\arraybackslash}p{(\columnwidth - 14\tabcolsep) * \real{0.1304}}
  >{\raggedleft\arraybackslash}p{(\columnwidth - 14\tabcolsep) * \real{0.1304}}@{}}
\caption{Overall four-model statistics under three feedback-enabled network modes}\label{tbl:model-comparable-summary}\tabularnewline
\toprule\noalign{}
\begin{minipage}[b]{\linewidth}\raggedright
Model
\end{minipage} & \begin{minipage}[b]{\linewidth}\raggedleft
$N$
\end{minipage} & \begin{minipage}[b]{\linewidth}\raggedleft
Wall avg(s)
\end{minipage} & \begin{minipage}[b]{\linewidth}\raggedleft
GPU avg(s)
\end{minipage} & \begin{minipage}[b]{\linewidth}\raggedleft
Comm avg(s)
\end{minipage} & \begin{minipage}[b]{\linewidth}\raggedleft
Comm/Wall
\end{minipage} & \begin{minipage}[b]{\linewidth}\raggedleft
Wall min(s)
\end{minipage} & \begin{minipage}[b]{\linewidth}\raggedleft
Wall max(s)
\end{minipage} \\
\midrule\noalign{}
\endfirsthead
\toprule\noalign{}
\begin{minipage}[b]{\linewidth}\raggedright
Model
\end{minipage} & \begin{minipage}[b]{\linewidth}\raggedleft
$N$
\end{minipage} & \begin{minipage}[b]{\linewidth}\raggedleft
Wall avg(s)
\end{minipage} & \begin{minipage}[b]{\linewidth}\raggedleft
GPU avg(s)
\end{minipage} & \begin{minipage}[b]{\linewidth}\raggedleft
Comm avg(s)
\end{minipage} & \begin{minipage}[b]{\linewidth}\raggedleft
Comm/Wall
\end{minipage} & \begin{minipage}[b]{\linewidth}\raggedleft
Wall min(s)
\end{minipage} & \begin{minipage}[b]{\linewidth}\raggedleft
Wall max(s)
\end{minipage} \\
\midrule\noalign{}
\endhead
\bottomrule\noalign{}
\endlastfoot
DeepSeek V4 Flash 0731 & 144 & 0.578756 & 0.024281 & 0.554475 & 95.8\% & 0.105609 & 2.172640 \\
Kimi K3 & 144 & 1.358810 & 0.137182 & 1.221628 & 89.9\% & 0.227998 & 3.554893 \\
DeepSeek V4 Pro 0813 & 144 & 1.417749 & 0.105370 & 1.312379 & 92.6\% & 0.342243 & 4.259608 \\
GLM-5.3 & 144 & 1.663011 & 0.090173 & 1.572838 & 94.6\% & 0.339574 & 5.102144 \\
\end{longtable}

\begin{center}
\small\bfseries DeepSeek V4 Flash 0731 \ensuremath{<} Kimi K3 \ensuremath{<} DeepSeek V4 Pro 0813 \ensuremath{<} GLM-5.3
\end{center}

Flash is 57.4\%, 59.2\%, and 65.2\% faster than Kimi K3, V4 Pro, and GLM-5.3, consistent with fewer layers and the lowest TP/EP proxies. GLM-5.3 has only 26.9\% as many parameters as Kimi K3 but a 22.4\% longer mean completion time; 34.3\% lower GPU time is outweighed by 28.7\% higher communication, reversing the parameter-count intuition.

Communication accounts for 89.9\%--95.8\% across models. Flash's max/min ratio is 20.6, above Kimi K3 (15.6), GLM-5.3 (15.0), and V4 Pro (12.4). Its shortest absolute time but largest relative variation shows that a short GPU baseline magnifies the relative effect of slow configurations.

\hypertarget{ux62d3ux6251ux6392ux5e8fux6a2aux5411ux5bf9ux6bd4}{%
\subsubsection{Cross-Model Topology Rankings}\label{ux62d3ux6251ux6392ux5e8fux6a2aux5411ux5bf9ux6bd4}}

Topology rankings in the 144-configuration comparable subset are:

\begin{longtable}[]{@{}
  >{\raggedright\arraybackslash}p{(\columnwidth - 2\tabcolsep) * \real{0.3333}}
  >{\raggedright\arraybackslash}p{(\columnwidth - 2\tabcolsep) * \real{0.6667}}@{}}
\caption{Marginal topology rankings for the four models}\label{tbl:topology-ranking}\tabularnewline
\toprule\noalign{}
\begin{minipage}[b]{\linewidth}\raggedright
Model
\end{minipage} & \begin{minipage}[b]{\linewidth}\raggedright
Topology ranking, best to worst
\end{minipage} \\
\midrule\noalign{}
\endfirsthead
\toprule\noalign{}
\begin{minipage}[b]{\linewidth}\raggedright
Model
\end{minipage} & \begin{minipage}[b]{\linewidth}\raggedright
Topology ranking, best to worst
\end{minipage} \\
\midrule\noalign{}
\endhead
\bottomrule\noalign{}
\endlastfoot
DeepSeek V4 Flash 0731 & Topology 6, Topology 1, Topology 2, Topology 5, Topology 4, Topology 3 \\
Kimi K3 & Topology 1, Topology 6, Topology 2, Topology 3, Topology 5, Topology 4 \\
GLM-5.3 & Topology 6, Topology 1, Topology 2, Topology 5, Topology 3, Topology 4 \\
DeepSeek V4 Pro 0813 & Topology 6, Topology 1, Topology 2, Topology 5, Topology 3, Topology 4 \\
\end{longtable}

Rankings are stable: Topologies~1, 2, and 6 occupy the top three for every model. Topology~6 ranks first for Flash, GLM-5.3, and V4 Pro and second for Kimi K3, for which Topology~1 leads. Topology~5 ranks fourth for three models and fifth for Kimi K3. Eliminating the external network, increasing independent GPU--NIC injection, and reducing intra-node NIC sharing can all create low-time regions, but their relative benefits depend on the model message sequence.

The lower rankings of Topologies~3--5 are model dependent. Topology~3 risks dual-CPU/UPI costs and a high inter-node fraction; Topology~4 risks eight GPUs sharing two NICs and TP/EP edges crossing PCIe domains. Message sizes and event phases determine which bottleneck enters the critical path first. Topology~3 rises to fourth for Kimi K3 while Topology~4 remains last; for the other models, Topology~5 is fourth and Topologies~3 and 4 occupy the final two positions.

Topology~5's ranking shows that four NICs and fully connected PCIe switches provide only theoretical capacity: if rank mapping concentrates inter-node edges on selected GPU-to-NIC egresses, unused links cannot shorten the slowest collective. Topology~1's high rank likewise does not imply that more nodes are universally better; ports, power, failure domains, and cost for its 32 single-GPU servers are absent from completion time, making it only a performance reference for maximal independent injection.

\hypertarget{tpep-ux7b56ux7565ux6392ux5e8fux6a2aux5411ux5bf9ux6bd4}{%
\subsubsection{Cross-Model TP/EP Rankings}\label{tpep-ux7b56ux7565ux6392ux5e8fux6a2aux5411ux5bf9ux6bd4}}

\begin{longtable}[]{@{}
  >{\raggedright\arraybackslash}p{(\columnwidth - 8\tabcolsep) * \real{0.2439}}
  >{\raggedleft\arraybackslash}p{(\columnwidth - 8\tabcolsep) * \real{0.1951}}
  >{\raggedleft\arraybackslash}p{(\columnwidth - 8\tabcolsep) * \real{0.1829}}
  >{\raggedleft\arraybackslash}p{(\columnwidth - 8\tabcolsep) * \real{0.1829}}
  >{\raggedleft\arraybackslash}p{(\columnwidth - 8\tabcolsep) * \real{0.1951}}@{}}
\caption{Mean simulation completion time for four TP/EP strategies}\label{tbl:parallel-wall}\tabularnewline
\toprule\noalign{}
\begin{minipage}[b]{\linewidth}\raggedright
Model
\end{minipage} & \begin{minipage}[b]{\linewidth}\raggedleft
TP2EP16 Wall(s)
\end{minipage} & \begin{minipage}[b]{\linewidth}\raggedleft
TP4EP8 Wall(s)
\end{minipage} & \begin{minipage}[b]{\linewidth}\raggedleft
TP8EP4 Wall(s)
\end{minipage} & \begin{minipage}[b]{\linewidth}\raggedleft
TP16EP2 Wall(s)
\end{minipage} \\
\midrule\noalign{}
\endfirsthead
\toprule\noalign{}
\begin{minipage}[b]{\linewidth}\raggedright
Model
\end{minipage} & \begin{minipage}[b]{\linewidth}\raggedleft
TP2EP16 Wall(s)
\end{minipage} & \begin{minipage}[b]{\linewidth}\raggedleft
TP4EP8 Wall(s)
\end{minipage} & \begin{minipage}[b]{\linewidth}\raggedleft
TP8EP4 Wall(s)
\end{minipage} & \begin{minipage}[b]{\linewidth}\raggedleft
TP16EP2 Wall(s)
\end{minipage} \\
\midrule\noalign{}
\endhead
\bottomrule\noalign{}
\endlastfoot
DeepSeek V4 Flash 0731 & 0.216811 & 0.450069 & 0.704890 & 0.943253 \\
Kimi K3 & 0.596076 & 0.982192 & 1.532131 & 2.324843 \\
GLM-5.3 & 0.702807 & 1.246371 & 1.973669 & 2.729197 \\
DeepSeek V4 Pro 0813 & 0.623923 & 1.071117 & 1.679061 & 2.296895 \\
\end{longtable}

All four models satisfy:

\begin{center}
\small\bfseries TP2EP16 \ensuremath{<} TP4EP8 \ensuremath{<} TP8EP4 \ensuremath{<} TP16EP2
\end{center}

Normalizing TP2EP16 to one compares relative sensitivity:

\begin{longtable}[]{@{}lrrrr@{}}
\caption{Completion time normalized to TP2EP16}\label{tbl:parallel-normalized}\tabularnewline
\toprule\noalign{}
Model & TP2EP16 & TP4EP8 & TP8EP4 & TP16EP2 \\
\midrule\noalign{}
\endfirsthead
\toprule\noalign{}
Model & TP2EP16 & TP4EP8 & TP8EP4 & TP16EP2 \\
\midrule\noalign{}
\endhead
\bottomrule\noalign{}
\endlastfoot
DeepSeek V4 Flash 0731 & 1.00 & 2.08 & 3.25 & 4.35 \\
Kimi K3 & 1.00 & 1.65 & 2.57 & 3.90 \\
GLM-5.3 & 1.00 & 1.77 & 2.81 & 3.88 \\
DeepSeek V4 Pro 0813 & 1.00 & 1.72 & 2.69 & 3.68 \\
\end{longtable}

Increasing TP2 to TP16 raises completion time by 3.68--4.35$\times$. Flash has the lowest absolute communication but the largest relative degradation; Kimi K3 grows least from TP2 to TP4 but still reaches 3.90$\times$ at TP16. TP AllReduce phases and physical paths therefore amplify more strongly than the cost of enlarging an EP Direct group.

Official weight footprints show that the largest checkpoint shard is 48.75~GB per rank, so all models satisfy the weight-residency criterion of at least 64~GB per accelerator. Once residency is satisfied, lower TP reduces AllReduce participation and serial phases while higher EP carries expert-dimension scaling; TP2EP16 consequently has the lowest strategy-mean completion time for every workload.

\hypertarget{ux7f51ux7edcux4e0eux62e5ux585eux63a7ux5236ux6392ux5e8fux6a2aux5411ux5bf9ux6bd4}{%
\subsubsection{Cross-Model Network and Congestion-Control Rankings}\label{ux7f51ux7edcux4e0eux62e5ux585eux63a7ux5236ux6392ux5e8fux6a2aux5411ux5bf9ux6bd4}}

Network-mode comparisons exclude Topology~6 and retain Topologies~1--5. Each mode and model covers five topologies, four TP/EP partitions, and two collective combinations ($5\times4\times2=40$ samples). Mean times for IB HPCC, RoCE HPCC, and RoCE DCQCN are:

\begin{longtable}[]{@{}
  >{\raggedright\arraybackslash}p{(\columnwidth - 10\tabcolsep) * \real{0.245}}
  >{\raggedleft\arraybackslash}p{(\columnwidth - 10\tabcolsep) * \real{0.145}}
  >{\raggedleft\arraybackslash}p{(\columnwidth - 10\tabcolsep) * \real{0.145}}
  >{\raggedleft\arraybackslash}p{(\columnwidth - 10\tabcolsep) * \real{0.145}}
  >{\raggedleft\arraybackslash}p{(\columnwidth - 10\tabcolsep) * \real{0.145}}
  >{\raggedleft\arraybackslash}p{(\columnwidth - 10\tabcolsep) * \real{0.175}}@{}}
\caption{Conditional means for IB HPCC, RoCE HPCC, and RoCE DCQCN}\label{tbl:cc-comparison}\tabularnewline
\toprule\noalign{}
\begin{minipage}[b]{\linewidth}\raggedright
Model
\end{minipage} & \begin{minipage}[b]{\linewidth}\raggedleft
IB HPCC(s)
\end{minipage} & \begin{minipage}[b]{\linewidth}\raggedleft
RoCE HPCC(s)
\end{minipage} & \begin{minipage}[b]{\linewidth}\raggedleft
IB relative to RoCE HPCC
\end{minipage} & \begin{minipage}[b]{\linewidth}\raggedleft
RoCE DCQCN(s)
\end{minipage} & \begin{minipage}[b]{\linewidth}\raggedleft
DCQCN relative increase
\end{minipage} \\
\midrule\noalign{}
\endfirsthead
\toprule\noalign{}
\begin{minipage}[b]{\linewidth}\raggedright
Model
\end{minipage} & \begin{minipage}[b]{\linewidth}\raggedleft
IB HPCC(s)
\end{minipage} & \begin{minipage}[b]{\linewidth}\raggedleft
RoCE HPCC(s)
\end{minipage} & \begin{minipage}[b]{\linewidth}\raggedleft
IB relative to RoCE HPCC
\end{minipage} & \begin{minipage}[b]{\linewidth}\raggedleft
RoCE DCQCN(s)
\end{minipage} & \begin{minipage}[b]{\linewidth}\raggedleft
DCQCN relative increase
\end{minipage} \\
\midrule\noalign{}
\endhead
\bottomrule\noalign{}
\endlastfoot
DeepSeek V4 Flash 0731 & 0.539281 & 0.539725 & -0.1\% & 0.732575 & +35.7\% \\
Kimi K3 & 1.251844 & 1.288947 & -2.9\% & 1.705585 & +32.3\% \\
GLM-5.3 & 1.593600 & 1.601189 & -0.5\% & 1.982651 & +23.8\% \\
DeepSeek V4 Pro 0813 & 1.361341 & 1.360284 & +0.1\% & 1.690913 & +24.3\% \\
\end{longtable}

\begin{center}
\small\bfseries IB HPCC \ensuremath{\approx} RoCE HPCC \ensuremath{<} RoCE DCQCN
\end{center}

IB HPCC--RoCE HPCC differences are much smaller than RoCE HPCC--DCQCN differences. Relative to RoCE HPCC, IB HPCC is 0.1\%, 2.9\%, and 0.5\% faster for Flash, Kimi K3, and GLM-5.3, and 0.1\% slower for V4 Pro. After equally aggregating these four per-model conditional means, IB HPCC is 0.9\% lower. In this 400-Gbps single-switch configuration, the 0.7- versus 1-µs endpoint-to-switch delay does not translate proportionally into workload completion time. Message serialization, shared-egress queues, synchronization, and the common HPCC feedback process jointly determine the critical path.

RoCE DCQCN is 23.8\%--35.7\% slower than RoCE HPCC for every model. Flash has the largest relative increase (35.7\%) because of its short baseline; Kimi K3 has the largest absolute increase (0.416637~s), consistent with accumulated feedback and recovery costs over a longer communication sequence.

GPU time is unchanged across modes for a given model, so all differences correspond to communication. This supports attribution to network transport and congestion feedback, but means alone cannot separate queue buildup, flow-control pauses, and straggler flows. Mechanistic interpretation requires qlen, pfc, and fct, and the percentages should not be extrapolated directly to production networks with different fabrics and traffic.

\hypertarget{collective-ux7b97ux6cd5ux6392ux5e8fux6a2aux5411ux5bf9ux6bd4}{%
\subsubsection{Cross-Model Collective-Algorithm Rankings}\label{collective-ux7b97ux6cd5ux6392ux5e8fux6a2aux5411ux5bf9ux6bd4}}

For consistency, this table uses the 144-configuration subset per model, with 72 configurations per algorithm:

\begin{longtable}[]{@{}
  >{\raggedright\arraybackslash}p{(\columnwidth - 8\tabcolsep) * \real{0.2778}}
  >{\raggedleft\arraybackslash}p{(\columnwidth - 8\tabcolsep) * \real{0.2083}}
  >{\raggedleft\arraybackslash}p{(\columnwidth - 8\tabcolsep) * \real{0.1944}}
  >{\raggedleft\arraybackslash}p{(\columnwidth - 8\tabcolsep) * \real{0.1528}}
  >{\raggedleft\arraybackslash}p{(\columnwidth - 8\tabcolsep) * \real{0.1667}}@{}}
\caption{Conditional means for Ring/Direct and DBT/Direct}\label{tbl:collective-comparison}\tabularnewline
\toprule\noalign{}
\begin{minipage}[b]{\linewidth}\raggedright
Model
\end{minipage} & \begin{minipage}[b]{\linewidth}\raggedleft
ring\_direct(s)
\end{minipage} & \begin{minipage}[b]{\linewidth}\raggedleft
dbt\_direct(s)
\end{minipage} & \begin{minipage}[b]{\linewidth}\raggedleft
Absolute increase (s)
\end{minipage} & \begin{minipage}[b]{\linewidth}\raggedleft
DBT relative increase
\end{minipage} \\
\midrule\noalign{}
\endfirsthead
\toprule\noalign{}
\begin{minipage}[b]{\linewidth}\raggedright
Model
\end{minipage} & \begin{minipage}[b]{\linewidth}\raggedleft
ring\_direct(s)
\end{minipage} & \begin{minipage}[b]{\linewidth}\raggedleft
dbt\_direct(s)
\end{minipage} & \begin{minipage}[b]{\linewidth}\raggedleft
Absolute increase (s)
\end{minipage} & \begin{minipage}[b]{\linewidth}\raggedleft
DBT relative increase
\end{minipage} \\
\midrule\noalign{}
\endhead
\bottomrule\noalign{}
\endlastfoot
DeepSeek V4 Flash 0731 & 0.408715 & 0.748797 & 0.340082 & 83.2\% \\
Kimi K3 & 1.190384 & 1.527237 & 0.336852 & 28.3\% \\
GLM-5.3 & 1.220223 & 2.105799 & 0.885576 & 72.6\% \\
DeepSeek V4 Pro 0813 & 1.030751 & 1.804748 & 0.773997 & 75.1\% \\
\end{longtable}

\begin{center}
\small\bfseries ring\_direct \ensuremath{<} dbt\_direct
\end{center}

Ring wins for every model, but DBT degradation ranges from 28.3\% for Kimi K3 to 83.2\% for Flash. GLM-5.3 has the largest absolute DBT cost (0.885576~s), approximately 2.3$\times$ its DCQCN--HPCC gap after excluding Topology~6; V4 Pro similarly shows a larger algorithm than congestion-control effect.

If DBT incurred only fixed startup overhead, degradation would be more uniform. The observed range implicates tree-edge message size, layer count, synchronization phases, and egress mapping. Kimi K3's smallest relative penalty does not make DBT preferable; its higher GPU/communication baseline only dilutes the extra cost, and Ring remains faster on average.

\hypertarget{ux63a8ux8350ux7ec4ux5408ux4e0eux7efcux5408ux6392ux5e8f}{%
\subsubsection{Low-Time Combinations and Overall Ranking}\label{ux63a8ux8350ux7ec4ux5408ux4e0eux7efcux5408ux6392ux5e8f}}

Restricting the comparable subset to ring\_direct gives:

\begin{longtable}[]{@{}
  >{\raggedright\arraybackslash}p{(\columnwidth - 12\tabcolsep) * \real{0.2410}}
  >{\raggedleft\arraybackslash}p{(\columnwidth - 12\tabcolsep) * \real{0.0482}}
  >{\raggedleft\arraybackslash}p{(\columnwidth - 12\tabcolsep) * \real{0.1446}}
  >{\raggedleft\arraybackslash}p{(\columnwidth - 12\tabcolsep) * \real{0.1325}}
  >{\raggedleft\arraybackslash}p{(\columnwidth - 12\tabcolsep) * \real{0.1446}}
  >{\raggedleft\arraybackslash}p{(\columnwidth - 12\tabcolsep) * \real{0.1446}}
  >{\raggedleft\arraybackslash}p{(\columnwidth - 12\tabcolsep) * \real{0.1446}}@{}}
\caption{Four-model statistics for the Ring/Direct subset}\label{tbl:ring-subset}\tabularnewline
\toprule\noalign{}
\begin{minipage}[b]{\linewidth}\raggedright
Model
\end{minipage} & \begin{minipage}[b]{\linewidth}\raggedleft
$N$
\end{minipage} & \begin{minipage}[b]{\linewidth}\raggedleft
Wall avg(s)
\end{minipage} & \begin{minipage}[b]{\linewidth}\raggedleft
GPU avg(s)
\end{minipage} & \begin{minipage}[b]{\linewidth}\raggedleft
Comm avg(s)
\end{minipage} & \begin{minipage}[b]{\linewidth}\raggedleft
Wall min(s)
\end{minipage} & \begin{minipage}[b]{\linewidth}\raggedleft
Wall max(s)
\end{minipage} \\
\midrule\noalign{}
\endfirsthead
\toprule\noalign{}
\begin{minipage}[b]{\linewidth}\raggedright
Model
\end{minipage} & \begin{minipage}[b]{\linewidth}\raggedleft
$N$
\end{minipage} & \begin{minipage}[b]{\linewidth}\raggedleft
Wall avg(s)
\end{minipage} & \begin{minipage}[b]{\linewidth}\raggedleft
GPU avg(s)
\end{minipage} & \begin{minipage}[b]{\linewidth}\raggedleft
Comm avg(s)
\end{minipage} & \begin{minipage}[b]{\linewidth}\raggedleft
Wall min(s)
\end{minipage} & \begin{minipage}[b]{\linewidth}\raggedleft
Wall max(s)
\end{minipage} \\
\midrule\noalign{}
\endhead
\bottomrule\noalign{}
\endlastfoot
DeepSeek V4 Flash 0731 & 72 & 0.408715 & 0.024281 & 0.384434 & 0.105609 & 0.894941 \\
DeepSeek V4 Pro 0813 & 72 & 1.030751 & 0.105370 & 0.925380 & 0.342243 & 2.300115 \\
Kimi K3 & 72 & 1.190384 & 0.137182 & 1.053202 & 0.227998 & 3.332118 \\
GLM-5.3 & 72 & 1.220223 & 0.090173 & 1.130050 & 0.339574 & 2.773054 \\
\end{longtable}

\begin{center}
\small\bfseries DeepSeek V4 Flash 0731 \ensuremath{<} DeepSeek V4 Pro 0813 \ensuremath{<} Kimi K3 \ensuremath{<} GLM-5.3
\end{center}

After removing DBT, V4 Pro outperforms Kimi K3, while Kimi K3 and GLM-5.3 differ by only 2.4\%. Their compositions differ: Kimi K3 has 0.047009~s more GPU time, and GLM-5.3 has 0.076848~s more communication, partially offsetting each other. V4 Pro is 13.4\% faster than Kimi K3, and Flash is 66.5\% faster than GLM-5.3.

Across dimensions, the common low-time region is TP2EP16 + ring\_direct; Topologies~6, 1, and 2 are the leading candidates. For inter-node networks, IB HPCC and RoCE HPCC have similar means, while RoCE HPCC outperforms RoCE DCQCN on average.

\hypertarget{ux7ec4ux5408ux654fux611fux5ea6ux6a2aux5411ux5bf9ux6bd4}{%
\subsubsection{Cross-Model Configuration Sensitivity}\label{ux7ec4ux5408ux654fux611fux5ea6ux6a2aux5411ux5bf9ux6bd4}}

Direct comparison of each model's minimum and maximum, without marginal averaging, exposes configuration sensitivity. Every minimum is Topology 6 + TP2EP16 + ring\_direct (Topology~6's network label has no inter-node interpretation), whereas worst-case combinations and sensitivity ratios differ substantially:

\begin{longtable}[]{@{}
  >{\raggedright\arraybackslash}p{(\columnwidth - 10\tabcolsep) * \real{0.1429}}
  >{\raggedright\arraybackslash}p{(\columnwidth - 10\tabcolsep) * \real{0.1429}}
  >{\raggedleft\arraybackslash}p{(\columnwidth - 10\tabcolsep) * \real{0.1905}}
  >{\raggedright\arraybackslash}p{(\columnwidth - 10\tabcolsep) * \real{0.1429}}
  >{\raggedleft\arraybackslash}p{(\columnwidth - 10\tabcolsep) * \real{0.1905}}
  >{\raggedleft\arraybackslash}p{(\columnwidth - 10\tabcolsep) * \real{0.1905}}@{}}
\caption{Minimum/maximum configurations and configuration sensitivity}\label{tbl:sensitivity}\tabularnewline
\toprule\noalign{}
\begin{minipage}[b]{\linewidth}\raggedright
Model
\end{minipage} & \begin{minipage}[b]{\linewidth}\raggedright
Minimum-time combination
\end{minipage} & \begin{minipage}[b]{\linewidth}\raggedleft
Minimum wall time (s)
\end{minipage} & \begin{minipage}[b]{\linewidth}\raggedright
Worst combination
\end{minipage} & \begin{minipage}[b]{\linewidth}\raggedleft
Maximum wall time (s)
\end{minipage} & \begin{minipage}[b]{\linewidth}\raggedleft
Sensitivity (max/min)
\end{minipage} \\
\midrule\noalign{}
\endfirsthead
\toprule\noalign{}
\begin{minipage}[b]{\linewidth}\raggedright
Model
\end{minipage} & \begin{minipage}[b]{\linewidth}\raggedright
Minimum-time combination
\end{minipage} & \begin{minipage}[b]{\linewidth}\raggedleft
Minimum wall time (s)
\end{minipage} & \begin{minipage}[b]{\linewidth}\raggedright
Worst combination
\end{minipage} & \begin{minipage}[b]{\linewidth}\raggedleft
Maximum wall time (s)
\end{minipage} & \begin{minipage}[b]{\linewidth}\raggedleft
Sensitivity (max/min)
\end{minipage} \\
\midrule\noalign{}
\endhead
\bottomrule\noalign{}
\endlastfoot
DeepSeek V4 Flash 0731 & Topology 6+\allowbreak{}TP2EP16+\allowbreak{}ring & 0.105609 & Topology 3+\allowbreak{}TP16EP2+\allowbreak{}dbt+\allowbreak{}DCQCN & 2.172640 & 20.6$\times$ \\
Kimi K3 & Topology 6+\allowbreak{}TP2EP16+\allowbreak{}ring & 0.227998 & Topology 3+\allowbreak{}TP16EP2+\allowbreak{}dbt+\allowbreak{}DCQCN & 3.554893 & 15.6$\times$ \\
GLM-5.3 & Topology 6+\allowbreak{}TP2EP16+\allowbreak{}ring & 0.339574 & Topology 3+\allowbreak{}TP16EP2+\allowbreak{}dbt+\allowbreak{}DCQCN & 5.102144 & 15.0$\times$ \\
DeepSeek V4 Pro 0813 & Topology 6+\allowbreak{}TP2EP16+\allowbreak{}ring & 0.342243 & Topology 3+\allowbreak{}TP16EP2+\allowbreak{}dbt+\allowbreak{}DCQCN & 4.259608 & 12.4$\times$ \\
\end{longtable}

Flash has the shortest absolute time but highest sensitivity (20.6$\times$); V4 Pro has the widest hidden dimension but lowest sensitivity (12.4$\times$). Sensitivity is therefore nonmonotonic in absolute completion time: a short GPU baseline magnifies slow configurations, whereas a high communication baseline reduces relative dispersion. This metric supplements mean rankings with within-design-space variability.

\hypertarget{ux540cux914dux7f6eux9010ux9879ux914dux5bf9ux4e0eux6a21ux578bux6392ux5e8fux7a33ux5b9aux6027}{%
\subsection{Matched-Configuration Analysis and Ranking Stability}\label{ux540cux914dux7f6eux9010ux9879ux914dux5bf9ux4e0eux6a21ux578bux6392ux5e8fux7a33ux5b9aux6027}}

\hypertarget{ux914dux5bf9ux76eeux7684ux4e0eux7edfux8ba1ux53e3ux5f84}{%
\subsubsection{Purpose and Statistical Basis}\label{ux914dux5bf9ux76eeux7684ux4e0eux7edfux8ba1ux53e3ux5f84}}

Section~4.5 compares marginal means but cannot establish whether model rankings persist under identical system configurations. We therefore align the four models on each (topology, TP/EP, network, algorithm) key, yielding 192 complete matches. Models are ranked by completion time within each match, and rank distributions and pairwise wins are counted. Controlling system configuration exposes model--algorithm interactions hidden by overall means.

The matched data use all 192 original observations per model, with CC\_MODE=0 for every IB CC0 run. Topology~6 modes are independent simulations but cannot compare IB and RoCE because no external network is present. This analysis concerns workload completion-time rankings only, not model capability, output quality, or per-token throughput.

\hypertarget{ux540dux6b21ux5206ux5e03}{%
\subsubsection{Rank Distribution}\label{ux540dux6b21ux5206ux5e03}}

\begin{longtable}[]{@{}lrrrrr@{}}
\caption{Rank distribution across 192 matched configurations}\label{tbl:rank-distribution}\tabularnewline
\toprule\noalign{}
Model & First & Second & Third & Fourth & Mean rank \\
\midrule\noalign{}
\endfirsthead
\toprule\noalign{}
Model & First & Second & Third & Fourth & Mean rank \\
\midrule\noalign{}
\endhead
\bottomrule\noalign{}
\endlastfoot
DeepSeek V4 Flash 0731 & 192 & 0 & 0 & 0 & 1.000 \\
Kimi K3 & 0 & 120 & 42 & 30 & 2.531 \\
DeepSeek V4 Pro 0813 & 0 & 69 & 118 & 5 & 2.667 \\
GLM-5.3 & 0 & 3 & 32 & 157 & 3.802 \\
\end{longtable}

Flash is fastest in all 192 matches, giving its shorter workload a stable absolute-time advantage. GLM-5.3 ranks fourth in 157 configurations but second or third in 35, so its highest overall mean does not make it slowest everywhere. Similar mean ranks for Kimi K3 and V4 Pro suggest that their ordering varies with algorithm and parallel strategy.

\hypertarget{ux4e24ux4e24ux80dcux8d1fux4e0eux6761ux4ef6ux6027ux7ffbux8f6c}{%
\subsubsection{Pairwise Wins and Conditional Reversals}\label{ux4e24ux4e24ux80dcux8d1fux4e0eux6761ux4ef6ux6027ux7ffbux8f6c}}

\begin{longtable}[]{@{}lrrr@{}}
\caption{Pairwise model comparisons by configuration}\label{tbl:pairwise-wins}\tabularnewline
\toprule\noalign{}
Model pair & First faster & Second faster & First win rate \\
\midrule\noalign{}
\endfirsthead
\toprule\noalign{}
Model pair & First faster & Second faster & First win rate \\
\midrule\noalign{}
\endhead
\bottomrule\noalign{}
\endlastfoot
Flash vs Kimi K3 & 192 & 0 & 100.0\% \\
Flash vs DeepSeek V4 Pro 0813 & 192 & 0 & 100.0\% \\
Flash vs GLM-5.3 & 192 & 0 & 100.0\% \\
Kimi K3 vs DeepSeek V4 Pro 0813 & 121 & 71 & 63.0\% \\
Kimi K3 vs GLM-5.3 & 161 & 31 & 83.9\% \\
DeepSeek V4 Pro 0813 vs GLM-5.3 & 185 & 7 & 96.4\% \\
\end{longtable}

Flash's lead is not tied to one network or collective setting. Kimi K3 and V4 Pro, however, reverse conditionally. Among 96 ring\_direct configurations, V4 Pro wins 60 and Kimi K3 36; among 96 dbt\_direct configurations, Kimi K3 wins 85 to 11. At TP16EP2 they are closer, leading in 23 and 25 configurations, respectively. Thus, the overall relation Kimi\ K3\ \textless{}\ DeepSeek\ V4\ Pro is an aggregate over algorithm and TP/EP distributions, not a configuration-invariant ordering.

Excluding Topology~6 leaves 160 external-network matches. Flash remains fastest in all 160; pairwise wins are 93:67 for Kimi K3 versus V4 Pro, 133:27 for Kimi K3 versus GLM-5.3, and 157:3 for V4 Pro versus GLM-5.3. Directions match the 192-configuration results, so the principal paired conclusions do not depend on Topology~6.

This motivates the paired analysis: Section~4.5 identifies which model is faster on average, whereas Section~4.6 establishes how often that ordering holds under identical configurations and when it reverses.

\hypertarget{ux7efcux5408ux7ed3ux679cux7a33ux5065ux4e3bux6548ux5e94ux6761ux4ef6ux6027ux4ea4ux4e92ux4e0eux8bc1ux636eux8fb9ux754c}{%
\subsection{Synthesis: Consistent Main Effects and Conditional Interactions}\label{ux7efcux5408ux7ed3ux679cux7a33ux5065ux4e3bux6548ux5e94ux6761ux4ef6ux6027ux4ea4ux4e92ux4e0eux8bc1ux636eux8fb9ux754c}}

\begin{figure}[p]
  \centering
  \includegraphics[width=\textwidth,height=0.82\textheight,keepaspectratio]{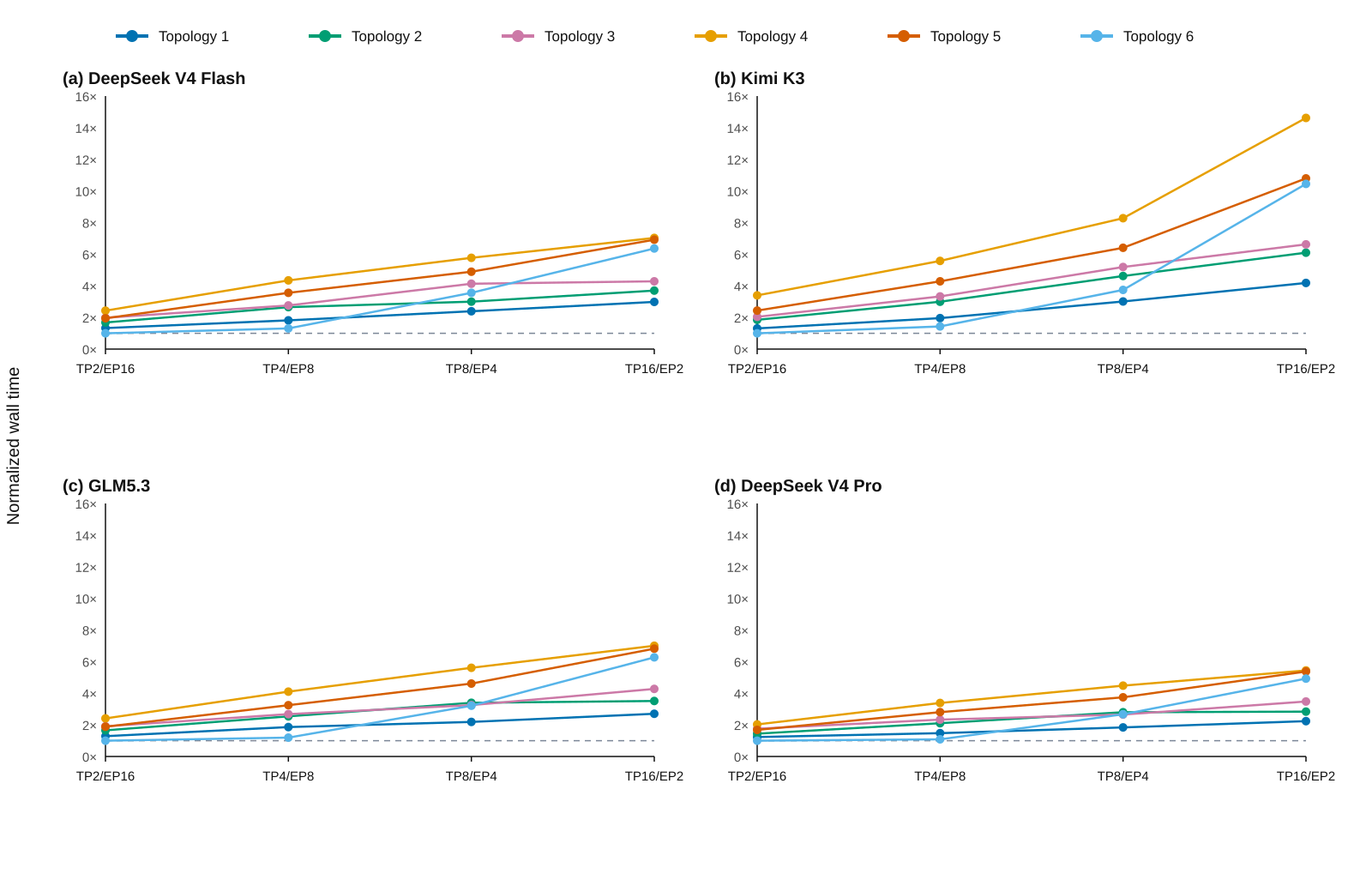}
  \caption{Interaction between topology and TP/EP strategy across the four models}
  \label{fig:topology-tp-interaction}
\end{figure}

\begin{figure}[tbp]
  \centering
  \includegraphics[width=\textwidth,height=0.72\textheight,keepaspectratio]{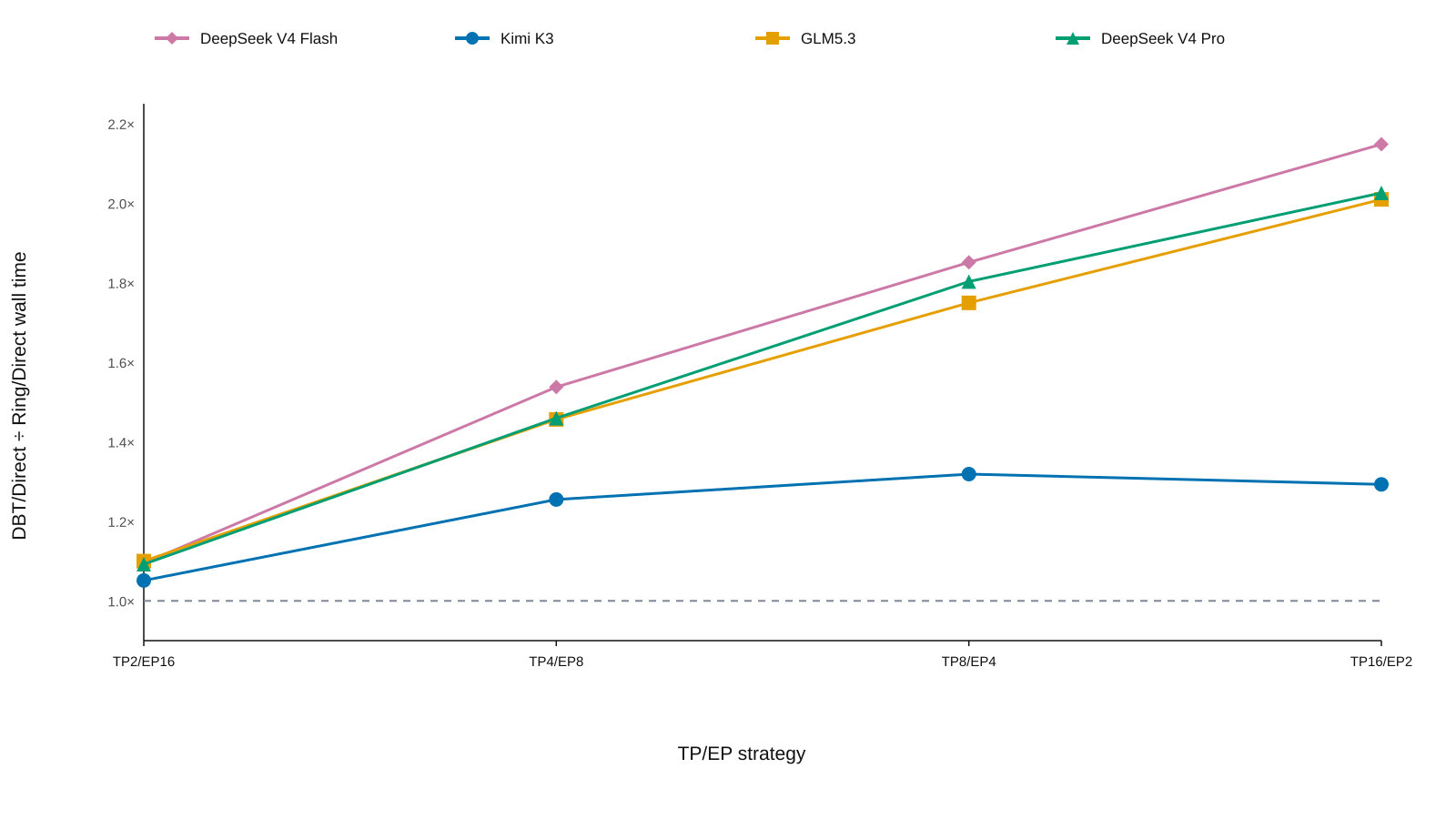}
  \caption{Interaction between collective algorithm and TP/EP strategy across the four models}
  \label{fig:collective-tp-interaction}
\end{figure}

\begin{figure}[tbp]
  \centering
  \includegraphics[width=\textwidth,height=0.72\textheight,keepaspectratio]{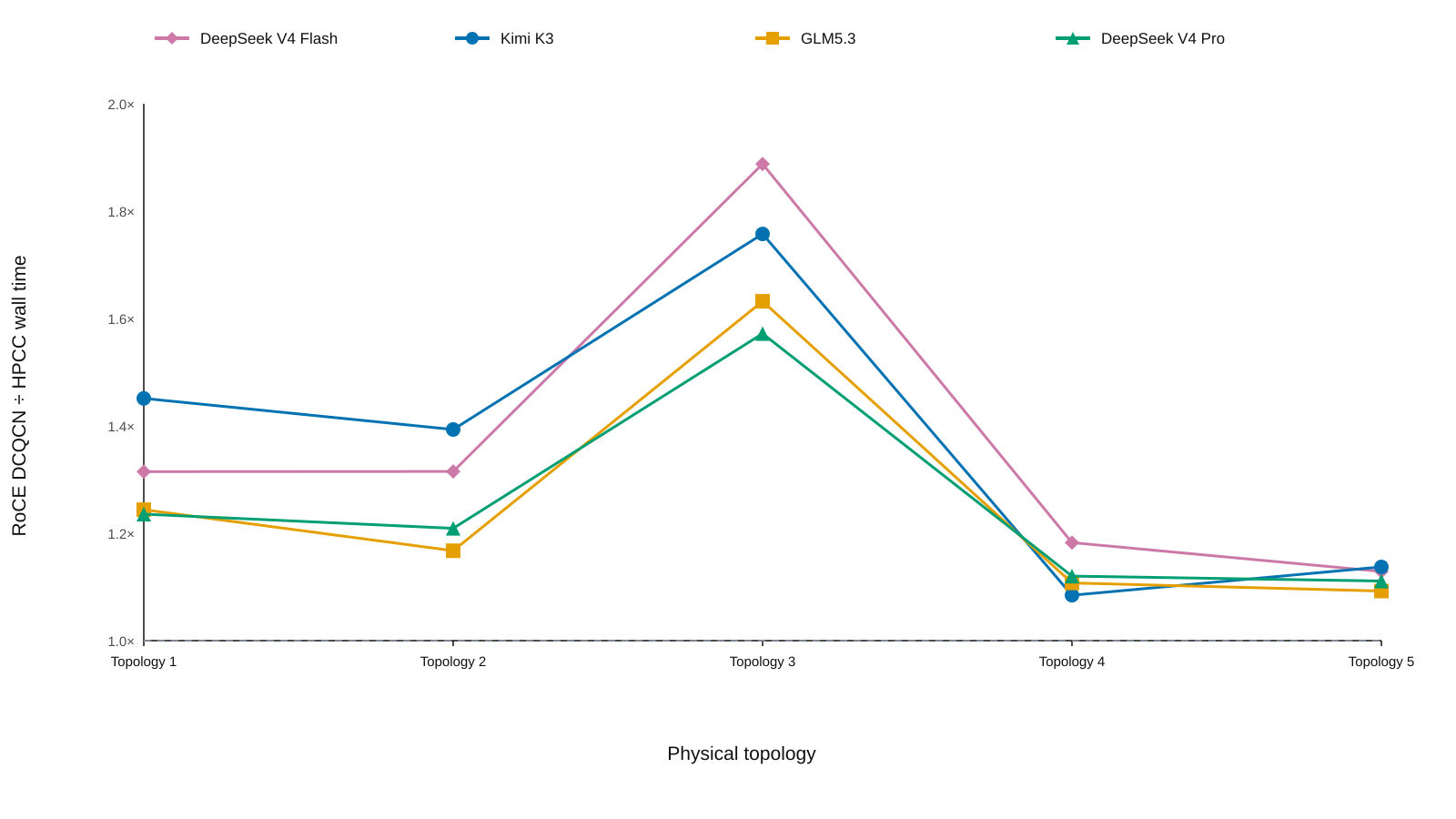}
  \caption{Interaction between congestion control and physical topology across the four models}
  \label{fig:cc-topology-interaction}
\end{figure}

The results fall into two classes: directionally consistent main effects across the four model configurations, and conditional interactions among topology, parallelism, collective algorithm, and congestion control. Consistency is defined within the present deterministic matrix and characterizes relative effects under common workload and system parameters. Table~\ref{tbl:robust-effects} summarizes the first class; Figures~\ref{fig:topology-tp-interaction}--\ref{fig:cc-topology-interaction} show three central interactions from the second.

\begin{longtable}[]{@{}
  >{\raggedright\arraybackslash}p{(\columnwidth - 6\tabcolsep) * \real{0.2500}}
  >{\raggedright\arraybackslash}p{(\columnwidth - 6\tabcolsep) * \real{0.2500}}
  >{\raggedright\arraybackslash}p{(\columnwidth - 6\tabcolsep) * \real{0.2500}}
  >{\raggedright\arraybackslash}p{(\columnwidth - 6\tabcolsep) * \real{0.2500}}@{}}
\caption{Directionally consistent main effects and their conditions}\label{tbl:robust-effects}\tabularnewline
\toprule\noalign{}
\begin{minipage}[b]{\linewidth}\raggedright
Factor
\end{minipage} & \begin{minipage}[b]{\linewidth}\raggedright
Common observation
\end{minipage} & \begin{minipage}[b]{\linewidth}\raggedright
Quantitative range
\end{minipage} & \begin{minipage}[b]{\linewidth}\raggedright
Conditions
\end{minipage} \\
\midrule\noalign{}
\endfirsthead
\toprule\noalign{}
\begin{minipage}[b]{\linewidth}\raggedright
Factor
\end{minipage} & \begin{minipage}[b]{\linewidth}\raggedright
Common observation
\end{minipage} & \begin{minipage}[b]{\linewidth}\raggedright
Quantitative range
\end{minipage} & \begin{minipage}[b]{\linewidth}\raggedright
Conditions
\end{minipage} \\
\midrule\noalign{}
\endhead
\bottomrule\noalign{}
\endlastfoot
TP/EP & TP2EP16\ \textless{}\ TP4EP8\ \textless{}\ TP8EP4\ \textless{}\ TP16EP2 & TP16EP2 is 3.68--4.35$\times$ TP2EP16 & DP=PP=1, EP Direct, 32 ranks, actual-weight capacity basis \\
TP collective & ring\_direct always outperforms dbt\_direct & DBT adds 28.3\%--83.2\% & Present ASTRA-sim DBT semantics and rank mapping only \\
Network/control & IB HPCC is close to RoCE HPCC; RoCE HPCC outperforms DCQCN & Excluding Topology 6: IB vs RoCE HPCC $-2.9$\%--$+0.1$\%; DCQCN adds 23.8\%--35.7\% & Present single-switch 400-Gbps queue/feedback settings only \\
Topology & Topologies 6, 1, and 2 rank consistently well & Topology 6 leads at low TP but joins the slow group at high TP; Topology 4 is usually slower & Depends strongly on TP/EP, placement, and used PCIe/NIC egresses \\
Model ranking & Flash is fastest in all 192 matches & Kimi K3 vs Pro pairwise wins: 121:71 & Synthetic-workload completion time, not model quality \\
\end{longtable}

TP/EP is the largest and most stable source of variation: TP2EP16 to TP16EP2 produces multi-fold increases for every model, and reduced computation cannot offset additional communication. Collective choice is second: Ring always beats DBT in marginal means, although its 28.3\%--83.2\% advantage varies with message sequence and model architecture. Network effects occur at two scales: IB HPCC versus RoCE HPCC differs by only $-2.9$\%--$+0.1$\%, whereas RoCE DCQCN adds 23.8\%--35.7\% over RoCE HPCC. This remains smaller than the high-TP amplification and is physically meaningful only for external-network topologies.

The next three figures cover topology $\times$ parallelism, collective $\times$ parallelism, and congestion control $\times$ topology. Ratios are computed from arithmetic means of completion time under each condition, not by averaging per-configuration ratios. Each configuration is one deterministic observation, so no confidence intervals are shown; the figures describe systematic interactions in this design space rather than stochastic population effects.

Topology cannot be ranked independently of parallelism. Figure~\ref{fig:topology-tp-interaction} fixes RoCE HPCC + ring\_direct and normalizes each model by its Topology~6/TP2EP16 time. At TP16EP2, Topology~6 grows to 4.93--10.44$\times$, above Topology~1's 2.24--4.18$\times$, and joins Topologies~4 and 5 in the slow group; Topology~4 reaches 5.44--14.61$\times$. Topology~6 loses its consolidation advantage at high TP without consistently becoming slowest. Removing inter-node communication eliminates external-network cost, not long or concentrated paths across intra-node PCIe switches.

Algorithm effects are not fixed multipliers. Figure~\ref{fig:collective-tp-interaction} computes mean DBT/Direct-to-Ring/Direct ratios within the RoCE HPCC/DCQCN subset. Flash, GLM-5.3, and V4 Pro grow from 1.09--1.10 at TP2EP16 to 2.01--2.15 at TP16EP2; Kimi K3 grows from 1.05 to 1.32 at TP8EP4, then declines to 1.29. Together with matched results, TP collective choice can reverse Kimi K3 versus V4 Pro, so theoretical phase complexity or a single-model mean is insufficient.

Network effects concentrate on topologies with shared egress or additional host paths. Figure~\ref{fig:cc-topology-interaction} excludes Topology~6, averages four TP/EP and two collectives within each model--topology--control cell, and computes DCQCN/HPCC. Topology~3 is most sensitive (1.57--1.89); Topologies~4--5 are 1.09--1.18, and Topologies~1--2 are 1.17--1.45. DCQCN's penalty is thus nonmonotonic in NIC or node count and jointly determined by UPI, shared egress, and logical-to-physical mapping. Congestion control regulates existing contention but cannot remove structural hot spots.

Overall, TP2EP16 + ring\_direct gives low completion time for every model and is the preferred starting point for this 32-GPU MoE system. Topology~6 benefits primarily at low TP; Topologies~1 and 2 rank consistently well among multi-node designs. With an external network, IB HPCC and RoCE HPCC means are close, and RoCE HPCC consistently outperforms RoCE DCQCN.

With fixed rank mapping and EP Direct semantics, reducing TP scope, using Ring, and avoiding concentration of logical edges on a few physical egresses lowers exposed communication. Preliminary Kimi K3 screening further showed that EP Direct reduces EP Ring's serial forwarding phases; the main study therefore fixes EP Direct and compares the mappings of TP Ring and TP DBT.

\hypertarget{ux8ba8ux8bbaux901aux4fe1ux7ed3ux6784ux4e0eux7269ux7406ux6620ux5c04ux7684ux8026ux5408ux673aux5236}{%
\section{Discussion: Coupling between Communication Structure and Physical Mapping}\label{ux8ba8ux8bbaux901aux4fe1ux7ed3ux6784ux4e0eux7269ux7406ux6620ux5c04ux7684ux8026ux5408ux673aux5236}}

\hypertarget{tp-ux96c6ux5408ux901aux4fe1ux7684ux6269ux5c55ux4ee3ux4ef7ux4e3bux5bfcux5e76ux884cux7b56ux7565ux6392ux5e8f}{%
\subsection{TP Collective Scaling Dominates Parallel-Strategy Rankings}\label{tp-ux96c6ux5408ux901aux4fe1ux7684ux6269ux5c55ux4ee3ux4ef7ux4e3bux5bfcux5e76ux884cux7b56ux7565ux6392ux5e8f}}

The observed TP2EP16\ \textless{}\ TP4EP8\ \textless{}\ TP8EP4\ \textless{}\ TP16EP2 arises from how TP and EP are implemented on the critical path. An ideal Ring AllReduce over $N$ ranks has $2(N-1)$ Reduce-Scatter/All-Gather phases and transfers approximately $2(N-1)M/N$ per rank. Although volume approaches only $2M$, serial phases, synchronization points, and edges crossing physical domains continue to grow. High TP therefore manifests less as proportional byte growth than as longer critical paths, repeated local-link occupancy, and waits for slow ranks.

EP behaves differently because expert exchange is approximated as Direct. With fixed per-layer token activation, a larger EP group primarily adds peers and does not necessarily increase per-rank payload proportionally; flows to different ranks can progress concurrently, so EP16 lacks TP16 Ring's serial-phase amplification. DeepSpeed-MoE, Tutel, and MegaBlocks show that expert layout, dynamic routing, load balance, and sparse kernels jointly govern MoE efficiency\cite{ref29,ref30,ref31}, while multidimensional inference sharding trades computation, memory, and communication\cite{ref32}. Our result is therefore that smaller TP groups are preferable under Direct EP, fixed mapping, and uniform routing---not that larger EP is universally better.

Under uniform routing, enlarging EP adds peers whose transfers can progress concurrently, whereas enlarging TP adds Ring phases and synchronization points. Low TP/high EP therefore forms a consistent low-time region across all four models.

\hypertarget{ux96c6ux5408ux901aux4fe1ux7b97ux6cd5ux7684ux5b9eux9645ux6027ux80fdux53d6ux51b3ux4e8eux5b9eux73b0ux8bedux4e49ux4e0eux7269ux7406ux6620ux5c04}{%
\subsection{Collective Performance Depends on Implementation Semantics and Mapping}\label{ux96c6ux5408ux901aux4fe1ux7b97ux6cd5ux7684ux5b9eux9645ux6027ux80fdux53d6ux51b3ux4e8eux5b9eux73b0ux8bedux4e49ux4e0eux7269ux7406ux6620ux5c04}}

DBT's logarithmic phase count can reduce startup latency abstractly but does not determine completion on hierarchical PCIe/RDMA topologies. Mapped tree edges may make concurrent branches share one PCIe uplink or NIC egress, and dependencies propagate local hot spots into collective-wide waits. Ring has more phases but regular neighbor traffic and stable per-phase volume, yielding more predictable occupancy for these large messages, fixed placement, and implementation. Horovod, Blink, TACCL, and BlueConnect likewise relate collectives to links, routing, and message size\cite{ref33,ref34,ref35,ref36}. Ring's advantage here does not contradict DBT's theoretical complexity; the theoretical benefit simply fails to shorten this mapped critical path.

ring\_direct and dbt\_direct vary only TP; EP is always Direct, so their difference reflects TP algorithm and mapping. Preliminary Kimi K3 screening compared EP Ring with EP Direct to select the main-matrix EP algorithm. All runs share Chakra groups, ASTRA-sim decomposition, and the NS-3 backend, ensuring common implementation semantics.

\hypertarget{ux6709ux6548ux901aux4fe1ux80fdux529bux7531ux94feux8defux5229ux7528ux65b9ux5f0fux800cux975eux8d44ux6e90ux603bux91cfux51b3ux5b9a}{%
\subsection{Effective Communication Capacity Depends on Link Use, Not Resource Count}\label{ux6709ux6548ux901aux4fe1ux80fdux529bux7531ux94feux8defux5229ux7528ux65b9ux5f0fux800cux975eux8d44ux6e90ux603bux91cfux51b3ux5b9a}}

Neither more NICs nor fewer nodes gives monotonic speedup, because nominal resources become effective only when collective paths use them. GPU-to-NIC affinity determines injection points; rank contention for PCIe uplinks and switch ports determines local queues; collective dependencies place the slowest logical edge on the global critical path. Physically present links unused by rank mapping or routing cannot shorten completion time.

This agrees with cross-layer optimization: TopoOpt jointly searches parallelism, communication, and topology, while Alpa, FlexFlow, and Unity co-optimize device meshes, operator partitioning, placement, and graph transformations\cite{ref37,ref38,ref39,ref40}. Topology~5 raises theoretical injection capacity with four NICs and connected PCIe switches, but hot ports dominate if logical edges concentrate on selected NICs. Topology~6 removes RDMA yet not repeated TP16 Ring occupancy across a few switches. Neither 0.1-µs intra-node links nor aggregate NIC bandwidth predicts time alone; port bytes, queue timelines, and collective critical paths must also be examined.

\hypertarget{ux62e5ux585eux63a7ux5236ux53eaux80fdux8c03ux8282ux6392ux961fux8fc7ux7a0bux800cux4e0dux80fdux6d88ux9664ux7ed3ux6784ux6027ux70edux70b9}{%
\subsection{Congestion Control Regulates Queues but Cannot Remove Structural Hot Spots}\label{ux62e5ux585eux63a7ux5236ux53eaux80fdux8c03ux8282ux6392ux961fux8fc7ux7a0bux800cux4e0dux80fdux6d88ux9664ux7ed3ux6784ux6027ux70edux70b9}}

IB CC0 is often faster because it omits feedback and convergence and is not a strict upper bound under real congestion. HPCC and DCQCN use different signals and rate updates\cite{ref21,ref22}; DCTCP, TIMELY, IRN, and Swift further show that ECN, RTT, PFC dependence, and loss recovery affect queues, burst tolerance, and FCT\cite{ref41,ref42,ref43,ref44}. Similar IB-like and RoCE HPCC means indicate limited critical-path impact from the modeled link-delay difference, while RoCE HPCC's advantage over DCQCN indicates less exposed waiting under the given buffers, ECN/PFC parameters, and bursts. These are not unconditional native-device rankings. Congestion control can regulate traffic already in shared queues, but cannot change mapping-generated edges or enlist unused NIC/PCIe links; faster feedback reduces buildup but cannot remove a structurally critical egress.

\hypertarget{ux4effux771fux5b8cux6210ux65f6ux95f4ux4e0eux771fux5b9eux8bf7ux6c42ux65f6ux5ef6ux4e4bux95f4ux5b58ux5728ux7cfbux7edfux6027ux8fb9ux754c}{%
\subsection{Simulation Completion Time Is Distinct from Request Latency}\label{ux4effux771fux5b8cux6210ux65f6ux95f4ux4e0eux771fux5b9eux8bf7ux6c42ux65f6ux5ef6ux4e4bux95f4ux5b58ux5728ux7cfbux7edfux6027ux8fb9ux754c}}

Simulation completion time is the makespan of an event DAG under simulated resources; it compares configurations but is not online end-to-end latency. Production inference adds queuing, prefill, per-token decode, continuous batching, KV-cache management, kernel launches, NCCL/RDMA overhead, and arrival/output-length variation. Orca and Sarathi-Serve show that scheduling changes the throughput--latency relation\cite{ref45,ref46}. These behaviors are not fully encoded in the synthetic ET, so rankings address relative network/parallelism differences under a fixed workload, not per-token latency or service-level objectives (SLOs).

Within output precision, all configurations satisfy Wall time $\approx$ GPU time + Comm time, indicating largely serialized computation and exposed communication with limited overlap. A communication fraction above 90\% refers to the event-DAG critical path, not raw network activity in a production GPU timeline. Daydream and Proteus show that absolute prediction requires real traces to calibrate dependencies and overlap\cite{ref47,ref48}. Runtime overlap of AllReduce or expert dispatch could reduce both absolute times and inter-strategy gaps.

\hypertarget{ux5b9eux8df5ux542fux793aux90e8ux7f72ux7a7aux95f4ux88c1ux526aux4e0eux5b9eux6d4bux6821ux51c6}{%
\section{Practical Implications: Design-Space Pruning and Measurement Calibration}\label{ux5b9eux8df5ux542fux793aux90e8ux7f72ux7a7aux95f4ux88c1ux526aux4e0eux5b9eux6d4bux6821ux51c6}}

Checkpoint-weight capacity provides a common 32-GPU residency basis, shifting server selection toward GPU--NIC affinity, PCIe-switch crossings, and critical-egress load. Topology~6 removes external-network cost at low/moderate TP, but its high-TP reversal demands better intra-node mapping; multi-node, low-GPU-count designs should reduce inter-node frequency and improve NIC/GPU binding. Such joint topology, parallelism, and routing decisions agree with cross-layer work\cite{ref37,ref38,ref39,ref40}.

Within the evaluated matrix, TP2EP16 + ring\_direct is a preferred candidate because the checkpoint weights fit across 32 GPUs and the configuration is consistently fast: smaller TP reduces AllReduce phases and scope, while larger EP uses Direct communication for concurrent expert dispatch and combine.

RoCE HPCC is the preferred simulated candidate over DCQCN under current parameters, but must be recalibrated for target NICs, switch buffers, ECN/PFC thresholds, and background traffic. Placement hot spots should be removed and NIC use balanced before feedback algorithms are tuned. IB CC0 isolates feedback overhead and should not predict absolute latency on a real uncongested network; protocol selection requires measurement\cite{ref21,ref22,ref41,ref42,ref43,ref44}.

We therefore propose a closed loop of simulation screening, link diagnosis, target-cluster calibration, and online validation. Simulation prunes poor topology/TP/EP choices; port bytes and queue timelines locate hot spots; a small request set calibrates prefill/decode, overlap, and runtime overhead; and throughput, TTFT, TPOT, and P95/P99 validate production behavior. ASTRA-sim is thereby used for controlled counterfactuals, while scheduling and service quality remain properties of the real runtime\cite{ref45,ref46,ref47,ref48}.

\hypertarget{ux7814ux7a76ux5c40ux9650ux4e0eux672aux6765ux5de5ux4f5c}{%
\section{Limitations}\label{ux7814ux7a76ux5c40ux9650ux4e0eux672aux6765ux5de5ux4f5c}}

\hypertarget{ux5de5ux4f5cux8d1fux8f7dux4e0eux8fd0ux884cux65f6ux62bdux8c61ux7684ux5c40ux9650}{%
\subsection{Workload and Runtime Abstractions}\label{ux5de5ux4f5cux8d1fux8f7dux4e0eux8fd0ux884cux65f6ux62bdux8c61ux7684ux5c40ux9650}}

Workloads are generated from model configurations and approximations rather than full-lifecycle Chakra runtime traces. They support controlled comparison but do not reproduce kernel scheduling, dynamic routing, KV-cache traffic, fragmentation, expert imbalance, or routing skew. Compute is approximated from peak throughput and memory bandwidth, and $T_{wall}\approx T_{GPU}+T_{comm}$ reveals limited modeling of overlap, launches, continuous batching, PP, and scheduling. Iteration-level scheduling, chunked prefill, and dynamic batching alter production critical paths\cite{ref45,ref46}; DAG accuracy depends on trace granularity and overlap\cite{ref47,ref48}. Rankings under a common abstraction are valid, but production TTFT, TPOT, and throughput cannot be reported directly.

\hypertarget{ux7f51ux7edcux4e0eux96c6ux5408ux901aux4fe1ux5b9eux73b0ux7684ux5c40ux9650}{%
\subsection{Network and Collective Implementations}\label{ux7f51ux7edcux4e0eux96c6ux5408ux901aux4fe1ux5b9eux73b0ux7684ux5c40ux9650}}

The single-switch external network isolates endpoint injection, GPU/NIC binding, and congestion control. ASTRA-sim decomposes Ring, DBT, and Direct into point-to-point transfers; NS-3 models RDMA headers, PFC, INT, and rate feedback. Per-flow policy maps local flows to fixed local links and inter-node flows to RDMA control paths. All algorithms and topologies use the same backend and parameters, so comparisons apply to these implementations. Prior work confirms that congestion signals, losslessness, and recovery jointly determine queues and FCT\cite{ref41,ref42,ref43,ref44}.

\hypertarget{ux7ed3ux8bba}{%
\section{Conclusion}\label{ux7ed3ux8bba}}

Using ASTRA-sim/NS-3, we conduct controlled simulations of four MoE configurations across six server topologies, four TP/EP strategies, two collective algorithms, and four network/control modes. Under uniform 32-way sharding, Kimi K3 has the largest checkpoint-weight requirement at approximately 48.75~GB per rank; all four configurations satisfy the weight-residency criterion of at least 64~GB per accelerator. This establishes a common 32-accelerator capacity basis.

Exposed communication contributes 89.9\%--95.8\% of mean completion time. TP16EP2 is 3.68--4.35$\times$ TP2EP16, and the evaluated dbt\_direct implementation adds 28.3\%--83.2\% over ring\_direct under the fixed rank mapping. Excluding Topology~6 and equally aggregating the four per-model conditional means, IB HPCC time is approximately 0.9\% below RoCE HPCC, while RoCE HPCC beats RoCE DCQCN for every model. Topology interacts strongly with parallelism: Topology~6 benefits from no external network at low TP but loses its advantage at high TP because intra-node PCIe paths concentrate; Topologies~1 and 2 retain low multi-node times through distributed NIC injection.

Every model attains its minimum at Topology 6 + TP2EP16 + ring\_direct. Strong multi-node configurations combine a small TP scope, regular Ring edges, and balanced GPU--NIC mapping. Within the evaluated workload, rank mapping, and simulator semantics, MoE inference performance is a joint outcome of model communication, parallelism, intra-server interconnect, and congestion control; adding GPUs, NICs, or nominal bandwidth alone does not guarantee improvement. Controlled ASTRA-sim/NS-3 exploration can identify critical paths and structural hot spots before deployment and provide quantitative guidance for topology and parallel configuration.

\FloatBarrier

\clearpage
\appendix
\hypertarget{ux5b9eux9a8cux62d3ux6251ux4e0eux7ec4ux5408}{%
\section{Experimental Topologies and Combinations}\label{ux5b9eux9a8cux62d3ux6251ux4e0eux7ec4ux5408}}

\begin{longtable}[]{@{}
  >{\raggedright\arraybackslash}p{(\columnwidth - 6\tabcolsep) * \real{0.2308}}
  >{\raggedright\arraybackslash}p{(\columnwidth - 6\tabcolsep) * \real{0.2308}}
  >{\raggedleft\arraybackslash}p{(\columnwidth - 6\tabcolsep) * \real{0.3077}}
  >{\raggedright\arraybackslash}p{(\columnwidth - 6\tabcolsep) * \real{0.2308}}@{}}
\caption{Node configurations and key interconnects for the six topologies}\label{tbl:appendix-topologies}\tabularnewline
\toprule\noalign{}
\begin{minipage}[b]{\linewidth}\raggedright
Topology
\end{minipage} & \begin{minipage}[b]{\linewidth}\raggedright
Per-server configuration
\end{minipage} & \begin{minipage}[b]{\linewidth}\raggedleft
Cluster scale
\end{minipage} & \begin{minipage}[b]{\linewidth}\raggedright
Key interconnect
\end{minipage} \\
\midrule\noalign{}
\endfirsthead
\toprule\noalign{}
\begin{minipage}[b]{\linewidth}\raggedright
Topology
\end{minipage} & \begin{minipage}[b]{\linewidth}\raggedright
Per-server configuration
\end{minipage} & \begin{minipage}[b]{\linewidth}\raggedleft
Cluster scale
\end{minipage} & \begin{minipage}[b]{\linewidth}\raggedright
Key interconnect
\end{minipage} \\
\midrule\noalign{}
\endhead
\bottomrule\noalign{}
\endlastfoot
Topology 1 & 1 CPU, 1 GPU, 1 external NIC & 32 servers & One 400-Gbps external link per server \\
Topology 2 & 1 CPU, 2 GPUs, 1 PCIe switch, 1 external NIC & 16 servers & Two GPUs share a PCIe switch and external egress \\
Topology 3 & 2 CPUs, 2 GPUs, 1 external NIC & 16 servers & GPUs attach to separate CPUs; 400-Gbps, 2-µs UPI \\
Topology 4 & 2 CPUs, 8 GPUs, 2 PCIe switches, 2 external NICs & 4 servers & Four GPUs and one NIC per switch; 400-Gbps, 0.1-µs inter-switch links \\
Topology 5 & 2 CPUs, 16 GPUs, 4 PCIe switches, 4 external NICs & 2 servers & Four switches form a fully connected K4, each with four GPUs and one NIC \\
Topology 6 & 2 CPUs, 32 GPUs, 8 PCIe switches, no NIC & 1 server & Eight switches form a fully connected K8; intra-node traffic only \\
\end{longtable}

For Topologies~1--5, each external link operates at 400~Gbps. Endpoint-to-switch propagation delay is 0.7~µs in the IB-like modes and 1~µs in the RoCE modes; Topology~6 has no external network.

\begin{figure}[p]
  \centering
  \begin{subfigure}[t]{0.485\textwidth}
    \centering
    \includegraphics[width=\linewidth,height=0.42\textheight,keepaspectratio]{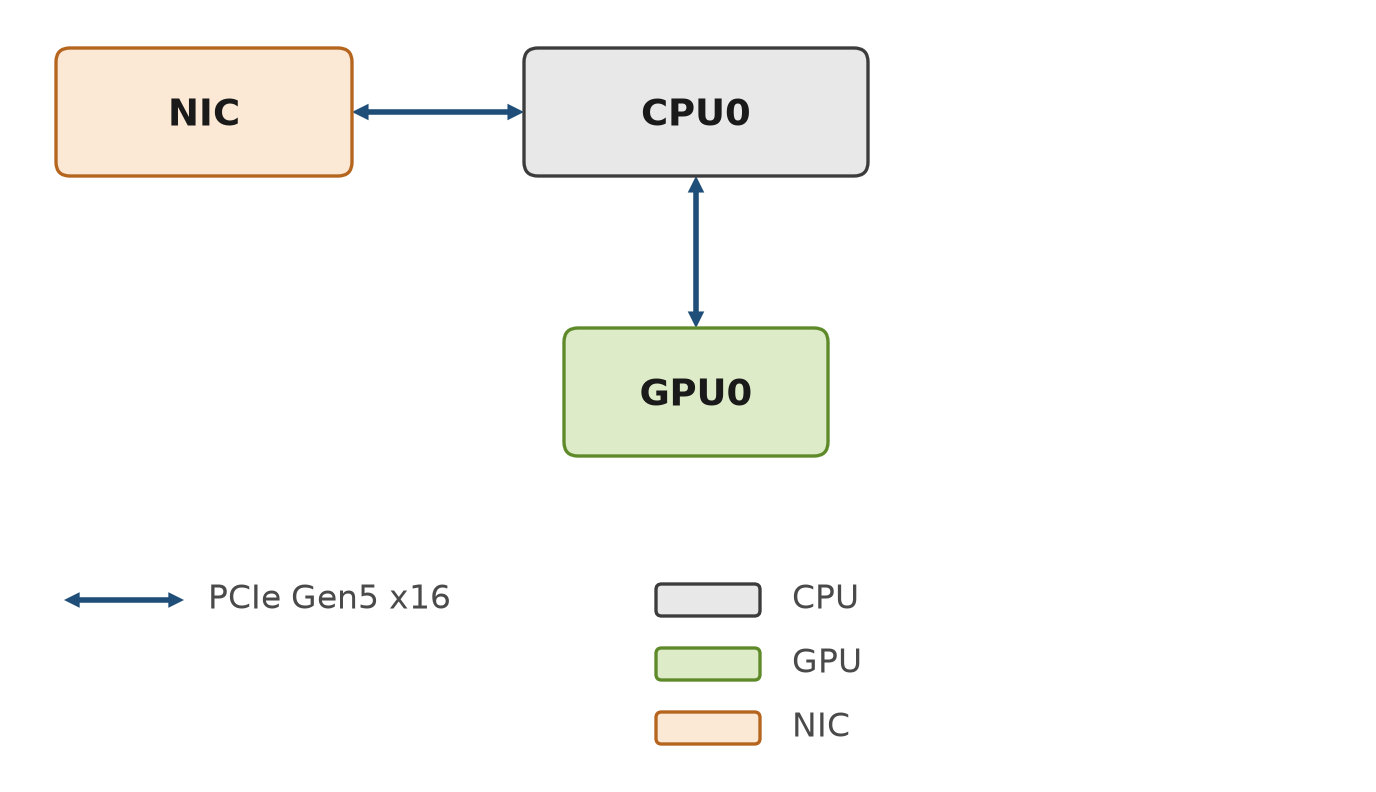}
    \caption{Intra-server organization}
    \label{fig:topology1-node}
  \end{subfigure}\hfill
  \begin{subfigure}[t]{0.485\textwidth}
    \centering
    \includegraphics[width=\linewidth,height=0.42\textheight,keepaspectratio]{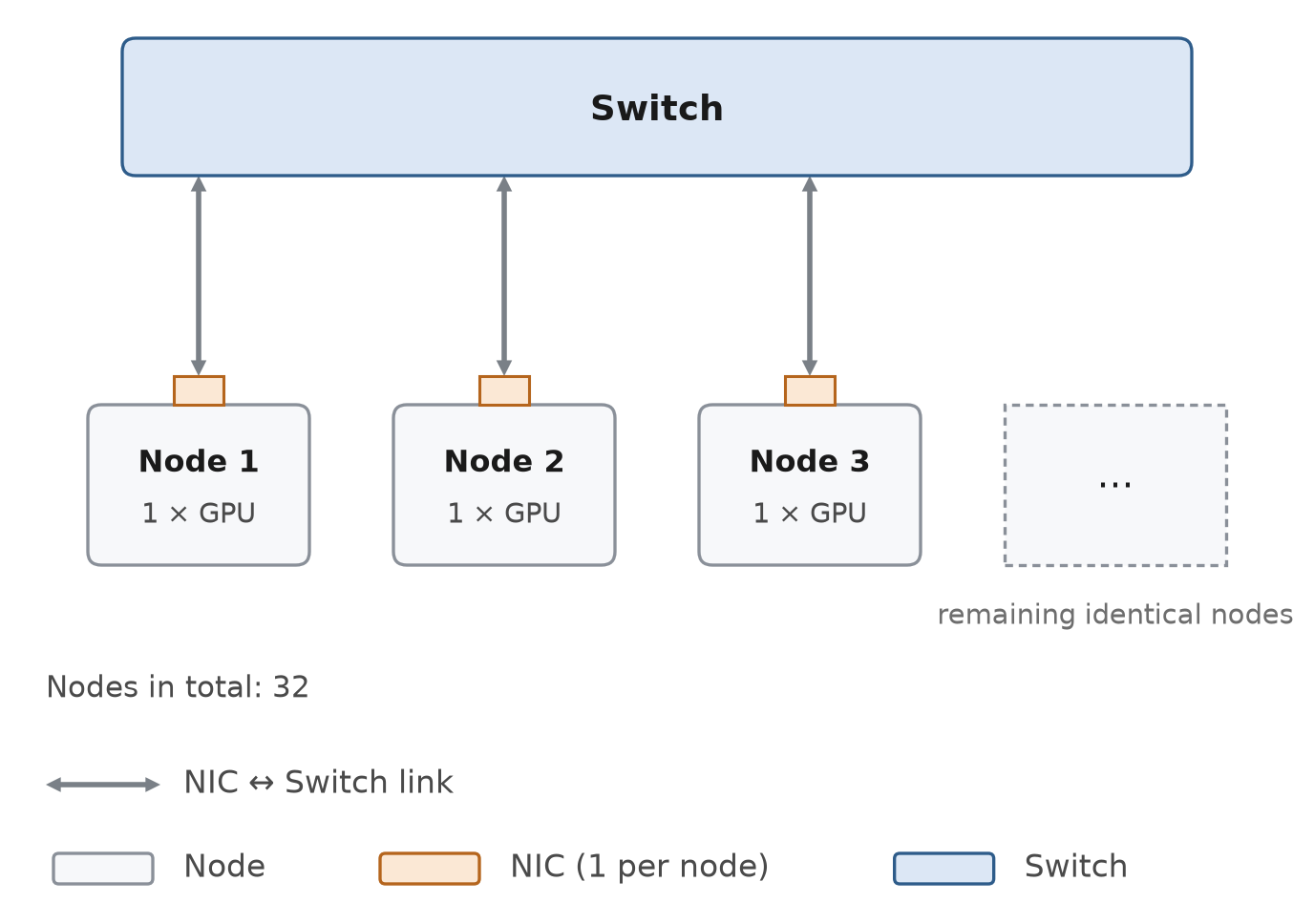}
    \caption{Cluster topology}
    \label{fig:topology1-cluster}
  \end{subfigure}
  \caption{Topology 1: 32-node cluster of single-CPU, single-GPU, single-NIC servers}
  \label{fig:topology1}
\end{figure}

\begin{figure}[p]
  \centering
  \begin{subfigure}[t]{0.485\textwidth}
    \centering
    \includegraphics[width=\linewidth,height=0.42\textheight,keepaspectratio]{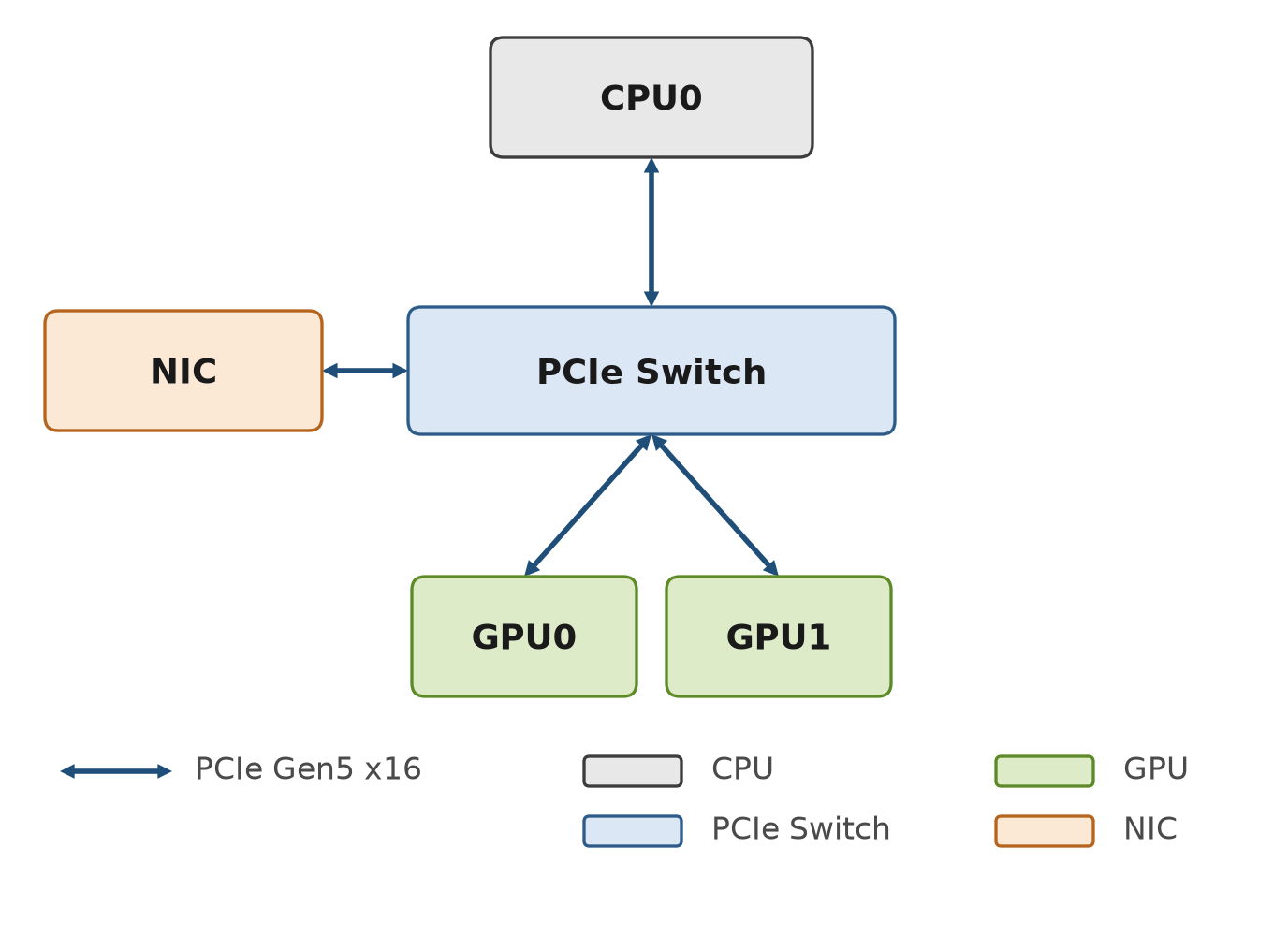}
    \caption{Intra-server organization}
    \label{fig:topology2-node}
  \end{subfigure}\hfill
  \begin{subfigure}[t]{0.485\textwidth}
    \centering
    \includegraphics[width=\linewidth,height=0.42\textheight,keepaspectratio]{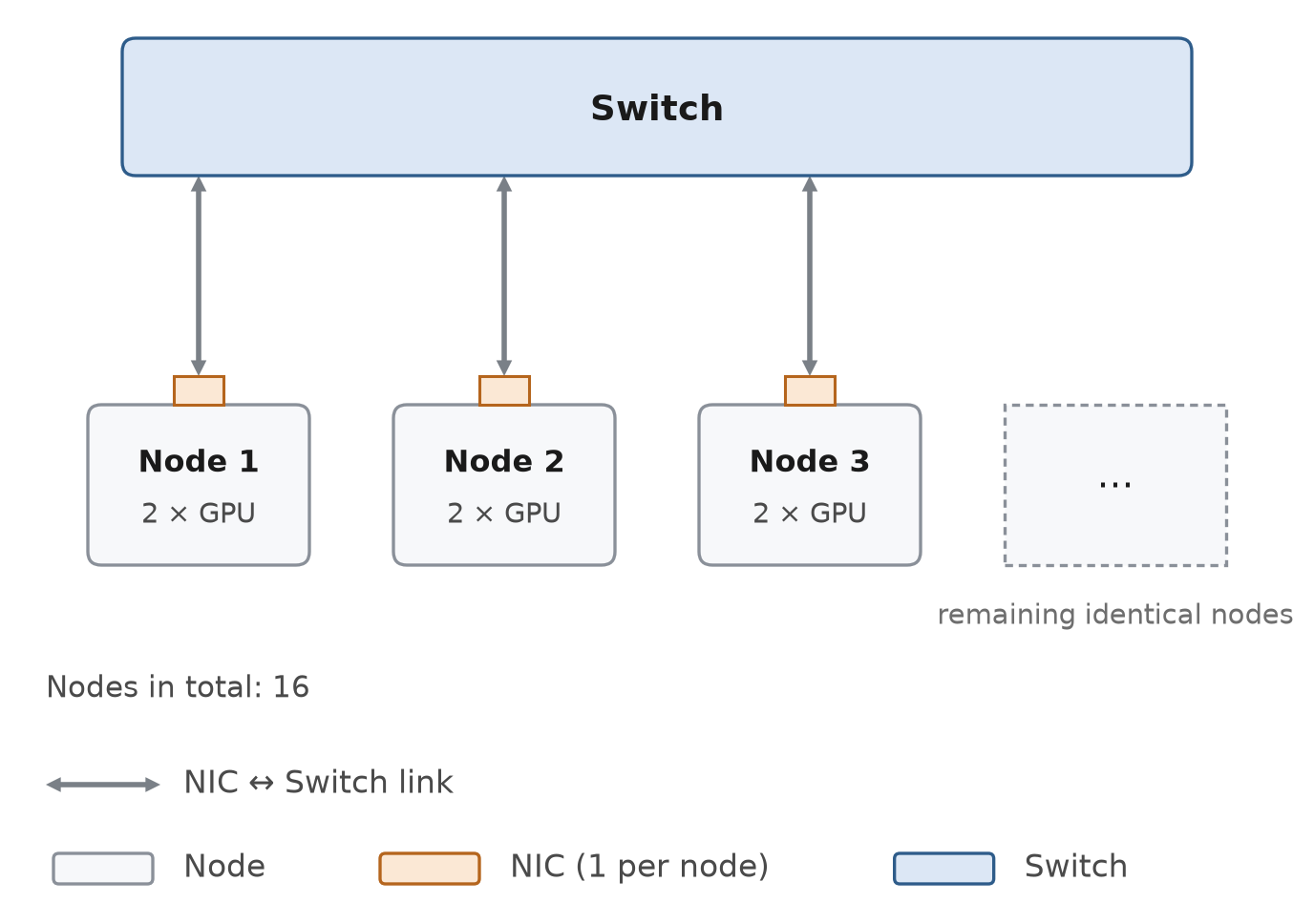}
    \caption{Cluster topology}
    \label{fig:topology2-cluster}
  \end{subfigure}
  \caption{Topology 2: 16-node cluster of single-CPU, dual-GPU PCIe-switch servers}
  \label{fig:topology2}
\end{figure}

\begin{figure}[p]
  \centering
  \begin{subfigure}[t]{0.485\textwidth}
    \centering
    \includegraphics[width=\linewidth,height=0.42\textheight,keepaspectratio]{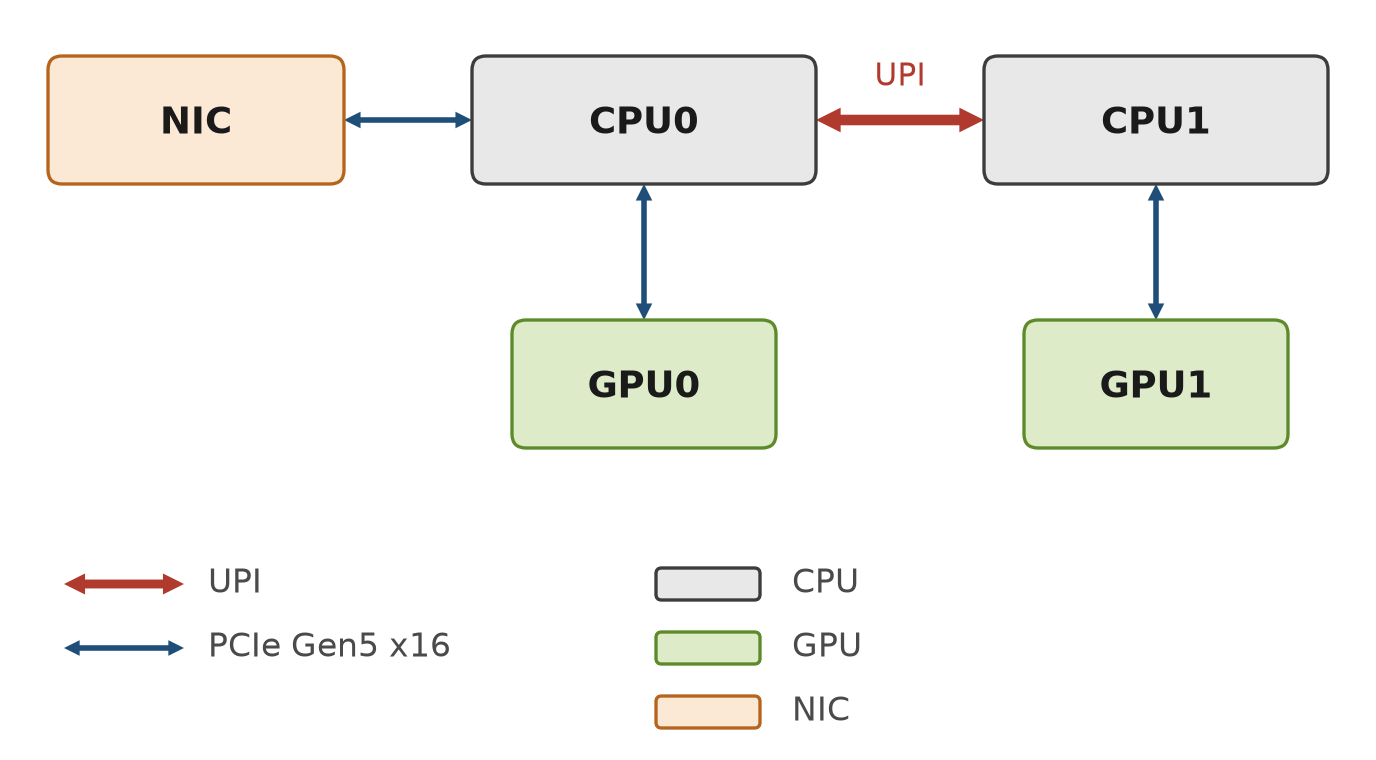}
    \caption{Intra-server organization}
    \label{fig:topology3-node}
  \end{subfigure}\hfill
  \begin{subfigure}[t]{0.485\textwidth}
    \centering
    \includegraphics[width=\linewidth,height=0.42\textheight,keepaspectratio]{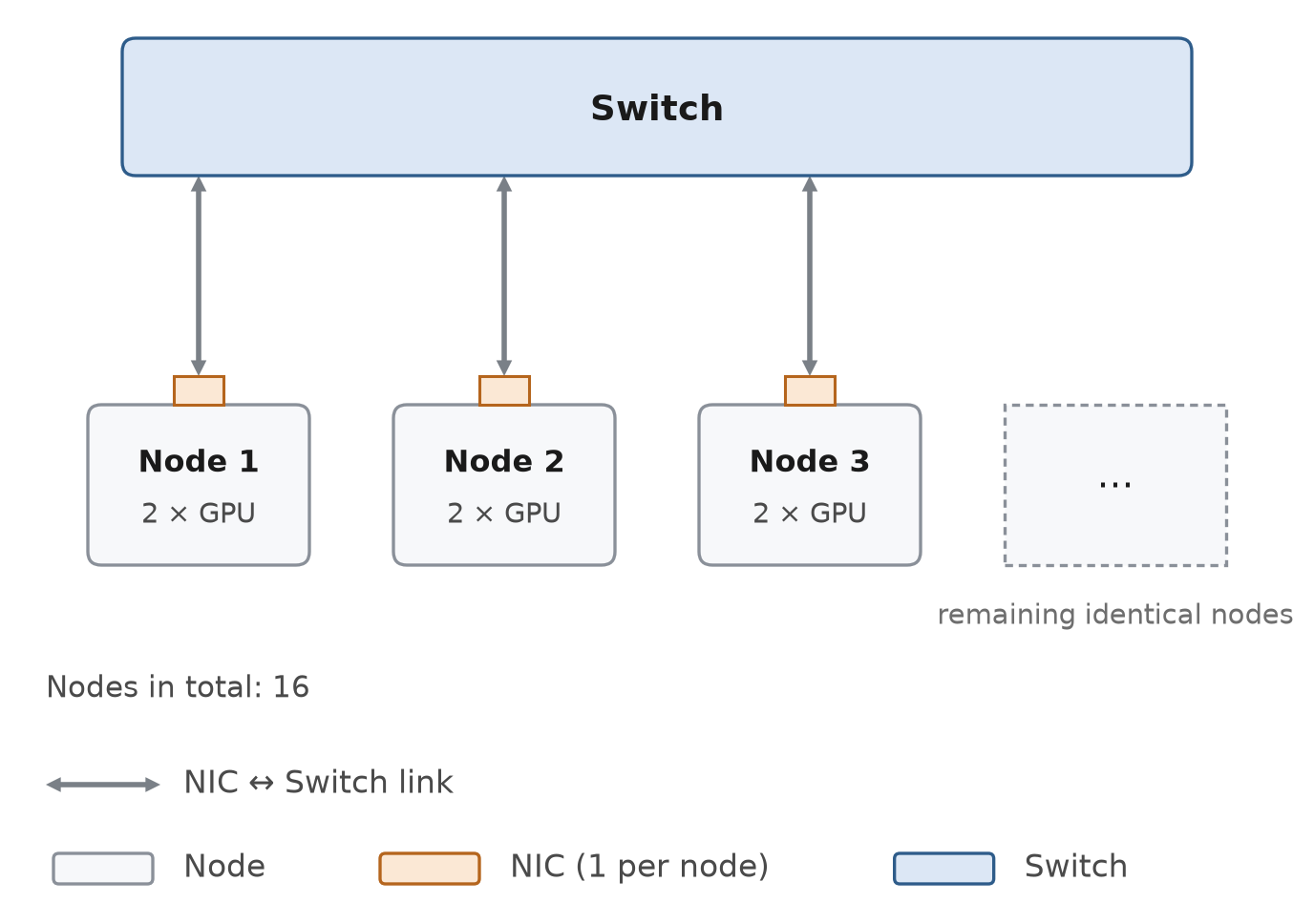}
    \caption{Cluster topology}
    \label{fig:topology3-cluster}
  \end{subfigure}
  \caption{Topology 3: 16-node cluster of dual-CPU, dual-GPU servers}
  \label{fig:topology3}
\end{figure}

\begin{figure}[p]
  \centering
  \begin{subfigure}[t]{0.485\textwidth}
    \centering
    \includegraphics[width=\linewidth,height=0.42\textheight,keepaspectratio]{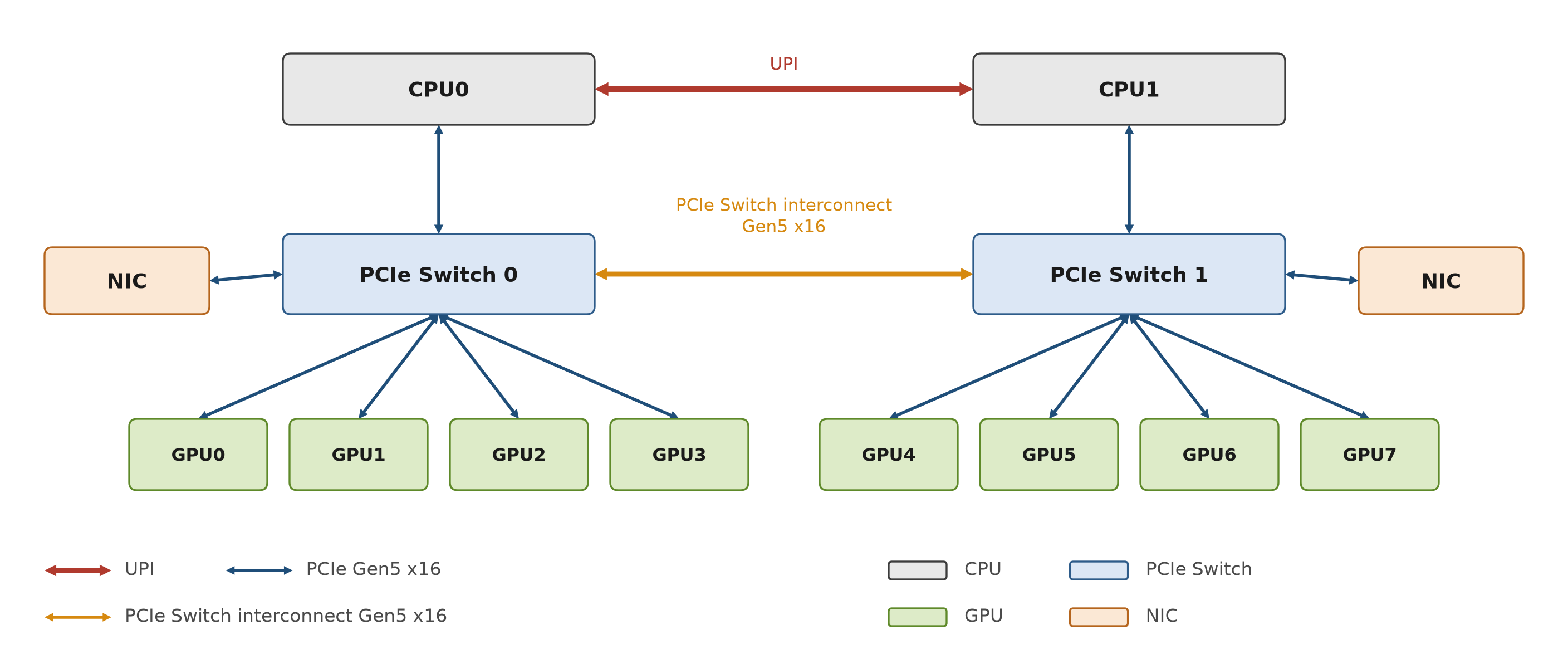}
    \caption{Intra-server organization}
    \label{fig:topology4-node}
  \end{subfigure}\hfill
  \begin{subfigure}[t]{0.485\textwidth}
    \centering
    \includegraphics[width=\linewidth,height=0.42\textheight,keepaspectratio]{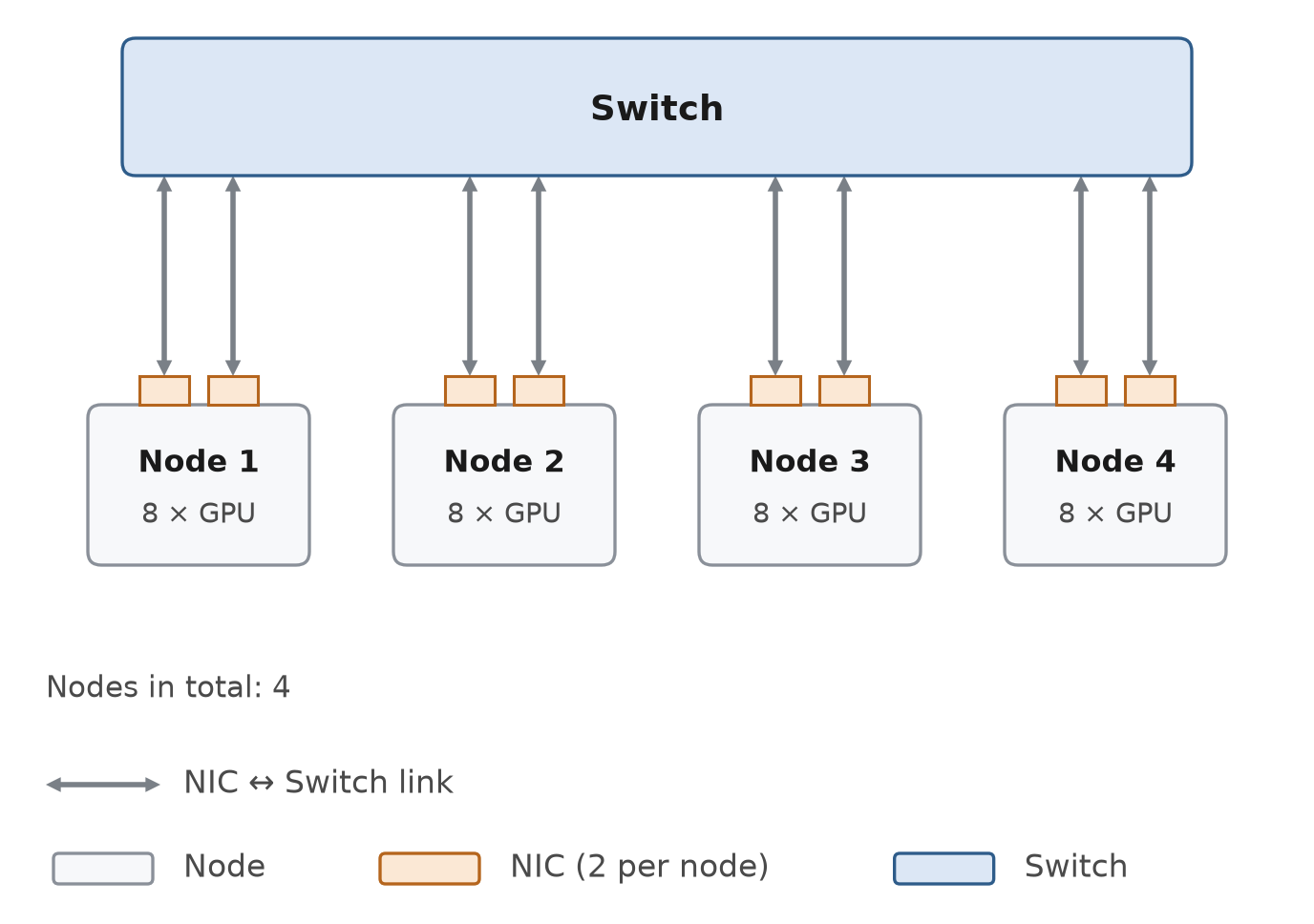}
    \caption{Cluster topology}
    \label{fig:topology4-cluster}
  \end{subfigure}
  \caption{Topology 4: four-node cluster of dual-CPU, eight-GPU, dual-NIC servers}
  \label{fig:topology4}
\end{figure}

\begin{figure}[p]
  \centering
  \begin{subfigure}[t]{0.485\textwidth}
    \centering
    \includegraphics[width=\linewidth,height=0.42\textheight,keepaspectratio]{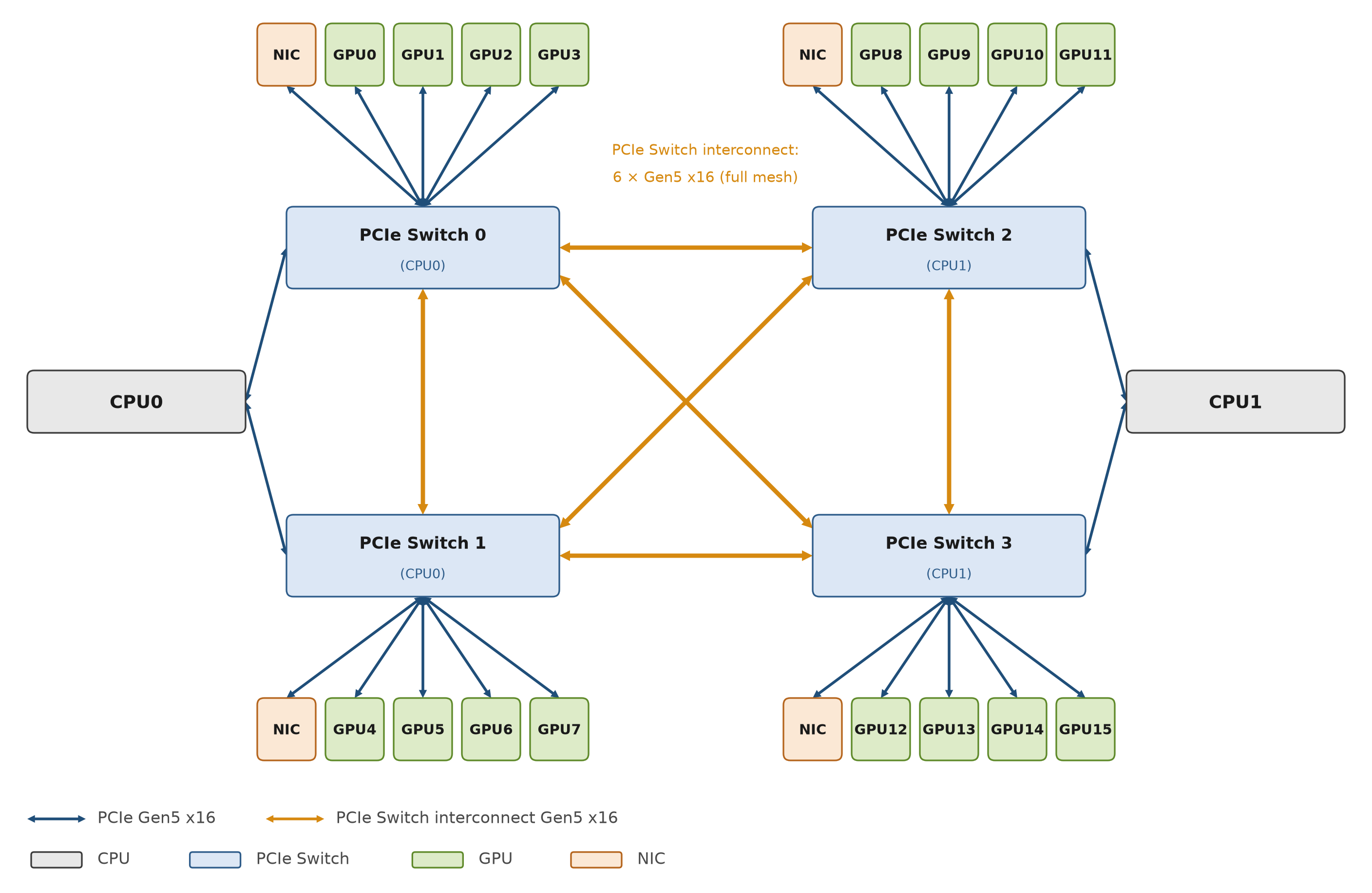}
    \caption{Intra-server organization}
    \label{fig:topology5-node}
  \end{subfigure}\hfill
  \begin{subfigure}[t]{0.485\textwidth}
    \centering
    \includegraphics[width=\linewidth,height=0.42\textheight,keepaspectratio]{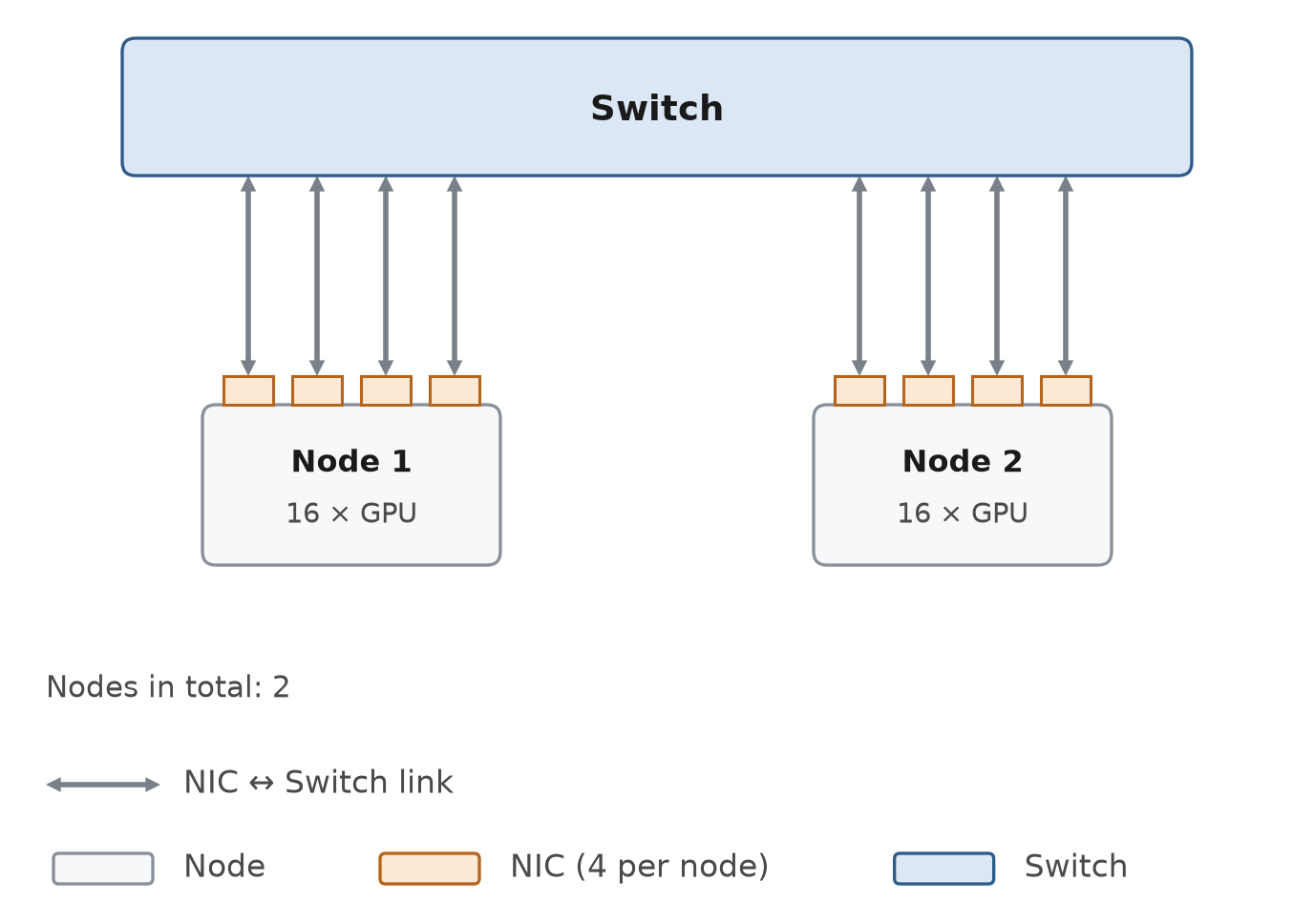}
    \caption{Cluster topology}
    \label{fig:topology5-cluster}
  \end{subfigure}
  \caption{Topology 5: two-node cluster of dual-CPU, 16-GPU, four-NIC servers}
  \label{fig:topology5}
\end{figure}

\begin{figure}[p]
  \centering
  \includegraphics[width=\textwidth,height=0.82\textheight,keepaspectratio]{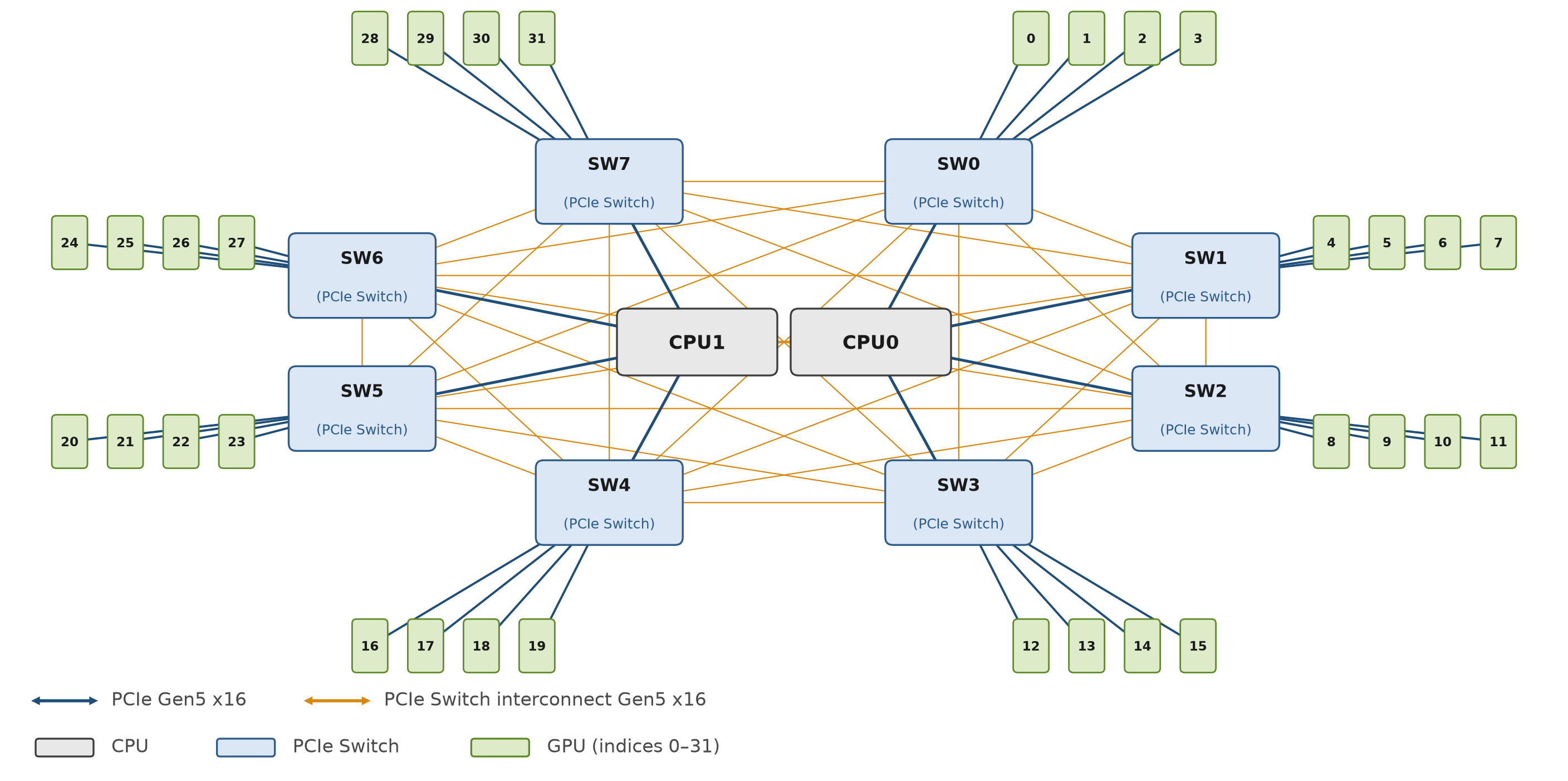}
  \caption{Topology 6: dual-CPU, 32-GPU server with eight fully connected PCIe switches}
  \label{fig:topology6}
\end{figure}

Each model includes the following combinations:

\begin{itemize}
\tightlist
\item
  Six topologies: Topologies 1--6;
\item
  Four TP/EP partitions: TP2EP16, TP4EP8, TP8EP4, and TP16EP2;
\item
  Four network/congestion modes: IB CC0, IB HPCC, RoCE HPCC, and RoCE DCQCN;
\item
  Two algorithm combinations: ring\_direct and dbt\_direct.
\end{itemize}

The theoretical number of combinations is:

\begin{center}
\small\bfseries 6 \ensuremath{\times} 4 \ensuremath{\times} 4 \ensuremath{\times} 2 = 192 configurations/model
\end{center}

Across four models:

\begin{center}
\small\bfseries 192 \ensuremath{\times} 4 = 768 configurations
\end{center}

\end{document}